\documentclass[1p]{elsarticle}

\usepackage[utf8]{inputenc}
\usepackage{bold-extra}
\usepackage{booktabs}
\usepackage{latexsym}
\usepackage{amssymb}
\usepackage{amsmath}
\usepackage{url}
\usepackage{graphicx}		
\usepackage{subfigure}
\usepackage{epsfig}		
\usepackage{epstopdf}
\usepackage{dcolumn}	
\usepackage{bm}		
\usepackage[dvipsnames]{xcolor}
\usepackage[normalem]{ulem}	
\usepackage{xspace}
\usepackage{blindtext}
\usepackage{enumitem}
   
\newcommand{\Smilei}{{\sc Smilei}\xspace}
\newcommand{\vE}{\mathbf{E}}
\newcommand{\vB}{\mathbf{B}}
\newcommand{\vJ}{\mathbf{J}}
\newcommand{\vx}{\mathbf{x}}
\newcommand{\vp}{\mathbf{p}}
\newcommand{\vv}{\mathbf{v}}
\newcommand{\vu}{\mathbf{u}}
\newcommand{\vF}{\mathbf{F}}
\newcommand{\python}{\textit{python}\xspace}

\begin{document}

\title{\Smilei: a collaborative, open-source, multi-purpose particle-in-cell code for plasma simulation}
\author[a]{J.~Derouillat}
\author[b]{A.~Beck}
\author[c]{F.~P\'erez}
\author[c]{T.~Vinci}
\author[d]{M.~Chiaramello}
\author[d,e,f]{A.~Grassi}
\author[g]{M.~Fl\'e}
\author[h]{G.~Bouchard}
\author[i]{I.~Plotnikov}
\author[j]{N.~Aunai}
\author[i,j]{J.~Dargent}
\author[d]{C.~Riconda}
\author[c]{M.~Grech\corref{mica}}

\cortext[mica] {Corresponding author.\\\textit{E-mail address:} mickael.grech@polytechnique.edu}

\address[a]{Maison de la Simulation, CEA, CNRS, Universit\'e Paris-Sud, UVSQ, Universit\'e Paris-Saclay, F-91191 Gif-sur-Yvette, France}
\address[b]{Laboratoire Leprince-Ringuet, \'Ecole polytechnique, CNRS-IN2P3, F-91128 Palaiseau, France}
\address[c]{Laboratoire d'Utilisation des Lasers Intenses, CNRS, \'Ecole Polytechnique, CEA, Universit\'e Paris-Saclay, UPMC Universit\'e Paris 06: Sorbonne Universit\'es, F-91128 Palaiseau Cedex, France }
\address[d]{Laboratoire d'Utilisation des Lasers Intenses, UPMC Universit\'e Paris 06: Sorbonne Universit\'es, CNRS, Ecole Polytechnique, CEA, Universit\'e Paris-Saclay, F-75252 Paris Cedex 05, France}
\address[e]{Dipartimento di Fisica Enrico Fermi, Università di Pisa, Largo Bruno Pontecorvo 3, I-56127 Pisa, Italy}
\address[f]{Istituto Nazionale di Ottica, Consiglio Nazionale delle Ricerche (CNR/INO), u.o.s.~Adriano Gozzini, I-56127 Pisa, Italy}
\address[g]{Institut du D\'eveloppement des Ressources en Informatique Scientifique, CNRS, F-91403 Orsay, France}
\address[h]{Lasers, Interactions and Dynamics Laboratory, CEA, CNRS, Universit\'e Paris-Saclay, DSM/IRAMIS, CEN Saclay, F-91191 Gif sur Yvette, France}
\address[i]{Institut de Recherche en Astrophysique et Plan\'etologie, Universit\'e de Toulouse, UPS-OMP, F-31400 Toulouse, France}
\address[j]{Laboratoire de Physique des Plasmas, Ecole Polytechnique, CNRS, UPMC, Universit\'e Paris-Sud, F-91128 Palaiseau, France}

\begin{abstract}

\Smilei is a collaborative, open-source, object-oriented (C++) particle-in-cell code.
To benefit from the latest advances in high-performance computing (HPC), 
\Smilei is co-developed by both physicists and HPC experts. 
The code's structures, capabilities, parallelization strategy and performances are discussed.
Additional modules (e.g. to treat ionization or collisions), benchmarks and physics highlights are also presented.
Multi-purpose and evolutive, \Smilei is applied today to a wide range of physics studies, from relativistic laser-plasma interaction to astrophysical plasmas.

\end{abstract}
\maketitle

\section*{Keywords}
\noindent Plasma kinetic simulation, Particle-In-Cell (PIC), High-performance computing, Laser-plasma interaction, Astrophysical plasmas

\section*{Program summary}

\begin{description}[itemsep=-1em,align=left]
\item[{\it Program title}:] \Smilei \\
\item[{\it Licensing provisions}:] CeCILL-B free software license \\
\item[{\it Programming language}:] C++11, Python 2.7\\
\item[{\it Repository}:] \url{https://github.com/SmileiPIC/Smilei}\\
\item[{\it References}:] \url{http://www.maisondelasimulation.fr/smilei}\\
\item[{\it Nature of the problem}:] The kinetic simulation of plasmas is at the center of various physics studies, from laser-plasma interaction to astrophysics. To address today's challenges, a versatile simulation tool requires high-performance computing on massively parallel super-computers.\\
\item[{\it Solution method}:] The Vlasov-Maxwell system describing the self-consistent evolution of a collisionless plasma is solved using the Particle-In-Cell (PIC) method. Additional physics modules allow to account for additional effects such as collisions and/or ionization. A hybrid MPI-OpenMP strategy, based on a patch-based super-decomposition, allows for efficient cache-use, dynamic load balancing and high-performance on massively parallel super-computers. \\
\end{description}

\section{Introduction}

The Particle-In-Cell (PIC) approach was initially developed for fluid dynamics studies~\cite{harlow1955}.
Having various advantages (conceptual simplicity, efficient implementation on massively parallel computers, etc.),
it has become a central simulation tool for a wide range of physics studies, 
from semiconductors to cosmology or accelerator physics, and in particular to plasma physics.
Today, the kinetic simulation of plasmas in various environments, from the laboratory to astrophysics, 
strongly relies on PIC codes~\cite{birdsall1985}.

In this paper, we present the new, open-source PIC code \Smilei.
It has been developed in a collaborative framework including physicists and high-performance computing (HPC) experts 
to best benefit from the new HPC architectures.

\Smilei's development was initially motivated by recent advances in ultra-high intensity (UHI) laser technology, 
and new projects aiming at building multi-petawatt laser facilities.
UHI laser-plasma interaction has indeed been successfully applied to probing matter under extreme conditions 
of temperature and pressure, opening the way to various promising applications such as charged-particle (electron and ion) 
acceleration~\cite{tajima1979, pukhov2002, mangles2004, geddes2004, faure2004,macchi2013},
ultra-bright light sources of unprecedented short duration~\cite{thaury2010}, 
or abundant electron-positron pair production~\cite{chen2009, sarri2015}.
This wide range of applications, as well as the associated deeper understanding of fundamental processes, lead to the creation 
of the {\it Centre Interdisciplinaire de la Lumi\`ere EXtr\^eme} (CILEX)\footnote{\url{http://goo.gl/kzJCjY}}~\cite{cros2014}.
This academic center will host, in the forthcoming years, the laser Apollon that will deliver ultra-short (15~fs), 
ultra-intense (beyond  $10^{22}~{\rm W/cm^2}$) laser pulses, corresponding to a record peak power of 10 PW.
This path toward the study of light-matter interaction at extreme intensities represents
a significant experimental and technological undertaking. New numerical tools have to be deployed as 
laser-plasma interaction, at intensities beyond $10^{22}~{\rm W/cm^2}$, is not only relativistic but also highly nonlinear and 
of quantum nature~\cite{dipiazza2012}.

Furthermore, a paradigm shift has occurred in HPC:
the number of cores available on massively parallel supercomputers has skyrocketed.
This tendency is progressing quickly but software development lags behind.
Today, most of the codes used by the plasma community face difficulties when confronted with these new challenges.
They can be overcome with a strong collaboration between physicists and HPC specialists.

In this context, a consortium of laboratories of the {\it Plateau de Saclay} decided to join their efforts in developing
the new PIC code \Smilei (for {\it Simulating Matter Irradiated by Light at Extreme Intensities}).
Intended as a multi-purpose and collaborative PIC code, \Smilei addresses a wide range of physics problems, 
from laser-plasma interaction to astrophysics.

This paper presents an overview of the code's principles, structure, performance and capabilities, 
as well as benchmarks and examples.
Section~\ref{secPICMethod} reviews the general PIC approach for simulating collisionless plasmas 
(the governing equations, and the associated numerical methods), and specifies the algorithms used in \Smilei.
The C++ object-oriented programming and polymorphism, highlighted in Sec.~\ref{secCodeStructure}, 
illustrates the multi-purpose, multi-physics and multi-geometry aspects of the code and its modularity and maintainability. 
We outline \Smilei's components, their interactions and the I/O management strategy.
Section~\ref{secParallelization} then presents the innovative parallelization strategy devised for \Smilei. 
In particular, the hybrid MPI-OpenMP (for synchronization in between distributed and shared memory processes)
and dynamic load balancing designs are built around ``patches'', 
which extend the notion of domain decomposition and improve data locality for faster 
memory access  and efficient cache use. 
The code performance on massively-parallel super-computers is then discussed.
The following Sec.~\ref{secAdditionalModules} describes additional modules (binary collisions, ionization, etc.), 
and Sec.~\ref{secInterface} explains the input interface and the output diagnostics.
Section~\ref{secPhysicsHighlights} features applications to different physical scenarii, the first two related to UHI laser-plasma interaction 
and the other two to astrophysics.
Finally, Sec.~\ref{secConclusions} concludes on \Smilei capabilities and perspectives.

\section{The Particle-In-Cell (PIC) method for collisionless plasmas}\label{secPICMethod}

\subsection{The Maxwell-Vlasov model} 

The kinetic description of a collisionless plasma\footnote{The PIC method can be applied to (fully or partially ionized) plasmas 
as well as beams of charged particles. For the sake of simplicity however, we will refer to all these states as plasmas.} 
relies on the Vlasov-Maxwell system of equations.
In this description, the different species of particles constituting the plasma are described by their respective distribution 
functions $f_s(t,\vx,\vp)$, where $s$ denotes a given species consisting of particles with charge $q_s$ and mass $m_s$, 
and $\vx$ and $\vp$ denote the position and momentum of a phase-space element. 
The distribution $f_s$ satisfies Vlasov's equation:
\begin{eqnarray}\label{eq_Vlasov}
\left(\partial_t  + \frac{\vp}{m_s \gamma} \cdot \nabla + \vF_L \cdot \nabla_{\vp} \right) f_s = 0\,,
\end{eqnarray}
where $\gamma = \sqrt{1+\vp^2/(m_s\,c)^2}$ is the (relativistic) Lorentz factor, $c$ is the speed of light in vacuum, and
\begin{eqnarray}\label{eq_LorentzForce}
\vF_L = q_s\,(\vE + \vv \times \vB)
\end{eqnarray}
 is the Lorentz force acting on a particle with velocity $\vv=\vp/(m_s\gamma)$.

This force follows from the existence, in the plasma, of collective electric [$\vE(t,\vx)$] and magnetic [$\vB(t,\vx)$] fields satisfying 
Maxwell's equations\footnote{It is important to stress that the electromagnetic fields considered here are macroscopic 
(mean) fields, and not microscopic fields. Therefore, the PIC simulation does not, in its standard form, accounts for particle collisions. 
Collisions are however introduced in an {\it ad hoc} module presented in Sec.~\ref{sec:collisions}.}:
\begin{subequations}\label{eq_Maxwell}
\begin{eqnarray}
\label{eq_BGauss} \nabla \cdot \vB &=& 0 \,,\\
\label{eq_Poisson} \nabla \cdot \vE &=& \rho/\epsilon_0 \,,\\
\label{eq_Ampere}\nabla \times \vB &=& \mu_0\, \vJ + \mu_0 \epsilon_0\,\partial_t \vE \,,\\
\label{eq_Faraday}\nabla \times \vE &=& -\partial_t \vB \,,
\end{eqnarray}
\end{subequations}
where $\epsilon_0$ and $\mu_0$ are the vacuum permittivity and permeability, respectively. 

The Vlasov-Maxwell system of Eqs.~\eqref{eq_Vlasov} -- \eqref{eq_Maxwell} describes the self-consistent dynamics of the plasma 
which constituents are subject to the Lorentz force, and in turn modify the collective electric and magnetic fields through their charge 
and current densities:
\begin{subequations}\label{eq_rhoJ}
\begin{eqnarray}
\rho(t,\vx) &=& \sum_s q_s\int\!d^3\!p f_s(t,\vx,\vp)\,,\\
\vJ(t,\vx) &=& \sum_s q_s\int\! d^3\!p\,\vv f_s(t,\vx,\vp)\,.
\end{eqnarray}
\end{subequations}

\subsection{Reference units}\label{secUnits}

\Smilei is a fully-relativistic electromagnetic PIC code. As such, it is convenient to normalize all velocities in the code to $c$.
Furthermore, charges and masses are normalized to $e$ and $m_e$, respectively, with $-e$ the electron charge and $m_e$ its mass. 
Momenta and energies (and by extension temperatures) are then expressed  in units of $m_e c$ and $m_e c^2$, respectively.

The normalization for time and space is not decided {\it a priori}. Instead, all the simulation results may
be scaled by an arbitrary  factor.
Denoting the ({\it a priori} unknown) time units by $\omega_r^{-1}$, distances are normalized to $c/\omega_r$.  
Electric and magnetic fields are expressed in units of $m_e c\,\omega_r/e$ and $m_e \omega_r/e$, respectively.
We define the units for number densities as $n_r = \epsilon_0 m_e \omega_r^2/e^2$, while charge and current 
densities are in units of $e\,n_r$ and $e\,c\,n_r$, respectively.
Note that this definition of $n_r$ is chosen for best simplification of the Vlasov-Maxwell equations, but does not 
correspond to the reference distance $c/\omega_r$ to the power of $-3$.

Let us now illustrate by two simple examples this choice of normalization.
When dealing with a plasma at constant density $n_{e}$, it is convenient to normalize times by introducing 
the electron plasma frequency $\omega_{pe} = \sqrt{e^2 n_e/(\epsilon_0 m_e)}$. Choosing $\omega_r=\omega_{pe}$, 
distances are now expressed in units of the electron skin-depth $c/\omega_{pe}$, while number densities are 
normalized to $n_e$, and the electric and magnetic fields are in units of $m_e\omega_{pe}c/e$ and $m_e \omega_{pe}/e$, respectively. 

In contrast, when considering the irradiation of a plasma by a laser with angular frequency $\omega_0$, it is convenient 
to use $\omega_r=\omega_0$. From this choice, it follows that distances are measured in units of $k_0^{-1}=c/\omega_0$, 
while the electric and magnetic fields are in units of $E_c = m_e c \omega_0/e$ and $m_e \omega_0/e$, respectively. 
Note that $E_c$ is the Compton field, which is widely used to measure the importance of relativistic effects in laser-plasma interaction.
In addition, number densities are expressed in units of $n_c = \epsilon_0 m_e \omega_0^2/e^2$, the well-known critical density 
delimiting plasmas that are transparent or opaque to an electromagnetic radiation with angular frequency $\omega_0$.

Table~\ref{tab_rormalisations} gives a list of the most common normalizations used in \Smilei. 
In what follows (and if not specified otherwise), all quantities will be expressed in normalized units.

\begin{table}
\centering
\begin{tabular}{l|l}
Units of velocity				& $c$\\
Units of charge					& $e $\\
Units of mass					& $m_e$\\
Units of momentum				& $m_e\,c$\\
Units of energy, temperature	& $m_e c^2$\\
Units of time					& $\omega_r^{-1}$ \\
Units of length					& $c/\omega_r$\\
Units of number density			& $n_r = \epsilon_0\,m_e\,\omega_r^2/e^2$\\
Units of current density		& $e\,c\,n_r$\\	
Units of pressure				& $m_e\,c^2\,n_r$\\
Units of electric field 		& $m_e\,c\,\omega_r/e$\\
Units of magnetic field			& $m_e\,\omega_r/e$\\
Units of Poynting flux			& $m_e\,c^3\,n_r/2$
\end{tabular}
\caption{List of the most common normalizations used in \Smilei. The value of $\omega_r$ is not defined {\it a priori}, but can be set {\it a  posteriori} as a scaling factor. 
For simulations requiring the use of ionization and/or collision modules (see Sec.~\ref{secAdditionalModules}), $\omega_r$ needs to be defined, in SI units, by the user.}
\label{tab_rormalisations}
\end{table}

\subsection{Quasi-particles and the PIC method}\label{sec:quasiPart}

The ``Particle-In-Cell'' method owes its name to the discretization of the distribution function $f_s$ as a sum of $N_s$ ``quasi-particles'' 
(also referred to as ``super-particles'' or ``macro-particles''):
\begin{eqnarray}\label{eq_fs_discretized}
f_s(t,\vx,\vp) = \sum_{p=1}^{N_s}\,w_p\,\,S\big(\vx-\vx_p(t)\big)\,\delta\big(\vp-\vp_p(t)\big)\,,
\end{eqnarray}
where $w_p$ is a quasi-particle ``weight'', $\vx_p$ is its position, $\vp_p$ is its momentum, $\delta$ is the Dirac distribution, 
and $S(\vx)$ is the shape-function of all quasi-particles. 
The properties of the shape-function used in \Smilei are given in~\ref{appendixA}.

In PIC codes, Vlasov's Eq.~\eqref{eq_Vlasov} is integrated along the continuous trajectories of these quasi-particles, 
while Maxwell's Eqs.~\eqref{eq_Maxwell} are solved on a discrete spatial grid, the spaces between consecutive grid points 
being referred to as ``cells''. 
Injecting the discrete distribution function of Eq.~\eqref{eq_fs_discretized} in Vlasov's Eq.~\eqref{eq_Vlasov}, 
multiplying the result by $\vp$ and integrating over all $\vp$ leads to:
\begin{eqnarray}
\nonumber \sum_{p=1}^{N_s}\,w_p\,\vp_p \cdot\left[\partial_{\vx_p} S(\vx-\vx_p) + \partial_{\vx} S(\vx-\vx_p)\right] \vv_p\\
\label{eq_trans1} +\sum_{p=1}^{N_s}\,w_p\,S(\vx-\vx_p)\, \left[ \partial_t \vp_p - q_s\,(\vE + \vv_p \times \vB) \right] = 0\,,
\end{eqnarray}
where we have introduced $\vv_p=\vp_p/(m_s \gamma_p)=d\vx_p/dt$ the $p^{th}$ quasi-particle velocity, and $\gamma_p=\sqrt{1+\vp_p^2/(m_s^2)}$ its Lorentz factor.
Considering all $p$ quasi-particles independently, and integrating over all (real) space $\vx$, the first term in Eq.~\eqref{eq_trans1} vanishes due to the 
properties of the shape-function (see~\ref{appendixA}) and one obtains that all quasi-particles satisfy the relativistic equations of motion:
\begin{eqnarray}
\frac{d\vx_p}{dt} &=& \frac{\vu_p}{\gamma_p}\,\\
\frac{d\vu_p}{dt} &=& r_s \, \left( \vE_p + \frac{\vu_p}{\gamma_p} \times \vB_p \right),
\end{eqnarray}
where we have introduced $r_s = q_s/m_s$ the charge-over-mass ratio (for species $s$),  
 $\vu_p = \vp_p/m_s$  the quasi-particle reduced momentum, and the fields interpolated at the particle position:
\begin{eqnarray}
\label{eq:interpE} \vE_p = \int d\vx\, S(\vx-\vx_p)\,\vE(\vx)\,,\\
\label{eq:interpB} \vB_p = \int d\vx\, S(\vx-\vx_p)\,\vB(\vx)\,.
\end{eqnarray}

Note that, because of the finite (non-zero) spatial extension of the quasi-particles (also referred to as quasi-particle size,~\ref{appendixA}), additional cells (called \textit{ghost cells}, see Sec.~\ref{secParallelization}) have to be added at the border of the simulation domain to ensure that the full quasi-particle charge and/or current densities are correctly projected onto the simulation grid.

In this Section, we present the general PIC algorithm, starting with the simulation initialization and then going through the PIC loop itself (see Tab.~\ref{tabAlgorithm}).

\begin{table}\label{tabAlgorithm}\small
\caption{ Summary of \Smilei's PIC algorithm.}
\centering
\begin{tabular}[t]{l}
\toprule
{\bf Initialization} \hspace{1cm} time step $n=0$, time $t=0$\\
\toprule
\\
{\bf Particle loading}	\hspace{2cm}	  $\forall p$, define $(\vx_p)^{n=0}$, $(\vu_p)^{n=-\tfrac{1}{2}}$\\
\\
{\bf Charge projection on grid} \hspace{.2cm} $\big[\forall p, (\vx_p)^{n=0}\big] \rightarrow \rho^{(n=0)}(\vx)$\\
\\
{\bf Compute initial fields}
\hspace{1cm}   - solve Poisson on grid: $\left[ \rho^{(n=0)}(\vx)\right] \rightarrow \vE^{(n=0)}_{\rm stat}(\vx)$ \\
\hspace{4.9cm}   - add external fields:  $\vE^{(n=0)}(\vx)=\vE^{(n=0)}_{\rm stat}(\vx)+\vE_{\rm ext}^{(n=0)}(\vx)$\\
\hspace{8.2cm}  $\vB^{(n=\tfrac{1}{2})}(\vx)=\vB_{\rm ext}^{(n=\tfrac{1}{2})}(\vx)$\\

\\
 \toprule
{\bf PIC loop:}  from time step $n$ to $n+1$, time $t=(n+1)\,\Delta t$\\
\toprule
\\
{\bf Restart charge \& current densities}\\
{\bf Save magnetic fields value} (used to center magnetic fields)\\
\\

{\bf Interpolate fields at particle positions} \hspace{0.3cm} $\forall p$, $[\vx_p,\vE^{(n)}(\vx), \vB^{(n)}(\vx)] \rightarrow {\vE}_p^{(n)}, {\vB}_p^{(n)}$\\
\\
{\bf Push particles} \hspace{.2cm} - compute new velocity $\forall p$, $\vp_p^{(n-\tfrac{1}{2})}\,\,\left[{\vE}_p^{(n)}, {\vB}_p^{(n)} \right]\,\,\vp_p^{(n+\tfrac{1}{2})}$\\
                      \hspace{2.9cm} - compute new position $\forall p$, $\vx_p^{(n)}\,\,\left[\vp_p^{(n+\tfrac{1}{2})}\right]\,\,\vx_p^{(n+1)}$\\
\\
{\bf Project current onto the grid}  using a charge-conserving scheme\\			
						\hspace{3cm}	 $\left[\forall p\,\, \,\vx_p^{(n)}, \vx_p^{(n+1)}, \vp_p^{(n+\tfrac{1}{2})}\right] \rightarrow {\vJ}^{(n+\tfrac{1}{2})}(\vx)$\\
\\
{\bf Solve Maxwell's equations}\\			
\hspace{3cm}- solve Maxwell-Faraday: ${\vE}^{(n)}(\vx) \,\, \left[{\vJ}^{(n+\tfrac{1}{2})(\vx)}\right] \,\, {\vE}^{(n+1)}(\vx)$\\
\hspace{3cm}- solve Maxwell-Ampère: ${\vB}^{(n+\tfrac{1}{2})}(\vx) \,\, \left[{\vE}^{(n+1)}(\vx)\right] \,\, {\vB}^{(n+\tfrac{3}{2})}(\vx)$\\
\hspace{3cm}- center magnetic fields: ${\vB}^{(n+1)}(\vx)=\tfrac{1}{2}\,\left({\vB}^{(n+\tfrac{1}{2})}(\vx)+{\vB}^{(n+\tfrac{3}{2})}(\vx)\right)$\\
\toprule
\end{tabular}
\end{table}

\subsection{Time- and space-centered discretization }

\begin{figure}\centering
\includegraphics[width=0.8\textwidth]{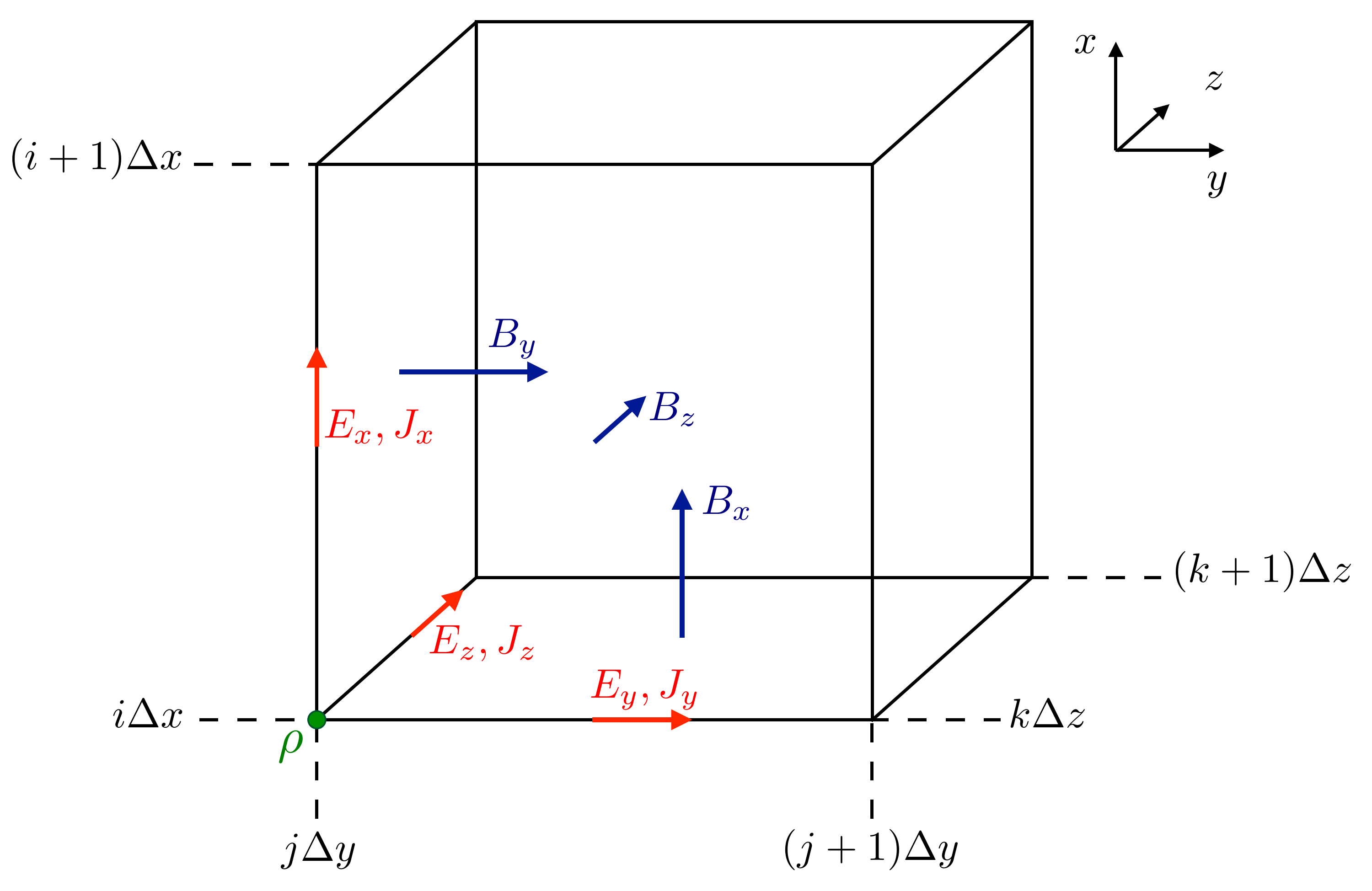}
\caption{Representation of the staggered Yee-grid. The location of all fields and current densities follows from the (rather standard) convention to define charge densities at the cell nodes.} \label{figYee}
\end{figure}

As will be discussed in Sec.~\ref{secMaxwellSolvers}, Maxwell's equations are solved here using the Finite Difference Time Domain (FDTD) approach~\cite{taflove2005} as well as refined methods based on this algorithm (for a review of these methods see~\cite{nuter2014}). In these methods, the electromagnetic fields are discretized onto a staggered grid, the Yee-grid, that allows for spatial-centering of the discretized curl operators in Maxwell's Eqs.~\eqref{eq_Ampere} and~\eqref{eq_Faraday}. Figure~\ref{figYee} summarizes at which points of the Yee-grid the electromagnetic fields, as well as charge and density currents,
are defined. Similarly, the time-centering of the time-derivative in Maxwell's Eqs.~\eqref{eq_Ampere} and~\eqref{eq_Faraday} is ensured by considering the electric fields as defined at integer time-steps $(n)$ and magnetic fields at half-integer time-steps $(n+\tfrac{1}{2})$. Time-centering of the magnetic fields is however necessary for diagnostic 
purposes, and most importantly when computing the Lorentz force acting on the quasi-particles.
It should also be noted, as will be discussed in Sec.~\ref{secPusher}, that a {\it leap-frog} scheme is used to advance the particles in time, so that their positions and velocities are defined at integer $(n)$ and half-integer $(n-\tfrac{1}{2})$ time-steps, respectively.

\subsection{Initialization of the simulation}

The initialization of a PIC simulation is a three-step process consisting in: (i) loading particles, (ii) computing the initial total charge and current densities onto the grid, and (iii) computing the initial electric and magnetic field at the grid points. In \Smilei, all three steps can be done either as a restart of a previous simulation (in which case the particles, charge and current densities and electromagnetic fields are directly copied from a file generated at the end of a previous simulation), or from a user-defined input file. In that case, the user defines the initial conditions of the particle, charge and current densities as well as the initial electromagnetic fields over the whole simulation domain.
 
In particular, the user prescribes spatial profiles for the number density $n_s$, the number of particle per cell $N_s$, the mean velocity $\vv_s$ and the temperature $T_s$ of each species $s$ at time $t=0$.
The particle loading then consists in creating, in each cell, $N_s$ particles with positions $\vx_p$ uniformly distributed within the cell (either randomly chosen or regularly spaced), and with momenta $\vp_p$ randomly sampled from a requested distribution\footnote{The user may select a zero-temperature distribution, a Maxwellian distribution, or Maxwell-J\"uttner distribution, i.e. the relativistic generalization of the maxwellian distribution~\cite{wright1975}. In the latter case, the method proposed in Ref.~\cite{zenitani2015} is used to ensure a correct loading of particles with a relativistic drift velocity.}. In \Smilei, a given numerical weight $w_p$ is assigned to each particle depending on the density associated to the cell it originates from:
\begin{eqnarray}
w_p = \frac{n_s\big(\vx_p(t=0)\big)}{N_s\big(\vx_p(t=0)\big)}\,.
\end{eqnarray}
This variable weighting is particularly beneficial when considering initially highly inhomogeneous density distributions.

Once all particles in the simulation domain have been created, the total charge and current densities $\rho(t=0,\vx)$ and $\vJ(t=0,\vx)$ are computed onto the grid using a direct projection technique (see~\ref{appendixA} for more details) that assigns to a grid point located at $\vx_i$ the total charge and or current contained in the cell surrounding it:
\begin{eqnarray}
\rho(t=0,\vx) = \sum_s\,q_s\,\sum_p\,w_p\int\!\! \,d\vx\,S\big(\vx-\vx_p(t=0)\big)\,P_D(\vx-\vx_i)\,,
\end{eqnarray}
where $P_D(\vx)=\Pi_{\mu=1}^D P(x^{\mu})$ ($D$ referring to the number of spatial dimensions) with $P(x)$ the crenel function such that $P(x^{\mu})=1$ if $\vert x^{\mu} \vert < \Delta\mu/2$ and $P(x^{\mu})=0$ otherwise, and $\Delta\mu$ is the cell length in the $\mu=(x,y,z)$-direction.

Then, the initial electric fields are computed from $\rho(t=0,\vx)$ by solving Poisson's Eq.~\eqref{eq_Poisson}.
In \Smilei, this is done using the conjugate gradient method~\cite{numericalRecipies}.
This iterative method is particularly interesting as it is easily implemented on massively parallel computers as it requires mainly local information exchange between adjacent  domains (see Sec.~\ref{secParallelization} for more information on domain decomposition for parallelization). 

External (divergence-free) electric and/or magnetic fields can then be added to the resulting electrostatic fields, provided they fulfill Maxwell's Eqs.~\eqref{eq_Maxwell}, and in particular Gauss and Poisson Eqs.~\eqref{eq_BGauss} and~\eqref{eq_Poisson}. 

\subsection{The PIC loop} \label{picLoop}

At the end of the initialization stage [time-step $(n=0)$], all quasi-particles in the simulation have been loaded and the electromagnetic fields have been computed over the whole simulation grid. The PIC loop is then started over $N$ time-steps each consisting in (i) interpolating the electromagnetic fields at the particle positions, (ii) computing the new particle velocities and positions, (iii) projecting the new charge and current densities on the grid, and (iv) computing the new electromagnetic fields on the grid.
In this section, we describe these four steps taken to advance from time-step $(n)$ to time-step $(n+1)$.\\

\subsubsection{Field interpolation at the particle}

At the beginning of time-step $(n)$, the particles velocities and positions are known at time-step $(n-\tfrac{1}{2})$ and $(n)$, respectively. For each particle $p$, the electromagnetic fields [at time-step $(n)$] are computed at the particle position using a simple interpolation technique:
\begin{eqnarray}
\vE_p^{(n)} = \int d\vx\, S\left(\vx-\vx_p^{(n)}\right) \vE^{(n)}(\vx)\,,\\
\vB_p^{(n)} = \int d\vx\, S\left(\vx-\vx_p^{(n)}\right) \vB^{(n)}(\vx)\,,
\end{eqnarray}
where we have used the time-centered magnetic fields $\vB^{(n)}=\tfrac{1}{2}[\vB^{(n+1/2) } + \vB^{(n-1/2)}]$.
Additional information on the field interpolation are given in~\ref{appendixA}.\\

\subsubsection{Particle pusher}\label{secPusher}

Knowing, for each quasi-particle, the electromagnetic fields at its position, the new particle momentum and position are computed using a (second order) leap-frog integrator.
In \Smilei, two different schemes have been implemented, the well-known Boris pusher~\cite{boris1970} and the one developed by J.-L. Vay~\cite{vay2008}. 
Both schemes compute the new particle momentum according to:
\begin{eqnarray}
\vu_p^{(n+\tfrac{1}{2})}=\vu_p^{(n-\tfrac{1}{2})} + r_s \Delta t \, \left[ E_p^{(n)} + \frac{\vv_p^{(n+\tfrac{1}{2})}+\vv_p^{(n-\tfrac{1}{2})}}{2} \times B_p^{(n)}\right],
\end{eqnarray}
as well as the new particle position:
\begin{eqnarray}
\vx_p^{(n+1)}=\vx_p^{(n)} + \Delta t \, \frac{\vu_p^{(n+\tfrac{1}{2})}}{\gamma_p},
\end{eqnarray}
where $\Delta t$ denotes the duration of a time-step.

The Boris pusher is a widely-used second-order leap-frog solver. 
However, Ref.~\cite{vay2008} shows that it introduces errors when calculating 
the orbits of relativistic particles in special electromagnetic field configurations (e.g. when the electric and magnetic contributions
cancel each other in the Lorentz force). Vay's solver proposes an alternative formulation of the leap-frog solver that prevents
such problems with an additional (albeit not large) computational cost.

\subsubsection{Charge conserving current deposition}

Charge deposition (i.e. charge and current density projection onto the grid) is then performed using the charge-conserving algorithm proposed by Esirkepov~\cite{esirkepov2001}.
The current densities in the dimensions of the grid (i.e., the $x$-direction for 1-dimensional simulations, both $x$- and $y$-directions for 2-dimensional simulations, and all three $x$-, $y$- and $z$-directions for 3-dimensional simulations) are computed from the charge flux through the cell borders (hence ensuring charge conservation) while the current densities along the other dimensions are performed using a simple projection. To illustrate this point, we take the example of current deposition in a 2-dimensional simulation. The current densities in the $x$- and $y$-directions associated to a particle with charge $q$ are computed as:

\begin{eqnarray}
(J_{x,p})_{i+\tfrac{1}{2},j}^{(n+\tfrac{1}{2})} = (J_{x,p})_{i-\tfrac{1}{2},j}^{(n+\tfrac{1}{2})} + q\,w_p\,\frac{\Delta x}{\Delta t}\,(W_x)_{i+\tfrac{1}{2},j}^{(n+\tfrac{1}{2})}\,\\
(J_{y,p})_{i,j+\tfrac{1}{2}}^{(n+\tfrac{1}{2})} = (J_{y,p})_{i,j-\tfrac{1}{2}}^{(n+\tfrac{1}{2})} + q\,w_p\,\frac{\Delta y}{\Delta t}\,(W_y)_{j,i+\tfrac{1}{2}}^{(n+\tfrac{1}{2})}\,
\end{eqnarray}
where $(W_x)^{(n+\tfrac{1}{2})}$ and $(W_y)^{(n+\tfrac{1}{2})}$ are computed from the particle present and former positions $x_p^{(n+1)}$ and $x_p^{(n)}$, respectively, using the method developed by Esirkepov.  The particle current in the $z$-direction (not a dimension of the grid) is, in this geometry, computed using the direct projection technique described in~\ref{appendixA}:
\begin{eqnarray}
(J_{z,p})_{i,j} = q w_r \vv_p\,\int\!\!d\vx\,S(\vx-\vx_p)\,P_D(\vx-\vx_{i,j})\,.
\end{eqnarray}
The charge density deposited by the particle can be obtained, if required e.g. for diagnostic purpose, using a similar direct projection.

The total charge and current densities henceforth gather the contributions of all quasi-particles of all species.
It is worth noting that, within a charge-conserving framework, charge densities are only projected on the grid for diagnostics purposes (as we will see in next paragraph, it is not used to advance the electromagnetic fields).

\subsubsection{Maxwell solvers}\label{secMaxwellSolvers}

Now that the currents are known at time-step $(n+\tfrac{1}{2})$, the electromagnetic fields can be advanced solving Maxwell's Eqs.~\eqref{eq_Maxwell}.
First, Maxwell-Ampère Eq.~\eqref{eq_Ampere} is solved, giving the advanced electric fields:
\begin{eqnarray}
\vE^{(n+1)} = \vE^{(n)} + \Delta t\, \left[\left(\nabla \times \vB\right)^{(n+\tfrac{1}{2})} - \vJ^{(n+\tfrac{1}{2})} \right]\,.
\end{eqnarray}
Then, Maxwell-Faraday Eq.~\eqref{eq_Faraday} is computed, leading to the advanced magnetic fields:
\begin{eqnarray}
\vB^{(n+\tfrac{3}{2})} = \vB^{(n+\tfrac{1}{2})} - \Delta t\, \left(\nabla \times \vE\right)^{(n+1)}\,.
\end{eqnarray}

Before discussing the discretization of the curl-operator in more details, it is worth noting that solving Eqs.~\eqref{eq_Ampere} and~\eqref{eq_Faraday} is sufficient to get a complete description of the new electromagnetic fields. Indeed, it can be shown that this conserves a divergence-free magnetic field if Gauss' Eq.~\eqref{eq_BGauss} is satisfied at time $t=0$. Similarly, Poisson's Eq.~\eqref{eq_Poisson} is verified as long as it is satisfied at time $t=0$ as long as the charge deposition algorithm fulfills the charge conservation equation:
\begin{eqnarray}
\partial_t \rho + \nabla \cdot \vJ = 0
\end{eqnarray}
This motivated the use of Esirkepov's projection scheme discussed in the previous paragraph.

We conclude this Section by discussing in more details the discretization of the curl-operators in Eqs.~\eqref{eq_Ampere} and~\eqref{eq_Faraday}.
To do so, let us focus on the equations for the electric and magnetic fields $E_x$ and $B_x$ discretized on the (staggered) Yee-grid:
\begin{eqnarray}
\frac{(E_x)_{i+\tfrac{1}{2},j,k}^{(n+1)} - (E_x)_{i+\tfrac{1}{2},j,k}^{(n)}}{\Delta t} = (J_x)_{i+\tfrac{1}{2},j,k}^{n+\tfrac{1}{2}} + \left(\partial_y B_z\right)_{i+\tfrac{1}{2},j,k}^{(n+\tfrac{1}{2})} - \left(\partial_z B_y\right)_{i+\tfrac{1}{2},j,k}^{(n+\tfrac{1}{2})} \,,\\
\frac{(B_x)_{i,j+\tfrac{1}{2},k+\tfrac{1}{2}}^{(n+\tfrac{3}{2})} - (B_x)_{i,j+\tfrac{1}{2},k+\tfrac{1}{2}}^{(n+\tfrac{1}{2})} }{\Delta t} =  \left(\partial^*_z E_y\right)_{i,j+\tfrac{1}{2},k+\tfrac{1}{2}}^{(n+1)} - \left(\partial^*_y B_z\right)_{i,j+\tfrac{1}{2},k+\tfrac{1}{2}}^{(n+\tfrac{1}{2})} \,.
\end{eqnarray}
The partial derivatives in space in both equations are discretized as follows. 
In the Maxwell-Ampère equation, the partial derivative in $x$ (similarly in $y$ and $z$) reads:
\begin{eqnarray}
\left(\partial_x F\right)_{i,j,k} = \frac{F_{i+\tfrac{1}{2},j,k}-F_{i-\tfrac{1}{2},j,k}}{\Delta x}\,,
\end{eqnarray}
and corresponds to the usual curl-operator discretization used in the FDTD method.
In the Maxwell-Faraday equation, the partial derivatives can be modified using an extended stencil (see Ref.~\cite{nuter2014} for a comparative study of different solvers). 
The spatial derivative in the $x$-direction (similarly in the $y$ and $z$ directions) reads:
\begin{eqnarray}
\left(\partial^*_x F\right)_{i,j,k} &=& \alpha_x\,\frac{F_{i+\tfrac{1}{2},j,k}-F_{i-\tfrac{1}{2},j,k}}{\Delta x} + \eta_x\,\frac{F_{i+\tfrac{3}{2},j,k}-F_{i-\tfrac{3}{2},j,k}}{\Delta x} \\
\nonumber &+& \beta_{xy} \left[ \frac{F_{i+\tfrac{1}{2},j+1,k}-F_{i-\tfrac{1}{2},j+1,k}}{\Delta x}  + \frac{F_{i+\tfrac{1}{2},j-1,k}-F_{i-\tfrac{1}{2},j-1,k}}{\Delta x} \right] \\
\nonumber &+& \beta_{xz} \left[ \frac{F_{i+\tfrac{1}{2},j,k+1}-F_{i-\tfrac{1}{2},j,k+1}}{\Delta x}  + \frac{F_{i+\tfrac{1}{2},j,k-1}-F_{i-\tfrac{1}{2},j,k-1}}{\Delta x} \right] \,,
\end{eqnarray}
the set of parameters $\alpha_x$, $\eta_x$, $\beta_{xy}$ and $\beta_{xz}$ depending of the type of solver used~\cite{nuter2014}, and the standard FDTD solver is recovered for $\alpha_x=1, \eta_x=\beta_{xy}=\beta_{xz}=0$.

Note that the FDTD solvers are subject to a Courant-Friedrich-Lewy (CFL) condition. For the standard solver, the CFL condition requires the time-step to be smaller than:
\begin{eqnarray}
\Delta t_{\rm CFL} = \sum_{\mu} \left(\Delta \mu^{-2}\right)^{-\tfrac{1}{2}}\,,
\end{eqnarray}
$\mu=(x,y,z)$ standing for the different spatial directions resolved in the simulation.

\subsubsection{Boundary conditions}

After having computed new quasi-particle positions and velocities, boundary conditions (BCs) are applied to each quasi-particle that may be located in a ghost cell, i.e. outside of the 'real' grid.
Quasi-particle species may have a different BC for each boundary of the simulation box:
the quasi-particles can either loop around the box (periodic), be stopped (momentum set to zero),
suppressed (removed from memory), reflected (momentum and position follow specular reflection rules) or thermalized.
In the latter case, the quasi-particle is set back inside the simulation box, and its new momentum is randomly sampled in a 
Maxwellian distribution~\cite{spohn1991} with a given temperature and drift velocity, both specified by the user. 

BCs are applied to the electromagnetic fields after Maxwell's equations have been solved.
Each boundary of the simulation box can feature a different BC.
First, injecting/absorbing BCs inspired from the ``Silver-M\"uller'' BC~\cite{barucq1997} are able to inject an electromagnetic wave (e.g. a laser) and/or to absorb outgoing electromagnetic waves.
In contrast, the reflective electromagnetic BC will reflect any outgoing electromagnetic wave reaching the simulation boundary. 
Lastly, periodic BCs are also available.

\newpage
\section{An evolutive, multi-purpose code}\label{secCodeStructure}

\Smilei's objectives are high performances, a large user community and support for a variety of applications.
Its C++ approach reflects these goals, providing structure to separate physics from computing aspects, 
to encourage their progress, to facilitate their maintainability and to ensure a multi-purpose capability.

\subsection{C++ elements and flow}
\Smilei's core program is written in the C++ language. Its multi-purpose and mature technology
ensures great flexibility and strong support for the new HPC machines.
Moreover, C++'s object-oriented programming provides an efficient way of structuring the code. 
Importantly, this eliminates a few bad habits such as passing large lists of parameters through
functions, or usage of global variables, inefficient in parallel computing.
Components can be constructed almost independently.
It offers a good separation between the purely computing/performance aspects and the physics calculations.

\begin{figure}[!b]
	\centering
	\includegraphics[width=10cm]{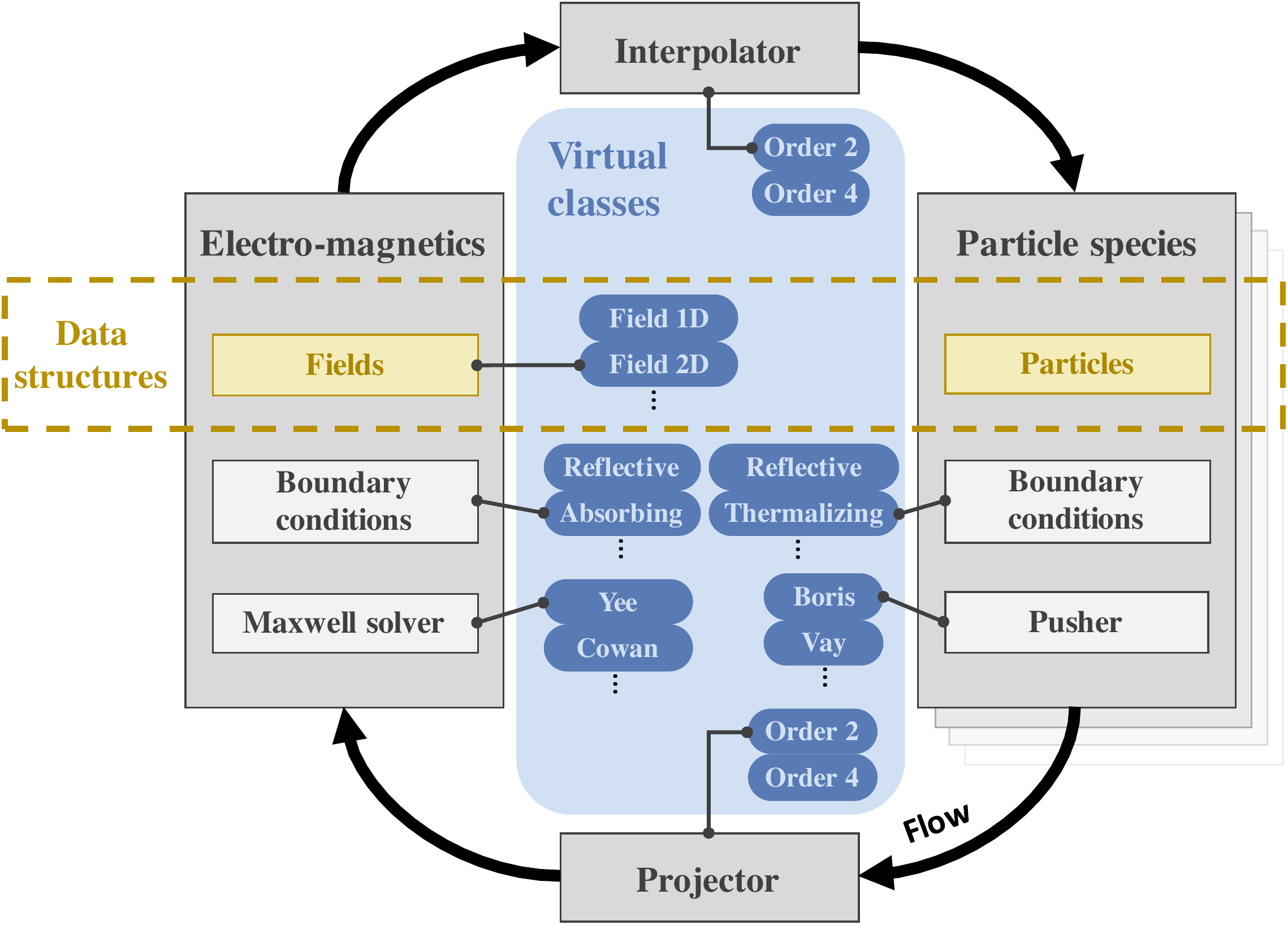}
	\caption{C++ flow, classes and data structure in \Smilei.
		\label{fig:cppstructure}}
\end{figure}

Figure~\ref{fig:cppstructure} shows the various elements of \Smilei's main code: C++ classes,
data structure, and the program flow. The main classes, namely ``particle species'' and
``electromagnetics'', are the counterparts of {\it particle} and {\it cell} in {\it Particle-in-cell},
respectively. The particle species class hold the particle object, which is the data structure for
the quasi-particles positions and momenta. It also contains operators on the quasi-particles such as the boundary
conditions and the pusher. On the other side, the electromagnetics class contains the fields, i.e. the
data structure for the electric and magnetic fields. Note that these fields also describe the
charge and current densities projected onto the grid. Electromagnetics also includes operators such as
the Maxwell solver and the boundary conditions for the fields.

Two additional operators are external to those structures because they operate between
particles and fields. The interpolator takes the field data and interpolates it at the particles
positions. The projector takes the particle data and projects it at the grid points.

\subsection{Polymorphism}
The C++ language supports the definition of {\it polymorphic} classes.
These classes contain functions, called {\it virtual functions},
that are selected at runtime among several options.
In other words, the behavior of an object is not decided {\it a priori}, but may be
defined during the simulation. 

\Smilei relies on C++ polymorphism to handle its multi-purpose ambition.
For instance, the basic polymorphic \verb!Field! class may be derived into different classes such as \verb!Field1D!, \verb!Field2D!, etc.
All these derived classes inherit
their functions from the base class, but they include different data
structures. In Fig.~\ref{fig:cppstructure}, examples of polymorphic ({\it virtual})
classes are highlighted.
Note that, in \Smilei, selecting the class from which each object
will be created is ensured by a ``factory design pattern''.

There are several advantages to polymorphism. First, it allows for straightforward
inheritance of properties between objects of similar structures. It also improves the
readability of the code by removing the complexity of all the multi-purpose
capabilities from the program flow. Lastly, it standardizes the form of the objects
for easier maintenance.
In these conditions, a single executable file can perform simulations in
various dimensions, interpolation orders, or physics components,
without the complexity of many code versions.

However, an excess of virtualization, or a large number of objects layers could have a significant computational cost.
For instance, the use of a virtual method to access a single data element 
(e.g., a single particle property) would have an unacceptable data access overhead.
This pitfall is avoided by passing the whole data structures to computational operators.
They are passed in their virtual form, then cast to the required class by the operator itself.

\subsection{Uncoupling operators from data}

An other fundamental ambition of the project is to provide an efficient tool of simulation
on current and future supercomputers whose architectures are in permanent evolution.
For instance, they may have complex memory hierarchy, whether distributed
or shared between several processors.
For ideal performances, the code must be adapted to these specific architectures.
Besides this multi-machine aspect, a multi-purpose code (able to simulate various physical
scenarii) may require a different optimization strategy depending on the subject of
each simulation.

For these two challenges, \Smilei's solution is based on its object-oriented design:
it consists in uncoupling the computing algorithms from the data formalism.
In all operators (solvers, interpolators, projectors, etc.), algorithms do not rely on raw data
but on wrappers (\verb!Field! and \verb!Particles!) which encapsulate and provide access to the data. 
Operators can thus be defined independently from the chosen data structure, provided
the ``protocol'' for accessing to the data is respected. 
As a consequence, performances can be optimized separately in operators and in the data structures.

Along the same principle, parallelism management tends to be decoupled from the physics calculations
by implementing different levels of parallelism, as detailed in Sec.~\ref{pb_omp}.

\subsection{HDF5 data management}
A significant amount of output data is generated by PIC simulations.
We examine here the representation of these data, focusing on the data access convenience and
performances on a large super-computer.

Classical output management would simply consist in gathering data on a ``master'' processor which writes everything out,
or in generating one file for each processor. The former technique is limited by the cost of communicating data and its memory overhead,
while the latter requires heavy post-processing. In both cases, the larger the simulation, the more expensive the overhead.

Parallel I/O libraries are optimized to avoid these pitfalls, and their development continuously improves their performances.
They can share and write data in parallel to a single file.
Famous examples are \textit{MPI-IO}\footnote{IBM Knowledge center at \url{http://www.goo.gl/XjUXzu}}, 
\textit{HDF5}  (\textit{Hierarchical Data Format}\footnote{\url{https://www.hdfgroup.org/HDF5}}) 
and \textit{NetCDF} (\textit{Network Common Data Form}\footnote{\url{http://www.unidata.ucar.edu/software/netcdf/docs/index.html}}).
Although no parallel I/O library is yet fully optimized for the most recent parallelism techniques,
they greatly enhance the simulations efficiency.

\textit{MPI-IO} has demonstrated good performances, but it generates unformatted data, thus requiring an additional effort from the user to access and analyse the simulation data.
In contrast, both \textit{HDF5} and \textit{NetCDF} rely on a structured data model, which is also open-source and widely used. {\it HDF5} also benefits from a large panel of open-source software for post-processing and visualization.
To sustain the required level of performance while maintaining its user-friendly and open-source approach
\Smilei currently uses \textit{HDF5}.

During preliminary studies done for the IDRIS {\it Grand Challenge} (see in Sec.~\ref{sec:physSBS}), 
\Smilei achieved a write bandwidth of 2.6~Gb/s on the Turing (BlueGene/Q) GPFS file system.  
The simulation domain consisted in a grid of size 30720 x 15360 cells, and 18 fields were written every 1755 timesteps (for a total of 135000 timesteps). 
The amount of 60~Gb of data was written in 24 seconds for each of the selected timesteps.

\newpage
\section{Parallelization}\label{secParallelization}

As high-performance computing (HPC) systems are evolving towards the exascale, 
there is an admitted risk that today's algorithms and softwares will be subpar, at best, for the upcoming architectures.
Manufacturers have been unable to improve the existing ``standard'' microprocessor technologies for the last decade.
Instead, the trend is oriented towards the multiplication of the number of computing units by several orders of magnitude.
This is achieved either using co-processors or massively multi-core processors.
In order to face this emerging complexity, codes must expose a tremendous amount of parallelism while conserving data locality and minimizing load imbalance.
In this Section, we first present the overall parallelization strategy chosen for \Smilei, and follow with accurate descriptions
of its elements.

\subsection{Strategy}
For the sake of generality, all fundamental computing items (cores, MPI processes, openMP threads, cuda threads, openCL work items, etc.)
will be referred to as computing elements (CE) in this subsection.

The difficulty in parallelizing a PIC code lies in the coupling between the grid and particle aspects of the code.
In a typical run, most of the load is carried by the particles.
It is therefore very tempting to distribute particles equally between CEs: benefits would be huge.
First, simplicity.
No particle communications are required because particles only interact with fields and are independent from each other.
Second, an almost perfect load balance is maintained at all times.
The drawback of this approach is that it implies that all CEs have access to a shared global array of grid quantities (fields and currents).
These accesses must be synchronized and require frequent global communications which, in practice, prevent any form of scalability above a couple hundreds of CEs.

A purely particle-based decomposition being impossible, we must apply a grid-based decomposition technique.
Domain decomposition is the technique used in all state-of-the-art PIC codes such as Osiris~\cite{fonseca2002} or Calder-Circ~\cite{lifschitz2009} in laser-plasma interaction
or Photon-Plasma~\cite{haugbolle2013} in astrophysics.

It has shown very good scalability but comes with a cost.
As most of the computational load is carried by particles, having a grid-based decomposition is inconvenient.
Its efficient implementation is more involved, and load balance is very difficult to achieve.
The biggest issue is that particles are volatile objects traveling throughout the entire domain, forcing 
(i) communications between CEs when particles cross their local domain boundary, and
(ii) random access to the grid at every interpolation and projection phases.
Communications are limited to neighbor domains and are not a fundamental threat to performance or scalability.
In contrast, the randomness of the particle positions is much more problematic.
Random access to the grid arrays breaks the principle of data locality paramount to the performance via a good cache use.
Conversely, a proper access to the data avoids multiple load operations when the same data is used several times.
And on top of that, if the access is well organized, {\it Single Instruction Multiple Data} (SIMD) operations can be executed thus accelerating the computation by a significant amount.

Most of the time, this issue is addressed by sorting particles. 
Different kind of algorithms can ensure that particles close to each other in space are also well clustered in memory.
Particles can be sorted at the cell level by a full count-sort algorithm every now and then during the simulation, or they can be subject to a more lax but more frequent sorting
as proposed in Ref.~\cite{stantchev2008}.
Note that the domain decomposition technique is already a form of sorting. 
Particles of a given sub-domain are naturally stored in a compact array of memory and attached to the grid portion they can interact with.
If each sub-domain is sufficiently small to fit in the cache, very good performances can be achieved. 
This approach was suggested in Refs.~\cite{decyk2011,germaschewski2016} and is the one used in \Smilei.
It consists in a very fine-grain domain decomposition referred to as ``patch-based'' decomposition where patches denote the very small sub-domains.
In addition, \Smilei still performs a very lightweight particle sorting within the patches, as in Ref.~\cite{stantchev2008}, in order to minimize cache misses.
It brings a convenient flexibility in the patches size without loss of performances as quasi-particles remain well sorted even if the patches are large.

\subsection{A patch-based MPI + openMP implementation}\label{implementation}

\Smilei uses the Message Passing Interface (MPI) to communicate data between distinct nodes of the distributed-memory architecture, and the Open Multi-Processing (openMP) interface to harmonize the computational load within each node with a reduced programming complexity.

This section shows that this hybrid MPI + openMP implementation of a patch-based decomposition naturally extends the pure MPI one described in Ref.~\cite{germaschewski2016}. It provides both scalability and dynamic load balancing.

\subsubsection{Patches distribution between MPI processes}\label{patches}
The first layer of parallelism in \Smilei is similar to the standard domain decomposition:
the simulation box is divided into sub-domains that can be treated in parallel.
In a standard ``traditional'' MPI approach, each MPI process handles one sub-domain.
But in \Smilei, the simulation box is divided into many more sub-domains than there are MPI processes.
They are called ``patches'' specifically to make this distinction: each MPI process handles many patches.
Note that the content of a patch is not different than that of a sub-domain: particles and a portion of the grid.

\begin{figure}
\centering
 \includegraphics[width=0.8\textwidth]{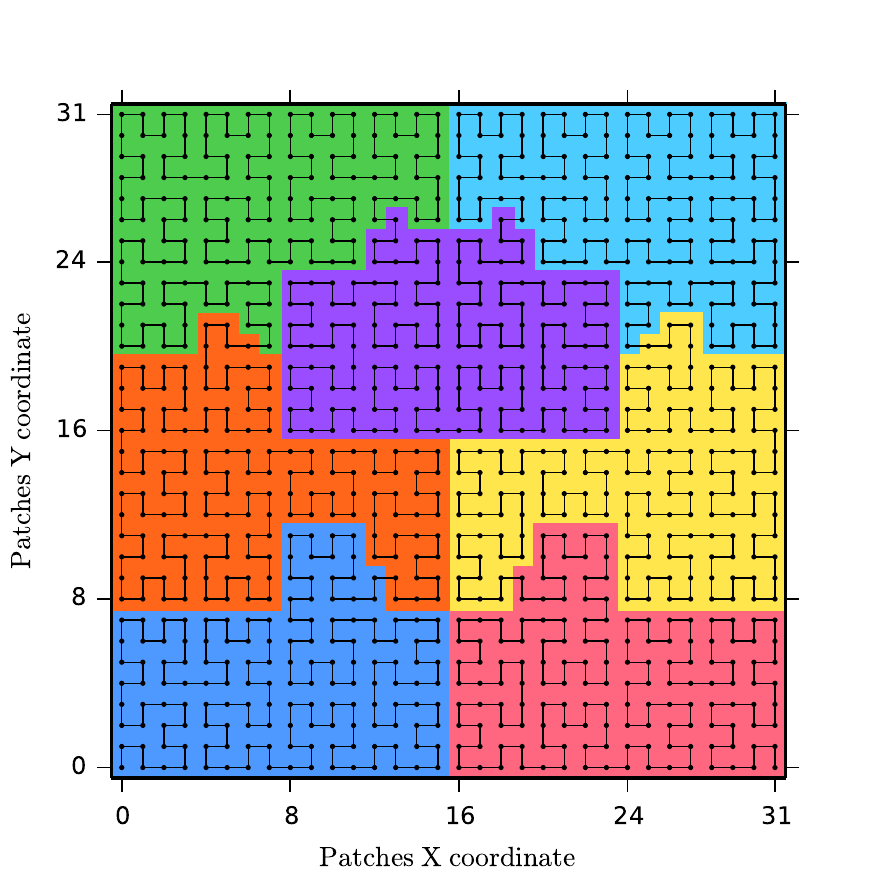}
 \caption{Example of a $32\times 32$ patches domain decomposition, shared between 7 MPI processes.
MPI domains are delimited by different colors.
The Hilbert curve (black line) passes through all the patch centers (black dots).
It starts from the patch with coordinates $(0,0)$ and ends at the patch with coordinates $(31,0)$.}
 \label{fig_hilbert}
\end{figure}

The obvious cost of this fine-grain domain decomposition is an additional, but necessary, synchronization between patches.
Synchronization between patches belonging to the same MPI process is very cheap.
It consists in a simple copy of a relatively small amount of ghost cells and exchange of particles in a shared memory system.
Synchronization becomes more expensive when it occurs between patches belonging to different MPI processes.
In that case, data has to be exchanged through the network between distributed memory systems via costly calls to the MPI library.
In order to limit this cost, we need a distribution policy of the patches between the different MPI processes which minimizes MPI calls.
This is achieved by grouping patches in compact clusters that reduce the interface between MPI sub-domains as much as possible. In addition, this policy must be flexible enough to support an arbitrary number of MPI processes and varying number of patches per process. In order to satisfy both compactness and flexibility, patches are ordered along a Hilbert space-filling curve~\cite{hilbert1891}.
An example of the Hilbert curve is given in Fig.~\ref{fig_hilbert}.
This curve is divided into as many segments as MPI processes and each process handles one of these segments.
The mathematical properties of the Hilbert curve guarantee that these segments form compact clusters of patches in space (see Fig.~\ref{fig_hilbert}) independently of their number or length.

\subsubsection{OpenMP parallelization and load balancing} \label{pb_omp}
Patch-based decomposition, in addition to its cache efficiency, is a very convenient way to expose a lot of local (inside MPI sub-domains) parallelism.
Each patch being independent, they can be easily treated in parallel by the threads owned by the MPI process.
Without this structure, the projection of particles might result in race conditions (threads overwriting each other's computation) and would require costly atomic operations.

In \Smilei, patches are treated by openMP threads. 
In practice, this allows the user to start the simulation with less (but larger) MPI domains than in a pure MPI implementation.
A similar level of computational performance is retained while decreasing the global amount of communications.
The number of macro-particles per patch may differ significantly and so does the computational load associated to each patch.
The use of the openMP dynamic scheduler therefore provides local load balancing at a reasonable cost.
If a thread is busy treating a patch with a lot of macro-particles, other threads will be able to handle the remaining lighter patches thus avoiding idle time
(see performance results in Sec.~\ref{MPIplusopenMP} and Fig.~\ref{openMP_extensibility}).

Patches also act as sorting structures. Indeed, quasi-particles of a given patch only interact with this patch's local grid.
Small patches therefore provide a finer-grain load balancing and optimized cache use at the cost of more inter-patch synchronization.
This cost is assessed in Sec.~\ref{sec_Npatches}.

\subsubsection{Load management}\label{loadmanagement}
The objective of load management is to harmonize the computational workload between CEs as homogeneously as possible,
in order to avoid idle, underloaded CEs waiting for overloaded CEs.
In \Smilei, the load is dynamically balanced.
Note that load balancing is not the only approach for load management:
it can also involve load-limiting techniques such as the k-means particle-merging algorithm implemented in Photon-Plasma~\cite{frederiksen2015}.

We have seen in Sec.~\ref{pb_omp} that openMP already provides some amount of load balancing at the node level, but it doesn't help managing the load between MPI processes.
\Smilei balances the load between MPI processes by exchanging patches (defined in Sec.~\ref{patches}). 
This technique is efficient because a single patch workload is much smaller than the total workload of a process.
The patch size defines the balance grain and the smaller the patches the smoother the balance.

This is yet another argument in favour of using patches as small as possible.
At this point, it becomes interesting to understand what limits the patch size.
The minimum size of a patch is dictated by the number of ghost cells used.
We consider reasonable that a patch must have more cells than ghost cells.
The number of ghost cells is defined by the order of Maxwell's equations discretization scheme, and by the shape function of the macro-particles.
A standard second-order Yee scheme, for instance, uses 4 ghost cells per dimension (2 on each side).
The minimum patch size in that case is therefore 5 cells per dimension.
This criteria also guarantees that ghost cells from non-neighbour patches do not overlap, which is convenient for the synchronization phases.
The influence of the patch size is illustrated in Sec.~\ref{perf}.

We have seen in Sec.~\ref{patches} that patches are organized along a Hilbert space-filling curve divided into as many segments of similar length as there are MPI processes.
Each process handles the patches located in its segment of the Hilbert curve.
Dynamically balancing the load simply consists in exchanging patches between neighbour MPI processes along the curve.
That is to lengthen or shorten the segments depending on how loaded they are.
When an MPI process is overloaded, it sends patches to its neighbours along the Hilbert curve; therefore its segment becomes shorter.
Inversely, an underloaded process will receive patches from its neighbours; its segment becomes longer.

The following describes the dynamic load-balancing algorithm (it is summarized in Table~\ref{BalanceAlgorithm}).
First, the computational load $L_p$ of each patch $p$ is evaluated as 
\begin{equation}
 L_p=N_{\rm part}+C_{\rm cell}\times N_{\rm cells}+C_{\rm frozen}\times N_{\rm frozen}
\end{equation}
where $N_{part}$ is the number of active particles in the patch, $N_{\rm cells}$ is the number of cells in the patch,
$N_{\rm frozen}$ is the number of frozen (immobile) particles in the patch,
and $C_{\rm cell}$ and $C_{\rm frozen}$ are user-defined coefficients representing the computational cost of
cells (mostly solving Maxwell equation) and frozen particles.
In most cases, the active particles are the major source of computational load.
By default \Smilei uses $C_{\rm cell}=1$ and $C_{\rm frozen}=0.1$.
The total computational load  is $L_\mathrm{tot} = \Sigma_p L_p$ and the optimal computational load per process $L_\mathrm{opt}=L_\mathrm{tot}/N_{\rm MPI}$, where $N_{\rm MPI}$ is the number of MPI processes.
The balancing algorithm proceeds to a new decomposition of the Hilbert curve so that each segment carries a load as close to $L_\mathrm{opt}$ as possible.
This balancing process is typically done every 20 iterations in order to follow the dynamics of the simulation.
Frequent and small corrections give superior performance than rare and dramatic adjustments (see Fig.~\ref{freq_dlb}).

The amplitude of the readjustment is limited in the ``movement limitation'' phase: each MPI process keeps at least one of its original patches.
This reduces the performance impact of strong, high-frequency, oscillatory variations of the load observed in certain cases.
Once the segments are defined, the actual exchange of data is done.
\begin{table}\small
\caption{\label{BalanceAlgorithm}Load balancing algorithm used in \Smilei.
After initialization, a segment of patches is attributed to each MPI rank.
Before the actual exchange of patches, the segment length can be modified by the movement limitation procedure to prevent too catastrophic balancing to occur in a single step.}
\begin{tabular}[t]{ll}
\toprule
{\bf Initialization} & \cr
\toprule
Compute the total computational load & $L_\mathrm{tot}=\sum_p L_p$ \cr
\hspace{2cm}($L_p$ = load of the patch $p$)                     &  \cr
Compute the optimal load                    &   $L_\mathrm{opt} = L_\mathrm{tot}/N_{\rm MPI}$\cr
\hspace{2cm}($N_{\rm MPI}$ = number of MPI processes.)                     &  \cr
Set current load to zero &  $L=0$ \cr
Set number of patches in current segment to zero & $N=0$ \cr
Set currently treated MPI rank to zero  &  $R=0$\cr
\toprule
{\bf Segment size calculation} & \cr
\toprule
\textcolor{blue}{Loop over patches $p$}		& \cr
\hspace{0.5cm}Add patch $p$ load to current load 	& $L\mathrel{+}= L_p$\cr
\hspace{0.5cm}Add one patch in the current segment           & $N\mathrel{+}= 1$ \cr
\hspace{0.5cm}If current load exceeds the optimal one         & If $L > L_\mathrm{opt}:$\cr
\hspace{0.5cm}and is closer to optimal than the previous  &   \hspace{0.5cm}If $L-L_\mathrm{opt} < L_\mathrm{opt}-(L-L_p):$\cr
\hspace{0.5cm}Set length of the $R^{th}$ segment to $N$       & \hspace{1cm} $S[R]=N$\cr
\hspace{0.5cm}Set current load to zero                       & \hspace{1cm} $L=0$\cr
\hspace{0.5cm}Else                                           & \hspace{0.5cm}Else:\cr
\hspace{0.5cm}Set length of the $R^{th}$ segment to $N-1$       & \hspace{1cm} $S[R]=N-1$\cr
\hspace{0.5cm}Set current load to patch $p$ load         & \hspace{1cm} $L=L_p$\cr
\hspace{0.5cm}Start treating the next MPI rank               & \hspace{0.5cm}$R\mathrel{+}=1$\cr
 \toprule
{\bf Movement limitation}		& \cr
\toprule
\textcolor{blue}{Loop over MPI ranks $R$}		& \cr
\hspace{0.5cm}Evaluate index of last patch of rank $R$  &  $P_{\rm last} = \sum_{r=0}^RS[r]$ \cr
\hspace{0.5cm}Evaluate index of former first patch of rank $R$  &  $P_{\rm min} = \sum_{r=0}^{R-1}S_{\rm old}[r]$ \cr
\hspace{0.5cm}Evaluate index of former last patch of rank $R+1$ &  $P_{\rm max} = \sum_{r=0}^{R+1}S_{\rm old}[r]-1$ \cr
\hspace{0.5cm}If last patch doesn't reach the minimum & If $P_{\rm last} < P_{\rm min}:$ \cr
\hspace{0.5cm}Increase current segment &   \hspace{0.5cm}$S[R]\mathrel{+}= P_{\rm min}- P_{\rm last}$\cr
\hspace{0.5cm}If last patch exceeds the maximum & If $P_{\rm last} > P_{\rm max}:$\cr
\hspace{0.5cm}Reduce current segment &   \hspace{0.5cm}$S[R]\mathrel{-}= P_{\rm last}- P_{\rm max}$\cr
 \toprule
\end{tabular}
\end{table}

\subsection{Performances and scaling}\label{perf}

This section illustrates the efficiency of the chosen parallelization strategy and gives some insight on the optimization of the numerical parameters available to the user.

\subsubsection{MPI}

\begin{figure}
\centering
 \includegraphics[width=0.6\textwidth]{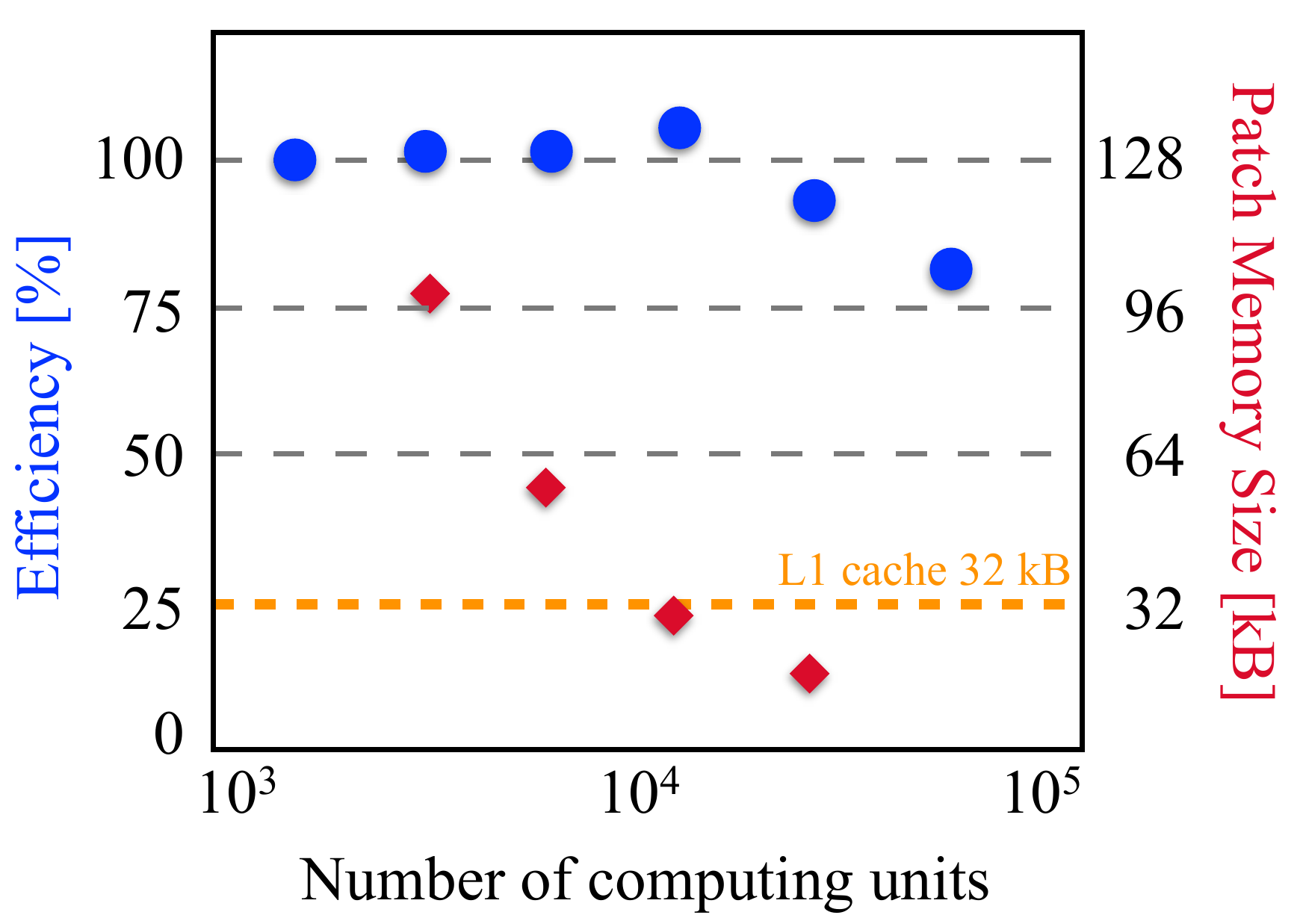}
 \caption{Pure MPI strong scaling of \Smilei in an homogeneous plasma case on the CINES/Occigen system. For this specific test case, the MPI domain size becomes smaller than the L1 cache around 20,000 cores.}
 \label{mpi_scaling}
\end{figure}

We study here the case of an MPI-only parallelization.  
The series of simulations presented here were performed on the CINES/Occigen system (Bull) and
focused a physics study devoted to Brillouin amplification of short laser pulses (see Sec.~\ref{sec:physSBS}), for which the plasma remains rather homogeneous
throughout the simulation.
Figure \ref{mpi_scaling} displays \Smilei's strong scaling for a pure MPI parallelization. 
The same simulation is run on different number of cores and a single MPI process is attached to each core.
As the number of cores increases, the size of the data handled by each core, or ``domain size'', decreases because the global domain is divided between all cores.
The efficiency remains close to 100\% as long as the domain size remains larger or equal to the L1 cache size.
For the specific global domain size used in this test, this occurs around 20,000 cores.
As the domain size approaches the L1 size, an improved cache-use slightly improves the performances.
Using a larger number of MPI processes then decreases the efficiency as the domain size becomes significantly smaller than the cache.
At this point, the system computing units occupation is too small to deliver proper performances, and the cost of additional MPI communications starts being significant.
Figure \ref{mpi_scaling} illustrates the fact that the pure MPI decomposition performs well in \Smilei up to the optimal (for this given simulation set-up) number of MPI domains.
There is no significant overhead due to MPI computations, their costs being much smaller than the computation in a standard case.  

In summary, MPI parallelization is good at handling homogeneous plasmas as long as the MPI domain sizes are not too small with respect to the L1 cache.

\subsubsection{MPI + openMP}\label{MPIplusopenMP}

In this section we present the performances achieved with the hybrid MPI+openMP parallelization described in section \ref{implementation}, when the plasma does not remain homogeneous.

\begin{figure}
\centering
 \includegraphics[width=\textwidth]{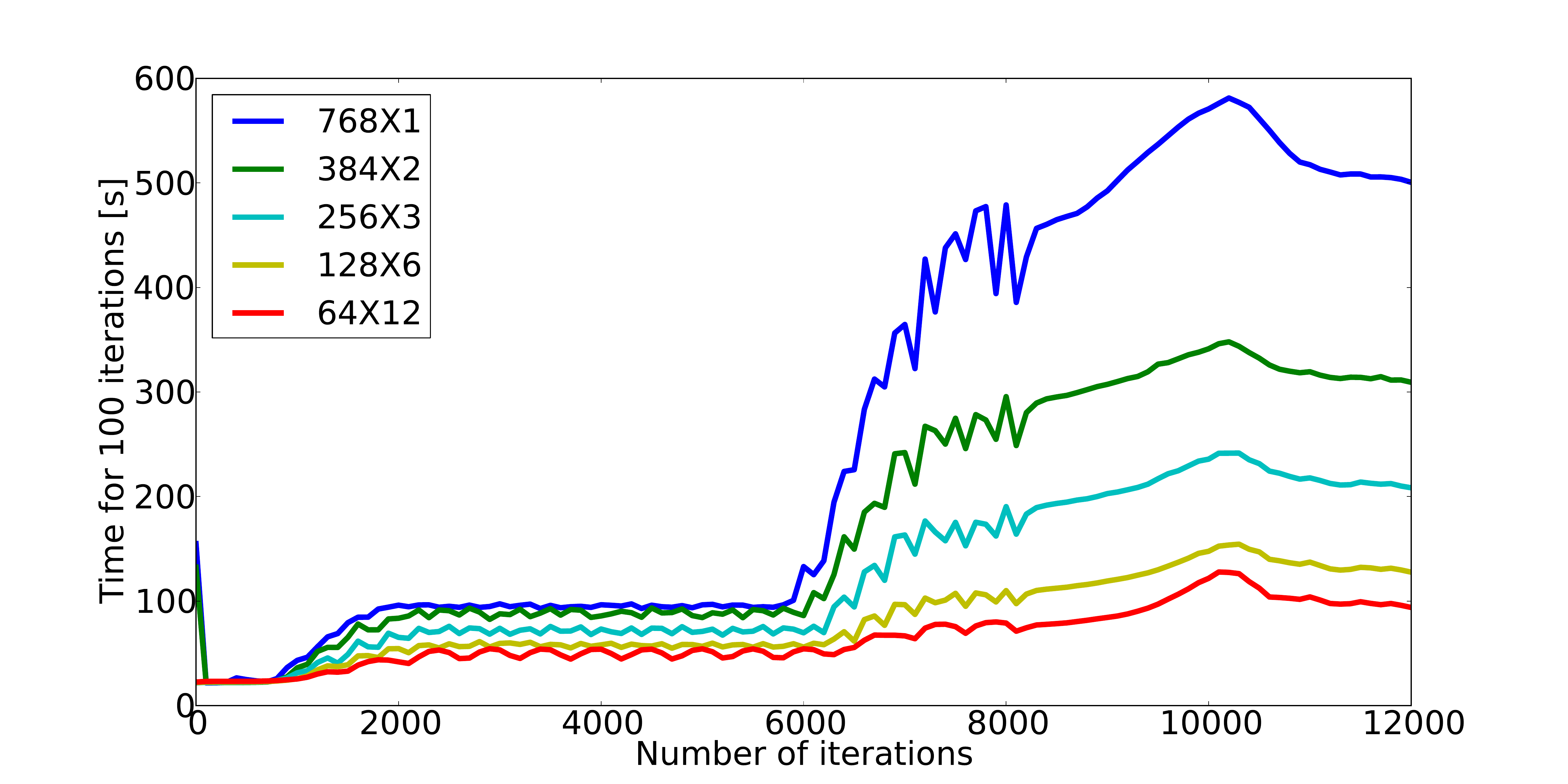}
 \caption{OpenMP load balancing effect.
 The plot displays the evolution of the wall-clock time necessary to complete 100 iterations as a function of the number of iterations already completed.
 The legend shows the total number of MPI processes and number of openMP threads per MPI process in the format MPI $\times$ openMP.}
 \label{openMP_extensibility}
\end{figure}

The case study is now, and until the end of the section, an ultra-high-intensity laser propagating in a plasma.
It is a typical laser wakefield acceleration case, well known for being strongly impacted by load imbalance~\cite{beck2016}.
It is a two-dimensional simulation consisting of $1024\times128$ patches (except in section \ref{sec_Npatches} where this parameter varies), each having $8\times5$ cells and 200 particles per cell. 
Each run ran on 32 nodes of the OCCIGEN system.
This represents 64 processors of 12 cores each for a total of 768 cores.
The plasma is initially homogeneous but load imbalance gradually builds up, then rises quickly after 6000 iterations before stabilizing.

Figure \ref{openMP_extensibility} shows the evolution of the wall-clock time necessary to complete 100 iterations as a function of the number of iterations already completed for different numerical settings.
The runs only differ by the number of openMP threads per MPI process and total number of MPI processes.
The total number of threads is kept constant and equal to 768 in order to have 1 thread per core.
The openMP dynamic scheduler is used in all cases.

Several interesting features can be noticed on figure \ref{openMP_extensibility}.
First, as long as the plasma is relatively homogeneous (first 1000 iterations), all runs perform similarly.
It means that the overhead for having a hybrid parallelization is negligible in this situation.
Later in the simulation, the pure-MPI case shows an extreme sensitivity to the load imbalance.
The wall-clock time spent to perform 100 iterations is almost multiplied by 20 with respect to the initial homogeneous plasma. 
Cases using more than one openMP thread per MPI process are much less sensitive to this effect.
And the more threads per MPI process, the smoother the performances.
This is perfectly in line with the local load balancing analysis given in section \ref{pb_omp}.
Nevertheless, even in the best case $64\times12$, a performance loss of a factor superior to 4 is still impacting the simulation.
This is explained by the fact that openMP can only balance the load within a given MPI domain.
Imbalance across MPI domains will keep slowing the simulation down.

Using more openMP threads, or equivalently more cores, per MPI process allows the use of larger MPI domains and therefore provides a better load balancing.
But the number of openMP threads is limited to the number of cores accessible on the shared memory system.
In our case, this is a single OCCIGEN node made of two processors of 12 cores each so up to 24 openMP threads could be used.
But going from 12 to 24 openMP threads per MPI process results in a drop of the performances because of the synchronization required between the two processors of the node.
The best performances are achieved when a single MPI process is given to each processor and when all cores of the processor are managed by the openMP scheduler.
The quality of the load balancing via the openMP dynamic scheduler thus directly depends on the size (in number of cores) of the processors composing the nodes.

\subsubsection{MPI + openMP + dynamic load balancing}

\begin{figure}
\centering
 \includegraphics[width=\textwidth]{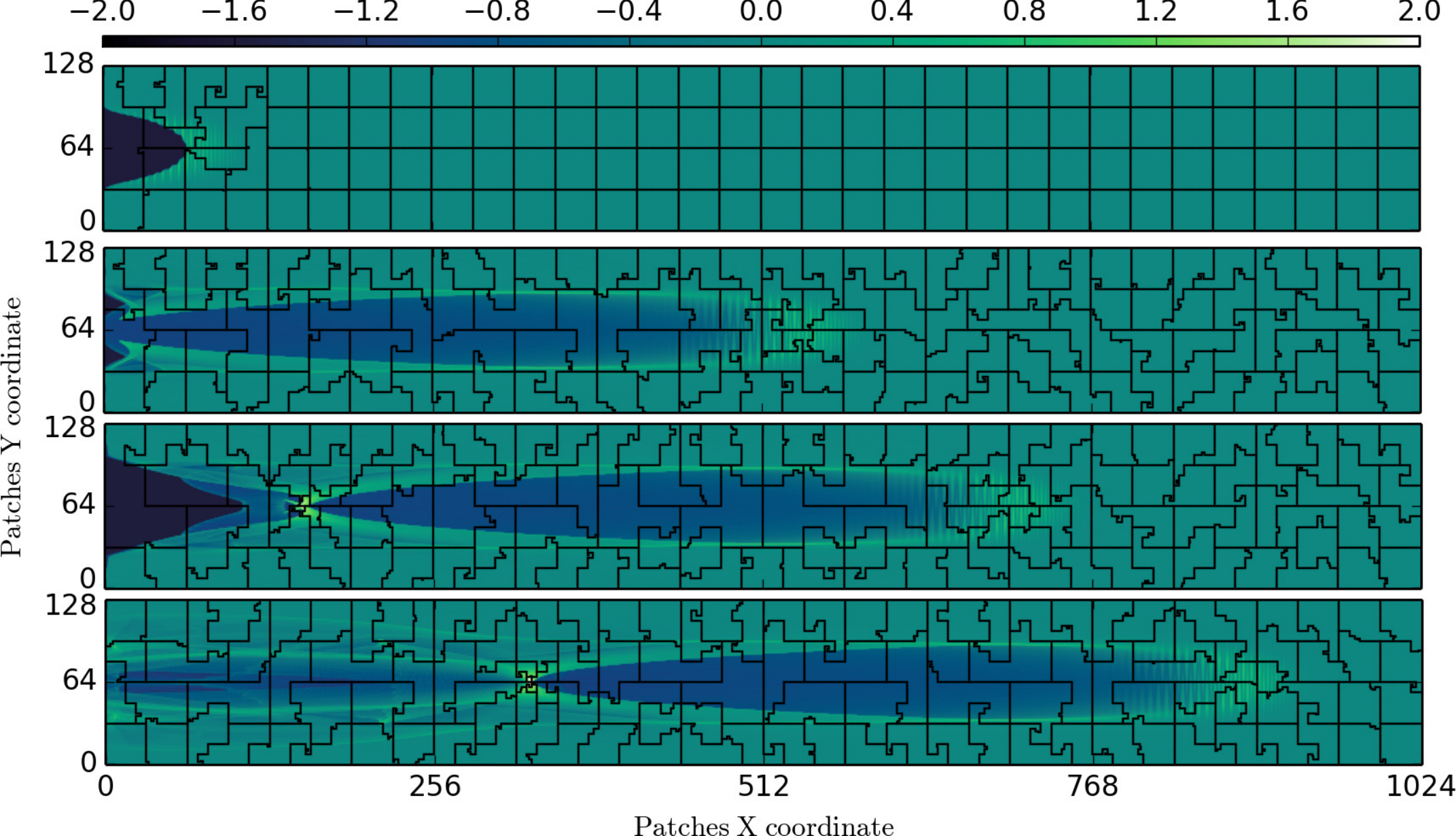}
 \caption{Evolution of the MPI domains shapes with respect to the computational load distribution.
 The colormap indicates the local imbalance $\rm I_{\rm loc}=\log_{\rm 10}\left(L_{\rm loc}/L_{\rm av}\right)$ where $\rm L_{\rm loc}$ is the local patch computational load
 and $\rm L_{\rm av}$ the average computational load.
 Black lines delimit the different MPI domains.
The laser enters an initially homogeneous plasma from the left side of the box and propagates towards the right.
The 4 panels show the entire simulation domain after 1600, 5480, 6820, 9080 iterations from top to bottom.}
 \label{mpi-evolution}
\end{figure}

This section presents results obtained with the dynamic load balancing (DLB) algorithm described in section \ref{loadmanagement}.
With DLB activated, the MPI domains are now capable of exchanging patches and therefore their shape evolves with respect to the computational load distribution.
Figure \ref{mpi-evolution} shows this distribution and the corresponding evolution of the shape of the MPI domains.
As expected, they tend to become smaller in areas where the computational load is high and, reversly, larger where the patches are underloaded.
The least loaded patches have approximately 1\% of the average patch load.
These under loaded patches are empty of particles and their computational load is limited to solving the maxwell equations.
On the opposite, the most loaded patches have almost 100 times as much computational load as the average and their load is completely dominated by particles.
Note that the computational load map looks very much like the density map for the simple reason that the computational load is mostly carried by the macro-particles
as in most PIC simulations.

\begin{figure}
\centering
 \includegraphics[width=\textwidth]{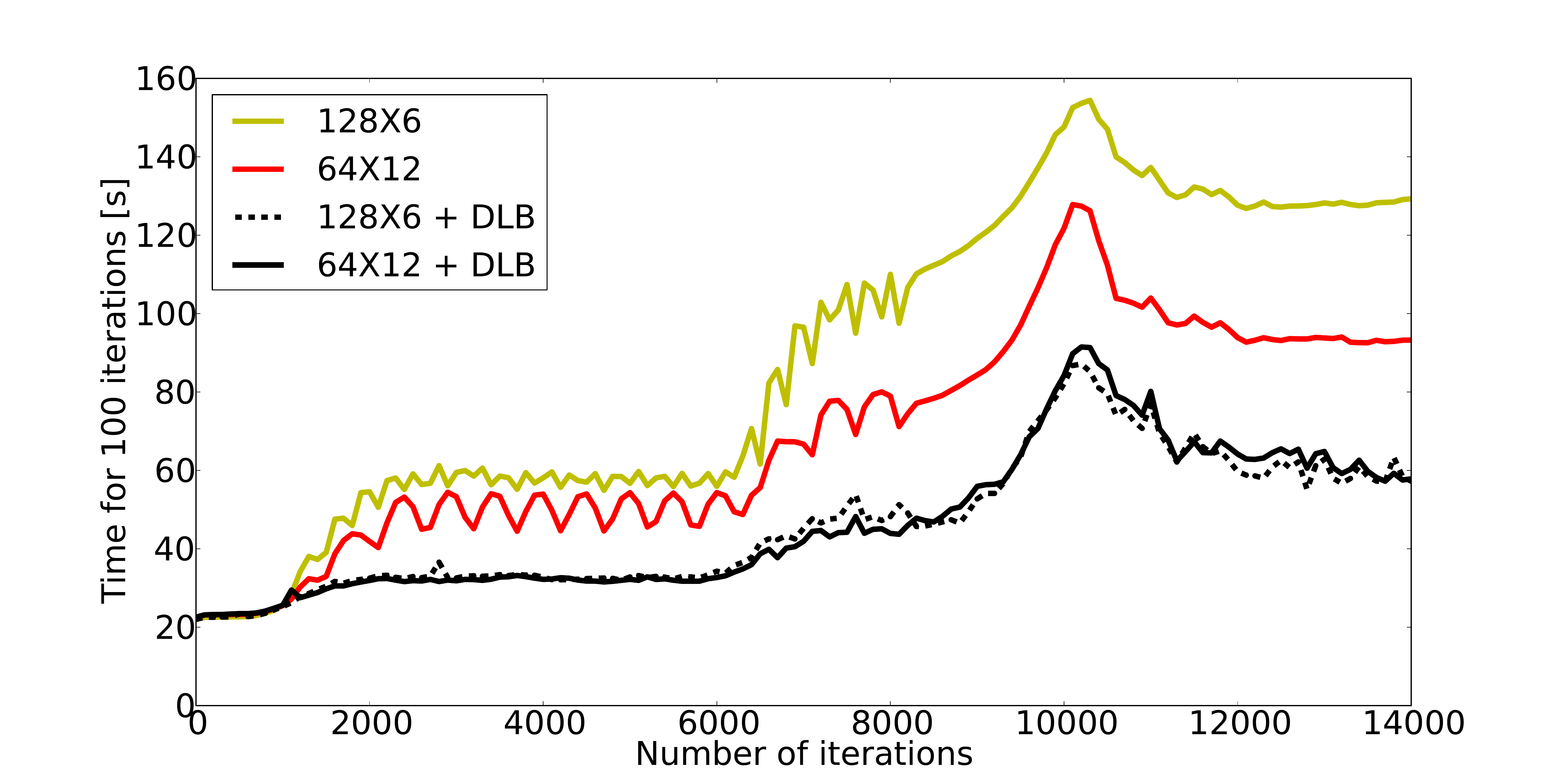}
 \caption{Dynamic load balancing (DLB) algorithm effect.
 The plot displays the evolution of the wall-clock time necessary to complete 100 iterations as a function of the number of iterations already completed.
 The legend shows the total number of MPI processes and number of openMP threads per MPI process in the format MPI $\times$ openMP.
 The red and yellow curves are replicas of figure \ref{openMP_extensibility}. }
 \label{dlb}
\end{figure}

Figure \ref{dlb} shows a performance comparison between the two best cases obtained in the previous section (without DLB) and the same cases with DLB activated.

The balancing here is done every 20 iterations and $C_{\rm cell}=2$.
No difference is observed during the balanced stage of the run (first 1000 iterations).
As expected, the cost of the balancing is negligible when actual balancing is not required.
In the imbalanced stage of the run, DLB provides an additional gain of almost 40\% with respect to the previous best case ``64$\times$12''.
A side benefit is also to reduce the dependency on the large number of openMP threads. 
Indeed, it appears that almost similar results are obtained with only 6 openMP threads when DLB is active.
As DLB balances the load between MPI processes, the local balancing via openMP becomes much less critical than before.
Note that the openMP parallelization remains necessary for an efficient fine grain balancing but it can be achieved with only a limited number of threads thus removing the dependency on a 
large shared memory hardware.

Note also that the cost of the imbalance is still significant in spite of all the efforts to balance the load.
The additional cost is mainly due to the imbalance of the particles communication cost which is not as well balanced as the computational cost of particles.

\subsubsection{Balancing frequency}

\begin{figure}
\centering
 \includegraphics[width=\textwidth]{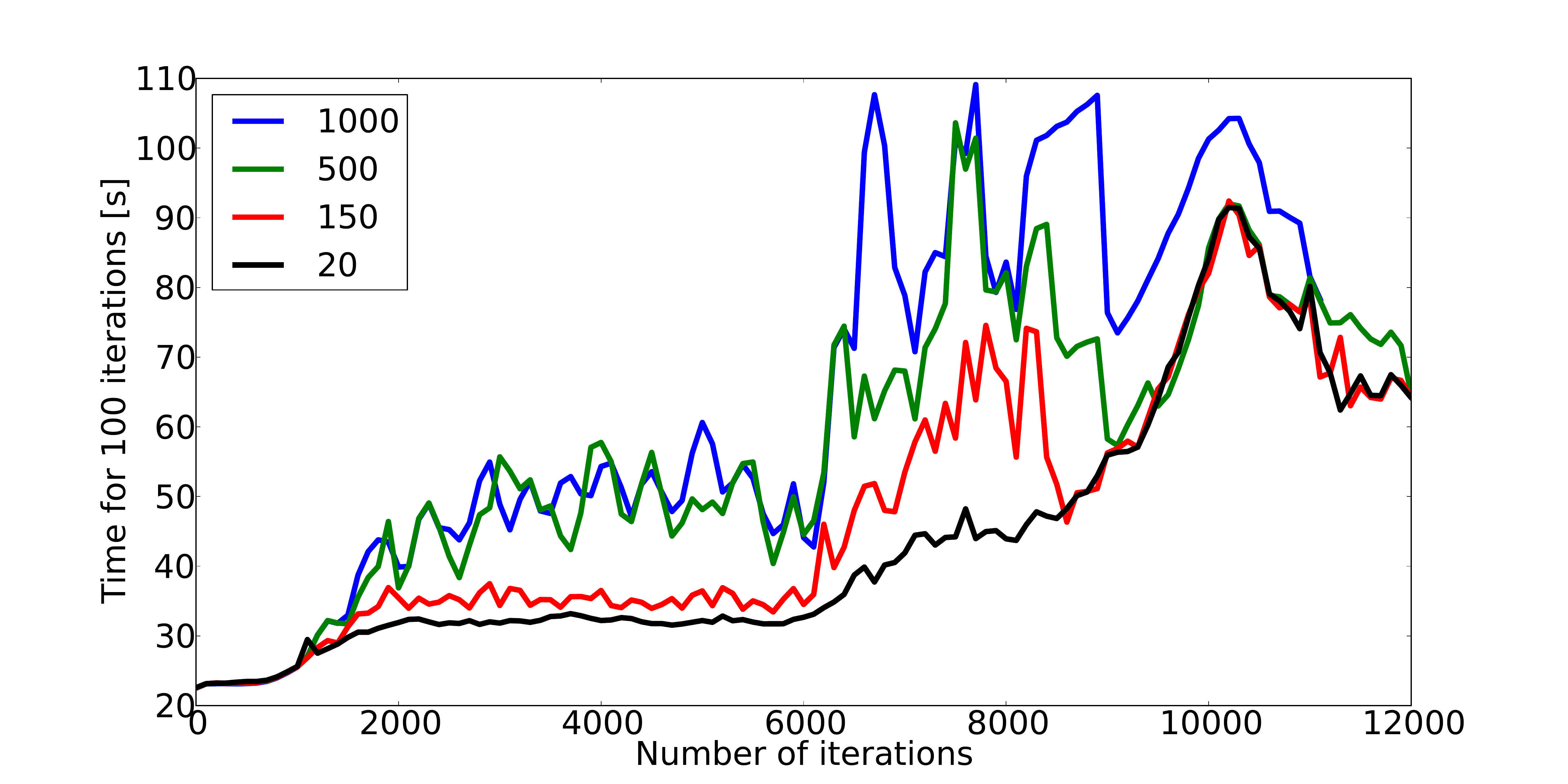}
 \caption{Evolution of the wall-clock time necessary to complete 100 iterations, as a function of the number of completed iterations, for four different values of $N_{\rm b}$ (the number of iterations between two load-balancing events).  The black curve is a replica of figure \ref{dlb}.}
 \label{freq_dlb}
\end{figure}

The load is balanced every $N_{\rm b}$ iterations. Figure \ref{freq_dlb} shows the influence of this parameter.
As expected, as $N_{\rm b}$ decreases, the load balance gets more accurate and the performances increase.
A more frequent balancing also means smaller adjustments each time.
Consequently, the overhead of load balancing remains low, even for low $N_{\rm b}$.
The cost of load balancing has a negative impact on performances only when $N_{\rm b}$ is much lower than the number of iterations over which imbalance builds up.

\subsubsection{Number of patches}\label{sec_Npatches}

The number of patches is an important parameter. 
It influences the quality of both the openMP implementation and DLB.
A large number of patches allows for finer-grain openMP parallelization and DLB, but also implies more ghost cells and synchronization costs.

\begin{figure}
\centering
 \includegraphics[width=\textwidth]{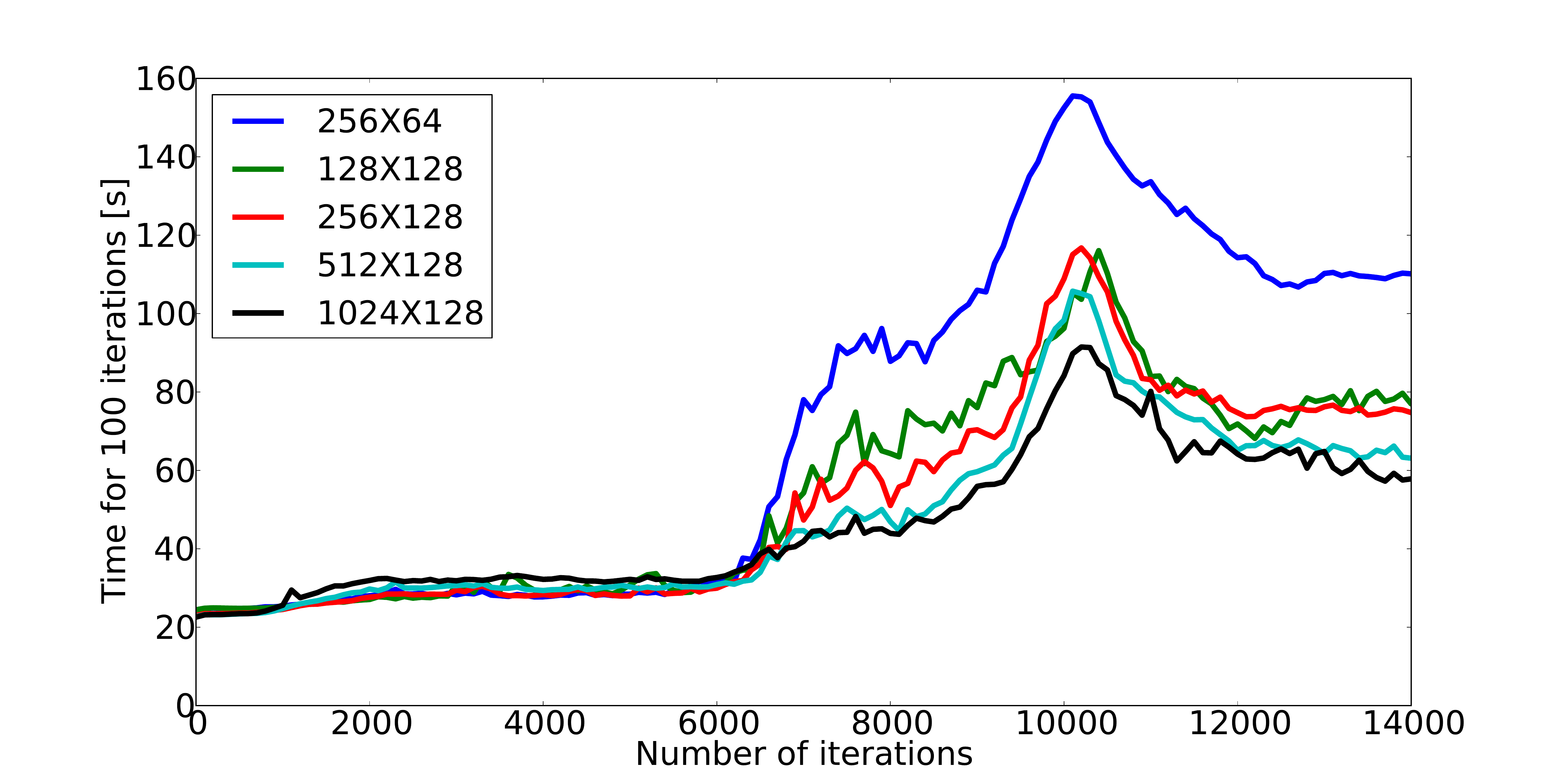}
 \caption{Evolution of the wall-clock time necessary to complete 100 iterations as a function of the number of iterations already completed, for various number of patches. The legend shows $N_{\rm patches}$ in the format $N_x\times N_y$ where $N_x$ and $N_y$ are respectively the number of patches in the $x$ and $y$ directions.
 The black curve is a replica of figure \ref{dlb}.}
 \label{fig_Npatches}
\end{figure}

Figure \ref{fig_Npatches} shows several interesting features.
First, while imbalance is weak (between iterations 2000 and 6000), having more patches noticeably costs additional synchronization. 
The cost of particles dynamics far outweighs the cost of synchronization, but this overhead is measurable.
On the other hand, in the second stage of the simulation where a strong imbalance kicks in (after iteration 6000), having smaller and more numerous patches is clearly beneficial.

Another interesting result comes from the comparison between the $256\times64$ and $128\times128$ cases.
Despite an equal number of patches, much better performances are achieved in the latter topology.
This can be explained by the way the Hilbert curve is generated.
The case of a ``square'' topology ($N_x=N_y$, $N_x$ being a power of 2) corresponds to the usual Hilbert curve.
If the number of patches is larger in one direction, the Hilbert curve is generated over a square of the smaller dimension's size, then repeated along the longer dimension.
The constraint on $N_x$ and $N_y$ being powers of 2 remains but they can be different.
The cost of this generalization is that the resulting space-filling curve loses some of its compactness.
This translates into additional synchronization cost.

\section{Additional modules}\label{secAdditionalModules}

To answer to the users various needs, additional modules have been implemented in \Smilei.

\subsection{Electric field and current density filters}\label{sec:filters}

Particle-in-Cell codes relying on the FDTD Maxwell solvers are known to face serious problems when dealing with 
relativistic beams of particles and/or relativistically drifting plasmas~\cite{godfrey1974} such as encountered in laser wakefield acceleration~\cite{lehe2014}
or in relativistic astrophysics simulation~\cite{nuter2016}.
Numerical dispersion indeed results in a spurious, direction-dependent reduction of the light waves velocity. 
As a result, ultra-relativistic particles may artificially catch up with the light waves giving rise to the grid-Cerenkov instability~\cite{godfrey1974}.
Various methods have been proposed to deal with this instability: in particular, time-filtering of the electric fields~\cite{greenwood2004} and spatial-filtering of the current density~\cite{vay2011}.

\Smilei specifically uses the Friedman time filter on the electric fields~\cite{greenwood2004}.
If required by the user, this filter consists in replacing the electric field in the Maxwell-Faraday solver by a time-filtered field:
\begin{eqnarray}
\vE^{(n)} = \left(1+\frac{\theta}{2}\right)\,\vE^{(n)} - \left(1-\frac{\theta}{2}\right)\,\vE^{(n-1)}+\frac{1}{2}(1-\theta)^2\,\bar{\vE}^{(n-2)}\,,
\end{eqnarray}
where $\bar{\vE}^{(n-2)}=\vE^{(n-2)}+\theta\,\bar{\vE}^{(n-3)}$, and the filtering parameter $\theta \in [0,1]$ is an input parameter defined by the user.

A multi-pass bilinear filter on the current density has also been implemented~\cite{vay2011}.
Each pass consists in a 3-points spatial averaging (in all spatial dimensions) of the current, so that the filtered current density (here defined at location $i$ on a one-dimensional grid) is recomputed as:
\begin{eqnarray}
J_i^f = \frac{1}{2}J_i + \frac{J_{i+1}+J_{i-1}}{4}\,.
\end{eqnarray}
Current filtering, if required by the user, is applied before solving Maxwell's equation, and the number of passes is an input parameter defined by the user.

Both methods can be used together or separately and have allowed to satisfactorily reduce the numerical grid-Cerenkov instability when dealing with relativistically drifting electron-positron plasmas in the framework of collisionless shock studies (see Sec.~\ref{sec:physShock}).

\subsection{Antennas}

\begin{figure}
	\centering
	\includegraphics[width=6cm]{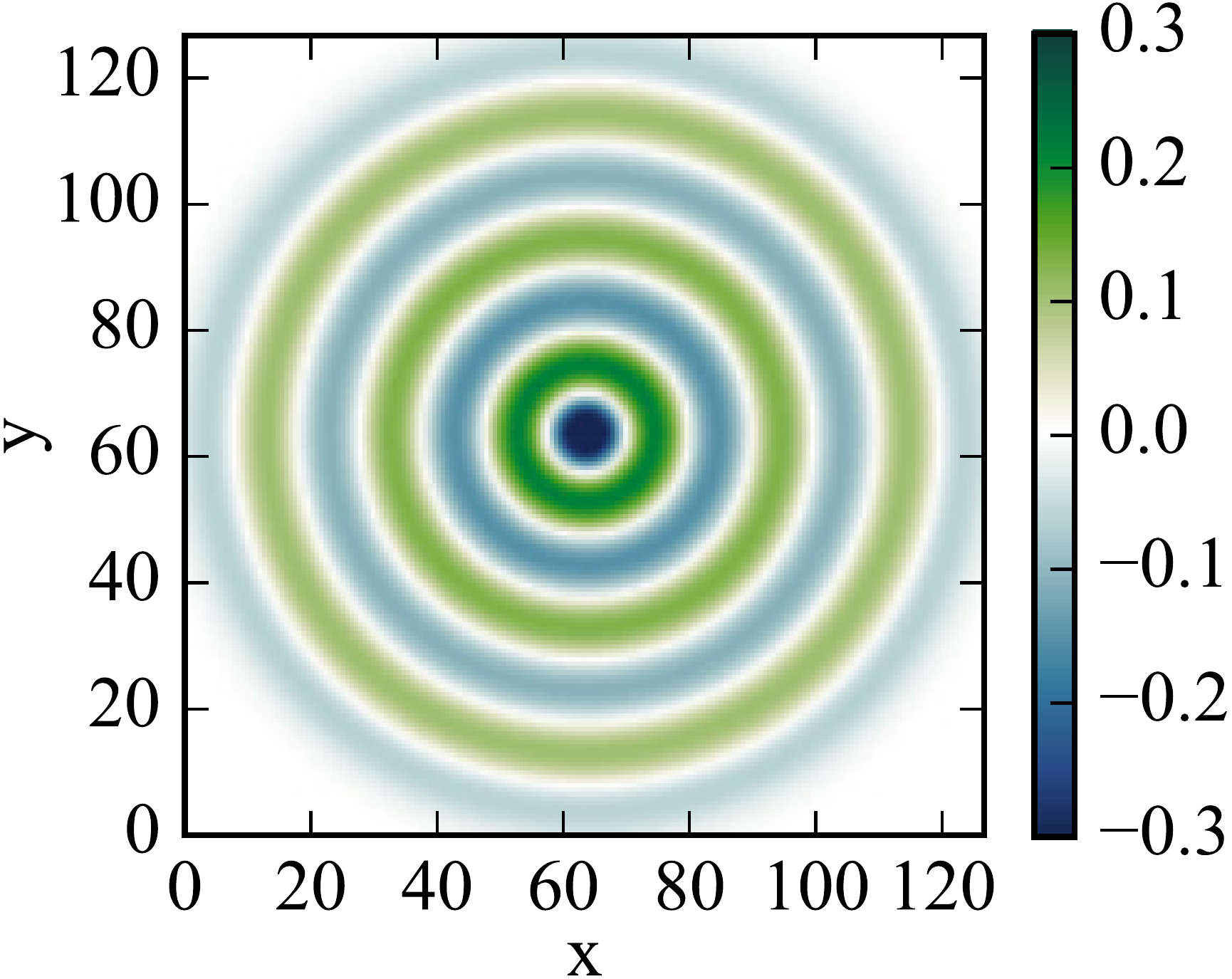}
	\caption{$E_z$ electric field (at $t=60$) generated by an oscillating $J_z$ source
		placed at the center of an empty box.
		\label{fig:antenna}}
\end{figure}

After the particle projection and before the Maxwell solver execution, custom additional
currents can be introduced. These additional, user-defined, currents are referred to as {\it antennas} in \Smilei. 
The user provides both spatial and temporal profiles for chosen currents $J_x$, $J_y$ and/or $J_z$.
Antennas may be used, for instance, to apply external electromagnetic sources anywhere in the box.
An example is provided in Fig.~\ref{fig:antenna}, showing the electric field induced by an oscillating
$J_z$ current applied within a small circular region in the center of an empty box.
A circular wave is generated from the antenna and propagates outwards.

\subsection{Field ionization}\label{sec:fieldIonization}

Field ionization is a process of particular importance for laser-plasma interaction in the ultra-high intensity regime.
It can affect ion acceleration driven by irradiating a solid target with an ultra-intense laser~\cite{nuter2011}, or can
be used to inject electrons through the accelerating field in laser wakefield acceleration~\cite{umstadter1996}. 
This process is not described in the standard PIC (Vlasov-Maxwell) formulation, and an {\it ad hoc} description needs to be implemented.
A Monte-Carlo module for field ionization has thus been developed in \Smilei, closely following the method proposed by Nuter {\it et al.}~\cite{nuter2011}.

\subsubsection{Physical model}

This scheme relies on the quasi-static rate for tunnel ionization derived in Refs.~\cite{perelomov1966,perelomov1967,ammosov1986}.
Considering an ion with atomic number $Z$ being ionized from charge state $Z^\star$ to $Z^\star+1 \le Z$ in an electric field $\mathbf{E}$ of magnitude $\vert E\vert$, the ionization rate reads:
\begin{eqnarray}\label{ionisationRate_qs}
\Gamma_{\rm Z^{\star}} = A_{n^\star,l^\star}\,B_{l,\vert m\vert}\,I_{Z^{\star}}\,\left( \frac{2 (2 I_{Z^{\star}})^{3/2}}{\vert E\vert} \right)^{2n^\star-\vert m \vert -1}\,\exp\!\left( -\frac{2 (2 I_{Z^{\star}})^{3/2}}{3 \vert E\vert}  \right)\,,
\end{eqnarray}
where $I_{Z^{\star}}$ is the $Z^{\star}$ ionization potential of the ion, $n^\star=(Z^\star+1)/\sqrt{2 I_{Z^{\star}}}$ and $l^\star=n^\star-1$ denote
the effective principal quantum number and angular momentum, 
and $l$ and $m$ denote the angular momentum and its projection on the laser polarization direction, respectively.
$\Gamma_{\rm qs}$, $I_{Z^{\star}}$ and $E$ are here expressed in atomic units\footnote{$\Gamma_{\rm qs}$ is in units of $\hbar/(\alpha^2 m_e c^2)$ with $\hbar$ the Planck constant and $\alpha$ the fine-structure constant.
$I_{Z^{\star}}$ is in units of $\alpha^2 m_e c^2$ (also referred to as Hartree energy) and $E$ is in unit of $\alpha^3 m_e^2 c^3/(e \hbar)$.}.
The coefficients $A_{n^\star,l^\star}$ and $B_{l,\vert m\vert}$ are given by:
\begin{subequations}
\begin{eqnarray}
A_{n^\star,l^\star}           \!\!\!&=&\!\!\! \frac{2^{2n^\star}}{n^\star\,\Gamma(n^\star+l^\star+1)\,\Gamma(n^\star-l^\star)},\\
B_{l,\vert m\vert} \!\!\!&=&\!\!\! \frac{(2l+1)(l+\vert m\vert)!}{2^{\vert m\vert} \vert m\vert! (l-\vert m\vert)!}\,,
\end{eqnarray}
\end{subequations}
where $\Gamma(x)$ is the gamma function. Note that considering an electric field $E=\vert E\vert\,\cos(\omega t)$ oscillating in time at the frequency $\omega$, averaging Eq.~\eqref{ionisationRate_qs} over a period $2\pi/\omega$ leads to the well-known cycle-averaged ionization rate:
\begin{eqnarray}\label{eq:ADKrate}
\Gamma_{\rm ADK} = \sqrt{\frac{6}{\pi}}A_{n^\star,l^\star}\,B_{l,\vert m\vert}\,I_{Z^{\star}}\,\left( \frac{2 (2 I_{Z^{\star}})^{3/2}}{\vert E\vert} \right)^{2n^\star-\vert m \vert -3/2}\,\exp\!\left( -\frac{2 (2 I_{Z^{\star}})^{3/2}}{3 \vert E\vert}  \right)\,.
\end{eqnarray}

In \Smilei, following Ref.~\cite{nuter2011}, the ionization rate Eq.~\eqref{ionisationRate_qs} is computed for $\vert m \vert=0$ only.
Indeed, as shown in Ref.~\cite{ammosov1986}, the ratio $R$ of the ionization rate computed for $\vert m\vert=0$ by the rate computed for $\vert m\vert=1$ is:
\begin{eqnarray}
R = \frac{\Gamma_{{\rm qs},\vert m \vert = 0}}{\Gamma_{{\rm qs},\vert m \vert = 1}} =  2\frac{(2\,I_{Z^{\star}})^{3/2}}{\vert E\vert} \simeq 7.91\,10^{-3} \,\,\frac{(I_{Z^{\star}}[\rm eV])^{3/2}}{a_0\,\hbar\omega_0[\rm eV]}\,,
\end{eqnarray}
where, in the practical units formulation, we have considered ionization by a laser with normalized vector potential $a_0=e\vert E\vert/(m_e c \omega_0)$, and photon energy $\hbar\omega_0$ in eV. Typically, ionization by a laser with wavelength $1~{\rm \mu m}$ (correspondingly $\hbar \omega_0 \sim 1~{\rm eV}$) occurs for values of $a_0\ll 1$ (even for large laser intensities for which ionization would occur during the rising time of the pulse) while the ionization potential ranges from a couple of eV (for electrons on the most external shells) up to a few tens of thousands of eV (for electrons on the internal shell of high-$Z$ atoms). As a consequence, $R\gg1$, and the probability of ionization of an electron with magnetic quantum number $\vert m \vert=0$ greatly exceeds that of an electron with $\vert m \vert = 1$.

Finally, it should be stressed that simulations involving field ionization (the same is true for those involving binary collisions and/or collisional ionization as detailed in the next two Secs.~\ref{sec:collisions} and~\ref{sec:collIonization}) cannot be arbitrarily scaled. The reference time normalization $\omega_r^{-1}$ needs to be prescribed. In \Smilei, this is done at initialization, the user having to define the reference angular frequency in SI units whenever one of this additional module is used.

\subsubsection{Monte-Carlo procedure}

In \Smilei, tunnel ionization is treated for each species (defined by the user as subject to field ionization) right after field interpolation and before applying the pusher. 
For all quasi-particles (henceforth referred to as quasi-ions) of the considered species, a Monte-Carlo procedure has been implemented that allows to treat multiple ionization events in a single timestep. It relies on the cumulative probability derived in Ref.~\cite{nuter2011}:
\begin{eqnarray}
F_k^{Z^{\star}} = \sum_{j=0}^k P_j^{Z^{\star}}\,,
\end{eqnarray}
to ionize from 0 to $k$ times a quasi-ion with initial charge state $Z^{\star}$ during a simulation timestep $\Delta t$, $P_j^{Z^{\star}}$ being the probability to ionize exactly $j$ times this ion given by:
\begin{equation}\label{eq:MultIonProb}
  P^i_k = \left\{
  \begin{array}{ll}
  \bar{P}^i
  &
  \quad\mathrm{if}\quad k=0
  \\
  \sum\limits_{p=0}^{k-1} R^{i+k}_{i+p} \left(\bar{P}^{i+k} - \bar{P}^{i+p}\right)
  \prod\limits_{j=0,j\ne p}^{k-1} R^{i+p}_{i+j}
  &
  \quad\mathrm{if}\quad 0<k<k_\mathrm{max}
  \\
  \sum\limits_{p=0}^{k-1} \left[ 1+R^{i+k}_{i+p}\left(\frac{\Gamma_{i+k}}{\Gamma_{i+p}}\bar{P}^{i+p} - \bar{P}^{i+k}\right) \right]
  \prod\limits_{j=0,j\ne p}^{k-1} R^{i+p}_{i+j}
  &
  \quad\mathrm{if}\quad k=k_\mathrm{max}\,,
  \end{array}
  \right.
\end{equation}
with $\bar{P}^i = \exp(-\Gamma_i\Delta t)$ the propability to {\it not} ionize an ion in initial charge state $i$, and $R_{\alpha}^{\beta}=(1-\Gamma_{\beta}/\Gamma_{\alpha})^{-1}$ with $\Gamma_i$ the $i^{th}$ ionization rate given by Eq.~\eqref{ionisationRate_qs}.

The Monte-Carlo scheme proceeds as follows.
A random number $r$ with uniform distribution between 0 and 1 is picked.
If $r$ is smaller than the probability $P_0^{Z^{\star}}$ to not ionize the quasi-ion, then the quasi-ion is not ionized during this time step.
Otherwise, we loop over the number of ionization events $k$, from $k=1$ to $k_{\rm max}=Z-Z^{\star}$ (for which $F_{k_{\rm max}}^{Z^{\star}}=1$ by construction), until $r<F_k^{Z^{\star}}$. At that point, $k$ is the number of ionization events for the quasi-ion. A quasi-electron is created with the numerical weight equal to $k$ times that of the quasi-ion, and with the same velocity as this quasi-ion. The quasi-ion charge is also increased by $k$.

Finally, to account for the loss of electromagnetic energy during ionization, an {\it ionization current} ${\bf J}_{\rm ion}$ is projected onto the simulation grid~\cite{nuter2011,mulser1998} such that
\begin{eqnarray}
{\bf J}_{\rm ion} \cdot {\bf E} = \Delta t^{-1}\,\sum_{j=1}^k I_{Z^{\star}+k-1}\,.
\end{eqnarray}

\subsubsection{Benchmarks}

In what follows, we present two benchmarks of the field ionization model implemented in \Smilei.
Both benchmarks consist in irradiating a thin (one cell long) neutral material (hydrogen or carbon)
with a short (few optical-cycle long) laser with wavelength $\lambda_0 = 0.8~{\rm \mu m}$.

\begin{figure}
	\includegraphics[width=0.9\textwidth]{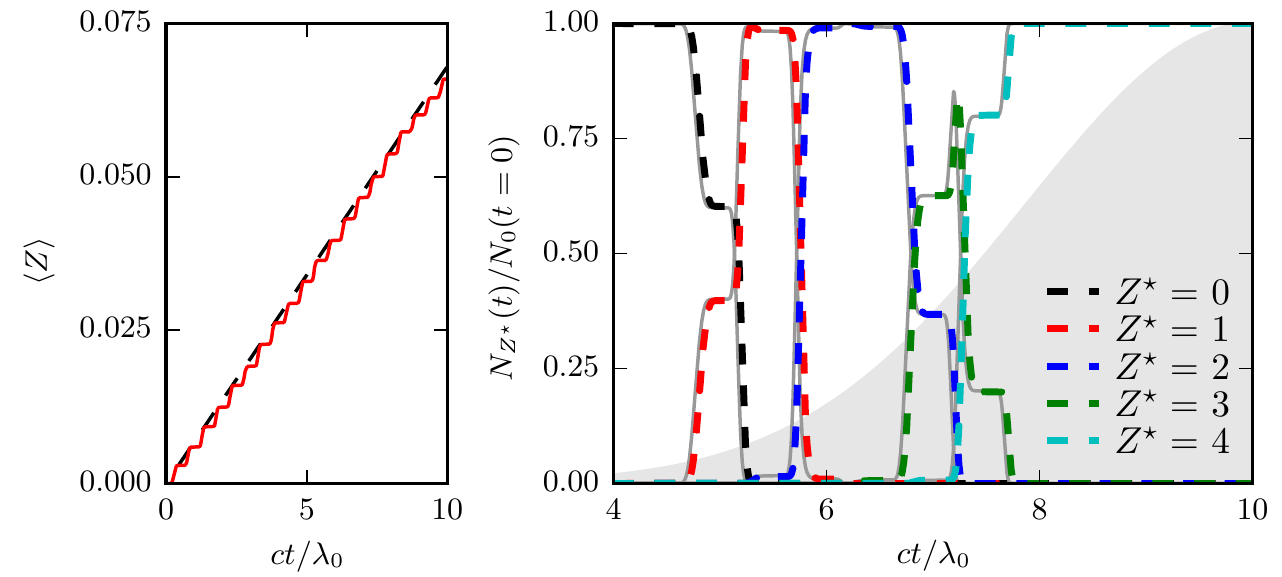}
	\caption{Results of two benchmarks for the field ionization model. Left: Average charge state of hydrogen ions as a function of time when irradiated by a laser. The red solid line corresponds to PIC results, the dashed line corresponds to theoretical predictions using the cycle-averaged ADK growth rate of Eq.~\eqref{eq:ADKrate}. Right: Relative distribution of carbon ions for different charge states as a function of time. Dashed lines correspond to PIC results, thin gray lines correspond to theoretical predictions obtained from Eq.~\eqref{eq:rateEqs}. The Gaussian gray shape indicates the laser electric field envelope.
		\label{Fig:tunnelIonization}}
\end{figure}
		
In the first benchmark, featuring hydrogen, the laser intensity is kept constant at $I_L = 10^{14}~{\rm W/cm^2}$, corresponding to a normalized vector potential $a_0 \simeq 6.81 \times 10^{-3}$, over 10 optical cycles. The resulting averaged ion charge in the simulation is presented as a function of time in Fig.~\ref{Fig:tunnelIonization}, left panel. It is found to be in excellent agreement with the theoretical prediction (dashed in Fig.~\ref{Fig:tunnelIonization}, left panel) considering the cycle averaged ionization rate $\Gamma_{\rm ADK} \simeq 2.55\times10^{12}~{\rm s^{-1}}$ computed from Eq.~\eqref{eq:ADKrate}.

The second benchmark features a carbon slab. The laser has a peak intensity $I_L = 5 \times 10^{16}~{\rm W/cm^2}$, corresponding to a normalized vector potential $a_0 \simeq 1.52 \times 10^{-1}$, and a gaussian time profile with full-width-at-half-maximum (FWHM) $\tau_L=5~\lambda_0/c$ (in terms of electric field). Figure~\ref{Fig:tunnelIonization}, right panel shows, as function of time, the relative distribution of carbon ions for different charge states (from 0 to $+4$). These numerical results are shown to be in excellent agreement with theoretical predictions obtained by numerically solving the coupled rate equations on the population $N_i$ of each level $i$:
\begin{eqnarray}\label{eq:rateEqs}
\frac{dN_i}{dt} = (1-\delta_{i,0}) \, \Gamma_{i-1}\,N_{i-1}  -  (1-\delta_{i,Z})\, \Gamma_{i}\,N_{i}\,,
\end{eqnarray}
with $\delta_{i,j}$ the Kroenecker delta, and $\Gamma_i$ the ionization rate of level $i$. Note also that, for this configuration, $\Delta t \simeq 0.04~{\rm fs}$ is about ten times larger than the characteristic time $\Gamma_{\rm ADK}^{-1} \simeq 0.006~{\rm fs}$ to ionize ${\rm C}^{2+}$ and ${\rm C}^{3+}$ so that multiple ionization from ${\rm C}^{2+}$ to ${\rm C}^{4+}$ during a single timestep does occur and is found to be correctly accounted for in our simulations.

\subsection{Binary collisions\label{sec:collisions}}

As detailed in Sec.~\ref{secPICMethod}, the PIC method aims at describing
the self-consistent evolution of a collisionless plasma by solving the coupled system of Vlasov-Maxwell Eqs.~\eqref{eq_Vlasov} -- \eqref{eq_Maxwell}.
Consequently, PIC codes must introduce additional modules to account for collisions.
In \Smilei, the effects of relativistic collisions have been implemented following the scheme described in Ref.~\cite{perez2012}.
It is based on Nanbu's approach \cite{nanbu1997}, with the addition of a few enhancements:
relativistic particles, low-temperature correction to the collision rate,
and variable Coulomb logarithm. We briefly review this scheme here and illustrate
it with typical applications.

Nanbu's theory first considers real particles, assuming that collisions occur many times during one time-step, and that each collision
introduces a deflection angle $\theta\ll 1$ (although the total deflection angle may be large). By simulating
a large number of Coulomb collisions, he finds that the total deflection angle $\left<\chi\right>$
is well described by a unique function of $s=\left<\theta^2\right> N/2$,
where $\left<\theta^2\right>$ is the expectation of
$\theta^2$ and $N$ is the number of collisions during one time-step ($N\gg 1$). He also provides a
probability density function $f(\chi)$ to pick randomly the deflection angle $\chi$ accumulated during
one time-step.

Colliding each quasi-particle with all other quasi-particles nearby would be time-consuming.
Instead, quasi-particles are randomly paired so that each collides with
only one other quasi-particle at a given time-step. After many time-steps, each of them
will have sampled the overall distribution of target particles.
This pairing follows Ref.~\cite{nanbu1998}.
It is split in two cases: {\it intra}-collisions, when a group of particles
collides with itself, and {\it inter}-collisions, when two distinct groups of particles collide.
Intra-collisions can occur, for example, within all the electrons in the plasma. In this case,
the group is split in two halves and one half is randomly shuffled to provide random pairs.
In the case of inter-collisions, the two halves are simply the two groups: only one group is randomly
shuffled. When the two halves do not contain the same number of particles, the extra particles
(not paired yet) are randomly assigned a companion particle from those which have already
been paired. Thus, one particle may participate in several pairs.

Whereas many codes naturally make quasi-particles collide within their own cell,
\Smilei makes them collide with all those in the same patch.
This is only accurate when the plasma parameters do not vary significantly within one patch,
but it greatly reduces the amount quasi-particle sorting required.

The parameter $s$ is normally calculated from the point of view of one particle traversing a
cloud of numerous target particles. However, this picture is broken by the quasi-particles
of the PIC code having variable weights.
To ensure a deflection angle common to both quasi-particles (in the center-of-mass frame), thus momentum conservation,
the parameter $s$ must be the same from both quasi-particles' point of views.
Unfortunately, this condition is not fulfilled when the weights or densities are different. 
Ref.~\cite{nanbu1998} provides a detailed solution consisting in
modifying $s$ by a factor which makes it symmetric when exchanging the quasi-particles in a pair.
This modification is later compensated by randomly picking quasi-particles
which will not actually undergo a deflection, so that energy and momentum are conserved in average.

In addition to these considerations, Ref.~\cite{perez2012} provides the relativistic expressions
of $s$ and $\chi$, specifying the relativistic changes of frames, and gives corrections
for low-temperature plasmas and a varying Coulomb logarithm. Note that these expressions,
just like those of the field ionization module, cannot be normalized to dimension-less equations
when using \Smilei's units: the value of the reference frequency $\omega_r$ must be specified
in the SI system of units.

\begin{figure}
	\includegraphics{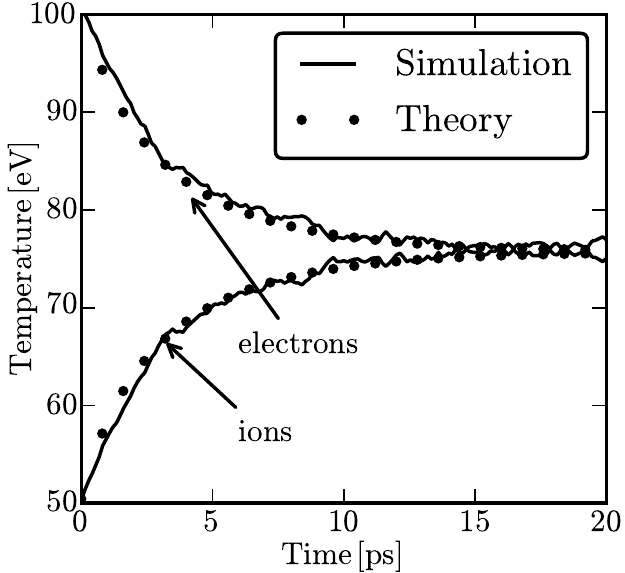}
	\includegraphics{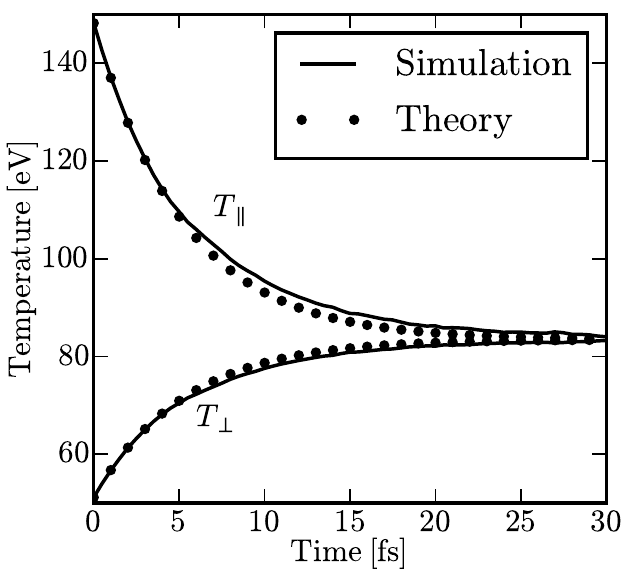}
	\caption{Left: thermalization by collisions between ions and electrons of an hydrogen plasma.
		Right: temperature isotropisation of an electron plasma.
		\label{fig:thermalisation}}
\end{figure}

As a first example of the possible effects of collisions, let us consider the thermalization between
ions and electrons: a fully-ionized hydrogen plasma of density $10^{22}$ cm$^{-3}$ is set with an
ion temperature of 50 eV and an electron temperature of 100 eV. The left panel of Fig.~\ref{fig:thermalisation}
shows the evolution of both temperatures due to the {\it e-i} collisions, well matched by the
theoretical solution taken from Ref.~\cite{NRL}. Note that, for a simpler comparison
between simulation and theory, the Coulomb logarithm was set to 5 and {\it e-e} and {\it i-i}
collisions were also applied to ensure maxwellian distributions of each species.

This example of the effect of {\it e-i} ({\it inter-}) collisions has the following counterpart for
{\it e-e} ({\it intra-}) collisions. We set an hydrogen plasma of the same density with
an anisotropic electron temperature: $T_\parallel=150$ eV and $T_\perp=50$ eV. 
The right panel of Fig.~\ref{fig:thermalisation} shows the evolution of both temperatures
due to the {\it e-e} collisions, again well matched by the theoretical solution in Ref.~\cite{NRL}.

\begin{figure}
	\centering
	\includegraphics{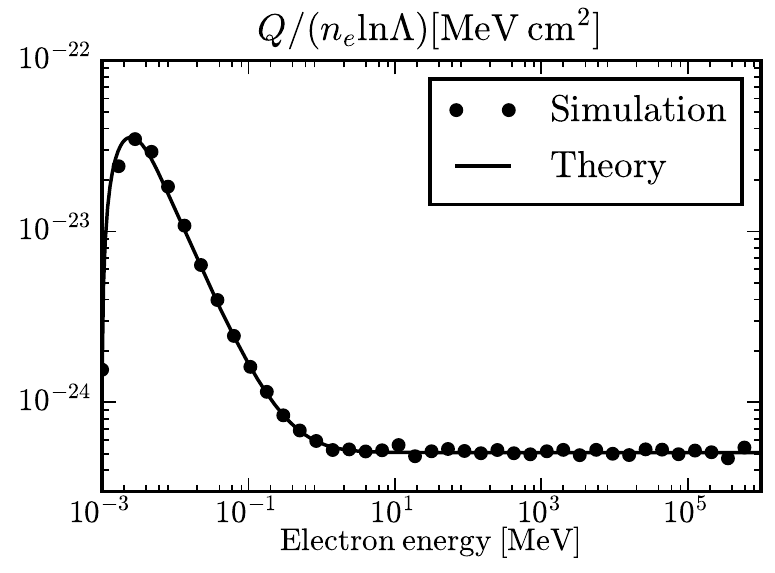}
	\caption{Stopping power $Q$ of a fully-ionized hydrogen plasma of density $n_e=10^{22}$ cm$^{-3}$
		and temperature 1 keV, divided by $n_e$ and by the Coulomb logarithm $\ln\Lambda$,
		as a function of the incident electron energy.
		\label{fig:stopping_power}}
\end{figure}

Another important consequence of Coulomb collisions is the slowing down of high-energy electrons 
passing through an ionized plasma (due to {\it e-e} collisions). We simulated this situation for
various electron energies traversing a fully-ionized hydrogen plasma, and present the resulting
stopping power in Fig.~\ref{fig:stopping_power}. It is in good agreement with theoretical
calculations from Ref.~\cite{frankel1979}.

\subsection{Collisional ionization}\label{sec:collIonization}

The collision module described in section~\ref{sec:collisions} hosts an electron-ion impact-ionization
model that makes use of the particle pairing to compute the ionization probability of each pair. The scheme
is identical to that of Ref.~\cite{perez2012} with the exception of a few improvements detailed in the following.

The overall approach consists in calculating quantities 
averaged over all orbitals of a given ion with atomic number $Z$ and charge $Z^\star$, instead of dealing
with each orbital individually. This greatly reduces the amount of random numbers to generate. In
this regard, this scheme is partially deterministic.

At the beginning of the simulation, the cross-section formulae from Ref.~\cite{kim2000}
are averaged over all the ions orbitals for each value of $Z^\star$ and for a given set of incident electron energies.
In addition to these tabulated average cross-sections $\overline{\sigma}$, the average energy
$\overline{e}$ lost by the incident electron,
and the average energy $\overline{w}$ transferred to the secondary electron, are tabulated at the same time.
For each particle pair that collides during the simulation, these tables are interpolated, providing an
ionization probability. When an ionization occurs, the incident electron energy is reduced by $\overline{e}$,
a new electron with energy $\overline{w}$ is created, and $Z^\star$ is incremented.

In Ref.~\cite{perez2012}, the ionization probabilities and the energy transfers assume that the ion frame
is the laboratory frame. To overcome this limitation, \Smilei introduces the following changes.
The electron Lorentz factor in the ion frame is calculated using the relativistic transformation
$\gamma_e^\star=\gamma_e\gamma_i-\vp_e\cdot\vp_i/(m_e m_i)$ and  
the probability for ionization can be expressed as:
\begin{equation}
\label{IonizationProba}
  P = 1-\exp\left( - v_e \overline\sigma n \Delta t \right) = 1-\exp\left( -V^\star \overline\sigma^\star n \Delta t \right)
\end{equation}
where $v_e$ is the electron velocity in the laboratory frame,
$n$ is the particle density in the laboratory frame,
$\overline\sigma^\star$ is the cross-section in the ion frame, and 
$V^\star=\sqrt{\gamma_e^{\star\,2}-1}/(\gamma_e\gamma_i)$.
If ionization occurs, the loss of energy $\overline{e}$ of the incident electron translates into
a change in momentum ${p_e^\star}' = \alpha_e p_e^\star$ in the ion frame, with
$\alpha_e=\sqrt{(\gamma_e^\star-\overline{e})^2-1}/\sqrt{\gamma_e^{\star2}-1}$.
To calculate this energy loss in the laboratory frame, we apply the relativistic transformation:
\begin{equation}
\label{IonizationQe}
\vp_e'=\alpha_e\vp_e+((1-\alpha_e)\gamma_e^\star-\overline{e})\frac{m_e}{m_i}\vp_i.
\end{equation}
A similar operation is done for calculating the momentum of the new electron in the laboratory frame:
it is created with energy $\overline{w}$ and its momentum is $p_w^\star = \alpha_w p_e^\star$ 
in the ion frame, with $\alpha_w=\sqrt{(\overline{w}+1)^2-1}/\sqrt{\gamma_e^{\star 2}-1}$.
In the laboratory frame, it becomes:
\begin{equation}
\label{IonizationQw}
\vp_w=\alpha_w\vp_e+(\overline{w}+1-\alpha_w\gamma_e^\star)\frac{m_e}{m_i}\vp_i.
\end{equation}
Finally, equations (\ref{IonizationProba}--\ref{IonizationQw}) ensure that all quantities are correctly
expressed in the laboratory frame.

\begin{figure}
	\centering
	\includegraphics[width=6cm]{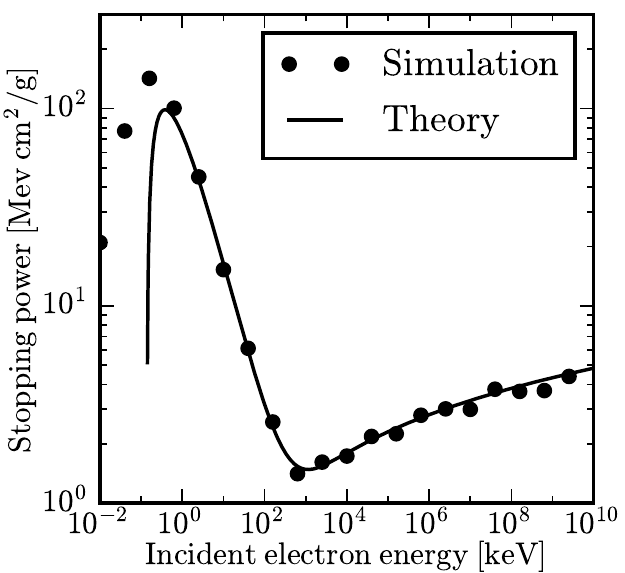}
	\includegraphics[width=6cm]{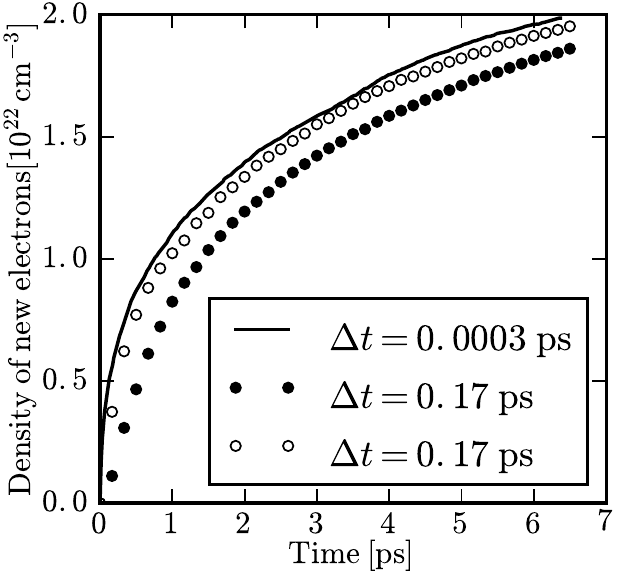}
	\caption{Left: inelastic stopping power of a cold aluminium plasma of density $10^{21}$ cm$^{-3}$
		as a function of the incident electron energy.
		Right: evolution of the secondary electron density caused by a beam of 1 MeV electrons
		traversing a cold zinc gas (both electrons and target have a density of $10^{21}$ cm$^{-3}$),
		for various simulation time-steps. The open circles correspond to the multiple-ionization
		scheme.
		\label{fig:ionization}}
\end{figure}

To test this first improvement, let us consider the inelastic stopping power caused by {\it e-i} collisions
when a test electron beam is injected in a cold, non-ionized Al plasma of ion density $10^{21}$ cm$^{-3}$.
Electrons of various initial velocities are slowed down by the ionizing collisions and their energy loss
is recorded as a function of time. The left panel of Fig.~\ref{fig:ionization} provides the corresponding stopping
power, compared to the theory of Rohrlich and Carlson \cite{rohrlich1954}. Knowing that this theory
is valid only well above the average ionization energy (here $\sim 200$ eV), the agreement is satisfactory.
At energies above $10^7$ keV (the ion rest mass), the center-of-mass frame is not that of the ions,
thus the agreement with the theory confirms the validity of our correction. 

Another modification has been added to the theory of Ref.~\cite{perez2012} in order to account for 
multiple ionization in a single time-step. This approach closely follows that presented for field ionization
in Sec.~\ref{sec:fieldIonization}, the only difference being in the computation of the ionization rates (described above).
It was tested and validated for a wide range of materials and incident electron energies.
An example is given in the right panel of Fig.~\ref{fig:ionization}, where a fast electron beam ionizes
a zinc plasma. The ionized electrons' density is plotted against time, for two vastly different time-steps.
With these parameters, the multiple-ionization scheme matches better the well-resolved case than the
single-ionization scheme.
We found that it takes a reduction of an order of magnitude in the time-step for the single-ionization approach to work 
as well as a the multiple-ionization scheme.
It therefore brings a significant accuracy improvement.

\section{User interface}\label{secInterface}

\subsection{Python input file}

End-users only need to know how to write an input file, or {\it namelist}.
Although the core of \Smilei is written in C++, the namelist is written in the \python language.
This has many advantages over the typical text-only inputs. Indeed, \python can process complex
operations that may be necessary to initialize the simulation. It can generate arbitrary numbers of
simulation elements at run-time, without the help of an external script (which would have to be
pre-processed). It supports thousands of additional packages, often helpful for specific physics
calculations. It is widely used and becoming a reference for all sorts of applications.
Very importantly, \python functions can be passed as arguments to \Smilei. For instance, a
density profile can be directly defined as a function of the coordinates.

When \Smilei is run, it starts a \python interpreter that parses the namelist line-by-line,
and executes all the \python commands. Throughout the initialization of the simulation elements
(particles, fields, diagnostics, etc.) the interpreter stays active. \Smilei gathers required data
from it, processes all required initialization steps, and finally closes the interpreter.
Note that, if a \python function needs to be evaluated throughout the simulation, 
the interpreter is kept active at all times. This happens, for instance, when defining a custom
temporal profile for a laser envelope.

\subsection{Diagnostics} \label{diags}

Data collection and analysis are performed by {\it diagnostics}.
They are not post-processing modules, but are part of the main code
and executed at runtime.
All of these diagnostics have the capability of being performed only
at user-defined times during the simulation.\\

{\bf Scalar diagnostic} -- The simplest diagnostic is called {\it scalars}:
it processes a large set of field and particle data, and
combines the results from all processors before writing out scalar quantities in a dedicated file.
Among these quantities, one can find the overall energy balance (with contributions from the 
different fields, particles, and losses at the boundaries), averaged particle quantities (charge,
energy, number of particles), and global field information (minima, maxima and Poynting flux
through boundaries).\\

{\bf Fields diagnostic} -- The diagnostic {\it fields} provides a direct copy of all the arrays in the code,
after concatenating them from all the processors.
Note that, in addition of the $\vE$ and $\vB$ fields, the particle densities and
currents are also written as they are projected on arrays at each time-step. Moreover, these data may be
temporally averaged over a number of time-steps requested by the user.\\

{\bf Probe diagnostics} -- The drawback of the diagnostic {\it fields} is that the whole arrays are written out.
To reduce the file space footprint, the {\it probes} have been implemented:
one {\it probe} corresponds to a series of points
at which locations the fields are interpolated and written in a dedicated file. This series of points can
be either regularly arranged in a line, in a rectangle (for a two-dimensional simulation), or in a
parallelepiped (for a three-dimensional simulation). The spatial separations between consecutive points
is defined by the user. Note that several {\it probes} can be added to a single simulation.\\

{\bf Trajectory diagnostics} -- Histories of individual quasi-particles are stored by the {\it tracking} diagnostic.
Each species of particles may be {\it tracked} independently, with custom output frequencies. In order to
follow individual particles, each tracked particle is assigned a unique number which is transported
throughout the simulation.\\

{\bf Particle distribution diagnostics} -- 
Tracking the position of all quasi-particles with a high frequency would be time- and memory-consuming.
To obtain digested data with flexible capabilities, the {\it particle diagnostic} has been implemented.
One diagnostic is defined by an arbitrary number of {\it axes}, which overall define a grid: all the quasi-particles in
the selected species deposit their weight in the grid cell they belong to (the cell size is unrelated to
the PIC grid). These {\it axes} are not necessarily
spatial ($x$, $y$ or $z$), but can also be one of $p_x$, $p_y$, $p_z$, $p$, $\gamma$, 
$v_x$, $v_y$, $v_z$, $v$ or the particle charge $q$. A large number of combinations can thus be
designed. For instance, using one axis $[x]$ will provide the density distribution {\it vs.} $x$; using
two axes $[x,y]$ will provide the two-dimensional density distribution {\it vs.} $x$ and $y$; using 
one axis $[p_x]$ will provide the $x$-momentum distribution; using two axes $[x,p_x]$ provides the
phase-space along $x$; using three axes $[x,y,\gamma]$ provides density maps at different energies;
using one axis $[q]$ provides the charge distribution.
Further versatility is possible by choosing which piece of data is deposited in each cell instead of the quasi-particle weight $w$.
For instance, depositing the product $w\,q\,v_x$ results in the $j_x$ current density and depositing $w\,v_x\,p_x$ results
in a component of the pressure tensor. A final feature of these {\it particle diagnostics} is the capability for
temporal averaging over an arbitrary number of time-steps.

\section{Physics highlights}\label{secPhysicsHighlights}

In this section, we present a few examples of simulations highlighting physics studies relying on \Smilei.
The first two are related to laser-plasma interaction studies, the latter two to astrophysics.

\subsection{High-harmonic generation and electron acceleration from intense femtosecond laser interaction with dense plasmas}

The interaction between an ultra-intense ($I>10^{18}~{\rm W/cm^2}$) femtosecond laser pulse
with a solid target generates a dense ``plasma mirror'' at its surface that reflects the laser
in a strongly non-linear manner. The temporal distortion of the reflected wave creates
a train of ultra-short attosecond pulses, associated, in the frequency domain,
to a comb of high-order harmonics. This scheme is considered as one of the best candidates
for attosecond light sources~\cite{thaury2010}. 
Recent experiments have shown that it also produces high-energy (relativistic) ultra-short
and very-high-charge (nC) electron bunches~\cite{thevenet2016}, of interest for electron injectors.

In what follows, we present a 2-dimensional \Smilei simulation
of laser-solid interaction, in conditions relevant to experiments at the UHI~100 laser facility\footnote{\url{http://iramis.cea.fr/slic/UHI100.php}}.
The laser pulse with wavelength $\lambda_0 = 0.8~{\rm \mu m}$ has a peak intensity
$I \simeq 2\times10^{19}~{\rm W/cm^2}$ (normalized vector potential $a_0=3$)
when focused to a $4\lambda_0$ waist, at $45^{\circ}$-indidence with p-polarization,
onto an overdense plasma slab. This overdense plasma mimics the solid target
considered fully ionized with a constant electron density $n_0=200\,n_c$
($n_c\simeq 1.7 \times 10^{21}~{\rm cm^{-3}}$ being the critical density),
$5\lambda_0$-thick, with an exponential pre-plasma of gradient length $0.1\,\lambda_0$
down to a cut-off density $n_{\rm c-off} = 0.05\,n_c$. The full box size is
$80\,\lambda_0 \times 60\lambda_0$ and the simulation time $150\,\lambda_0/c$.
The cell size is $\Delta x = \Delta y = \lambda_0/256$ (for a total of $25600\times26880$ cells)
and the timestep is $c\Delta t = \lambda_0/384 \simeq 0.95\,\Delta t_{\rm CFL}$.
Eight to 49 quasi-particles are set in each cell, for a total of $\sim 1.4$ billions
of quasi-particles in the entire box. They are frozen (not moved) until $t=50\,\lambda_0/c$,
i.e. until the laser pulse reaches the target.

\begin{figure}\centering
\includegraphics[width=0.7\textwidth]{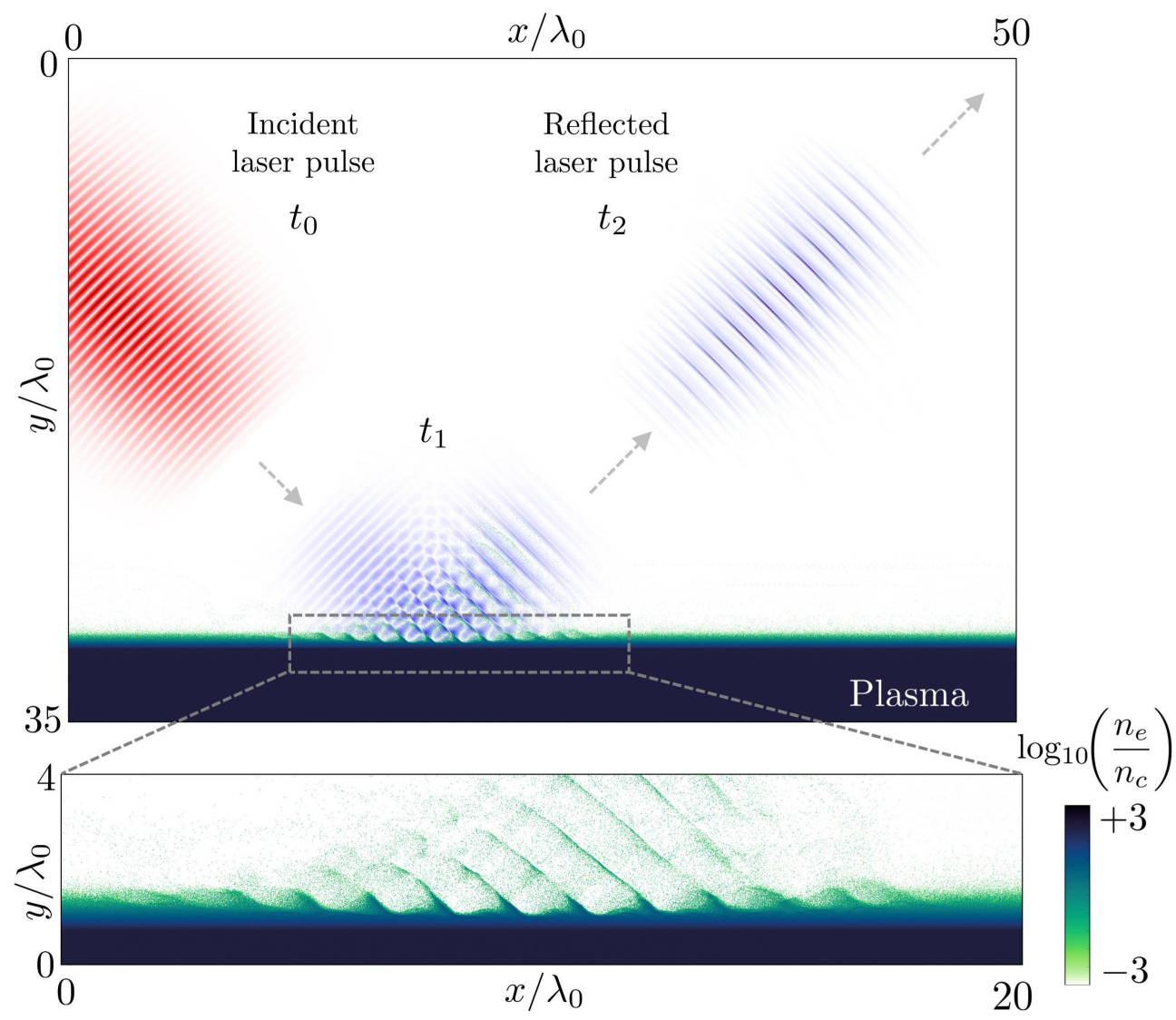}
\caption{Setup and results of a laser-solid interaction simulation. Top: laser magnetic field $B_z$ snapshots at three different times: $t_0$ before interaction with the plasma, $t_1$ during interaction and $t_2$ after reflection by the plasma mirror. The dark-scale region represents the plasma electron density at time $t_1$. Bottom: close-up of the interaction region showing the plasma electron density at $t_1$, during interaction.} \label{fig:hhg1}
\end{figure}

Figure~\ref{fig:hhg1} presents the simulation set-up and a summary of the results obtained. The top panel represents half of the simulation box in the $y$-direction, and the laser field is reported at three different times. The reflected laser pulse (at time $t_2$) shows a different spectral content than the incident pulse (at time $t_0$). The plasma electron density is shown in black. A close-up view of the interaction region is given in the bottom panel, illustrating the electron bunches being pulled out from the plasma surface.

\begin{figure}\centering
\includegraphics[width=0.7\textwidth]{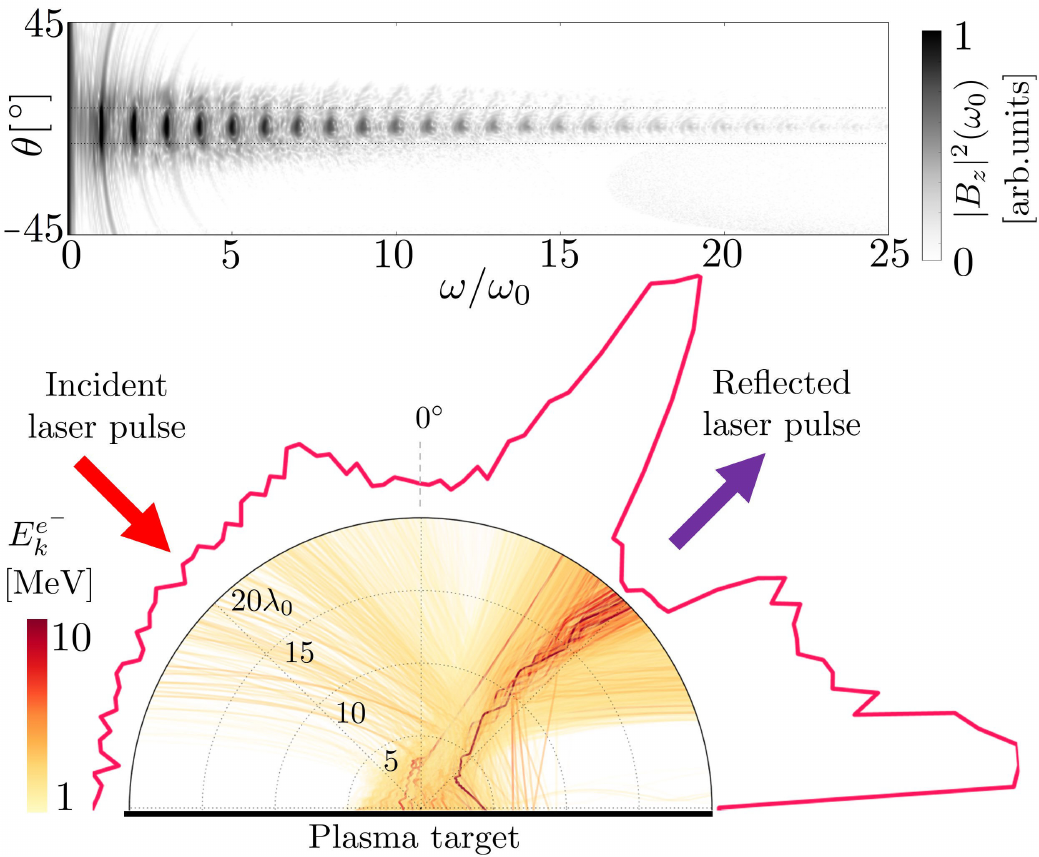}
\caption{Top: angular distribution of high-harmonics generated in short-pulse laser-plasma interaction. Bottom: typical trajectories of electrons ejected from the target and accelerated in the reflected laser field. The color scale denotes the electron kinetic energy. The pink curve is a histogram of the electron momentum angle when they reach a "detector" located at a distance of $20\,\lambda_0$ from the focal spot.} \label{fig:hhg2}
\end{figure}

Fourier analysis of the reflected laser magnetic field $B_z$ in space and time provides the angular distribution of the frequency spectrum of the reflected light. High  harmonics (up to order 16) are observed as seen in the top panel of Fig.~\ref{fig:hhg2}. In addition, electron acceleration was observed, as highlighted in the bottom panel of Fig.~\ref{fig:hhg2}, showing the trajectories of electrons ejected from the target. The most energetic electrons (with energies up to 10~MeV) are found to propagate in the direction of the reflected light pulse. The angular histogram also shows that the momenta of the escaping energetic electrons are mostly directed along two directions which are close to the reflected laser direction. This is consistent with vacuum electron acceleration suggested in Ref.~\cite{thevenet2016}. 

This simulation was run on the CINES/Occigen (Bullx) machine using 256~MPI $\times$ 14 OpenMP threads for about 10700 CPU-hours. 
Considering only the simulation time during which particles are not frozen, the characteristic time to push a particle (complete time to run one full PIC loop divided by the product of the number of particles by the number of timesteps) is of the order of $0.717~{\rm \mu s}$, 25\% of which were devoted to diagnostics.

\subsection{Short laser pulse amplification by stimulated Brillouin scattering}\label{sec:physSBS}

The generation of short high-intensity laser pulses is limited by the damage threshold of solid optics materials~\cite{stuart1995,ristau2014}, 
but such limitations could be overcome using a plasma as an amplifying medium.
This can be achieved by coupling, in a plasma, a long energetic "pump" pulse of moderate intensity and a short counter-propagating "seed" pulse
of initially low intensity. Energy transfer from the pump to the seed thanks to the excitation of a plasma wave can then be obtained~\cite{forslund1975,cohen1979}.
In what follows, we focus on stimulated Brillouin scattering (SBS) amplification, where the excited waves are ion-acoustic waves.

In the case of a pump with intensity $I_p  \gtrsim 10^{15} {\rm W/cm^2}$ (with the laser wavelength $\lambda_0=1~{\rm \mu m}$), SBS amplification operates in its ``strong-coupling'' regime \cite{andreev2006,weber2013,chiaramello2016,chiaramello2016b}. 
This scheme is particularly robust with respect to plasma inhomogeneities and does not require any frequency shift of the seed pulse.

Multi-dimensional kinetic simulations are required to describe the competing processes (spontaneous Raman scattering, filamentation, saturation), to study the non-linearities intervening in the amplification mechanism,  and to optimize the resulting phase front (for later focusing \cite{fuchs2014,wilson2016}), but they appear very challenging as inherently multi-scale. We present here two 2-dimensional \Smilei simulation of short-pulse SBS amplification
in conditions close to actual experiments~\cite{lancia2010,lancia2016}. 
The simulation box size is $\rm{1024\ \mu m \times 512 \ \mu m}$ and the grid cells are $\rm{33 \ nm}$ in both directions, resulting in $\rm{30720\times 15360 }$ cells. 
The simulation lasts $\rm{10~ps}$ with a timestep of  $\rm{7.3 \times 10^{-2}~fs}$ (over $\rm{135000}$ timesteps in total). Respectively 25 and 16 billions of quasi-particles have been set in each simulation.

The first simulation corresponds to typical present-day experiments. 
The pump has a $\cos^2$-temporal profile with duration 4.2~ps FWHM and maximum intensity $I_p = 10^{15}{\rm W/cm}^2$ and propagates along the $x$-direction towards $x>0$. The counter-propagating seed has a $\cos^2$-temporal profile with duration 0.5~ps FWHM and initial intensity $I_s = 10^{15} {\rm W/cm}^2$. 
Both the seed and pump lasers have a transverse gaussian profile with $130~{\rm \mu m}$ FWHM (in terms of intensity). 
The plasma has a gaussian density profile over all the simulation box, with a maximum (central) electron density $n =0.1~n_c$, where $n_c$ the critical density for both the laser pump and seed ($n_c \simeq 1.1\times 10^{21}~{\rm cm^{-3}}$ at $\lambda_0=1~{\rm \mu m}$).  

\begin{figure}
    \centering
    \includegraphics[width=\textwidth]{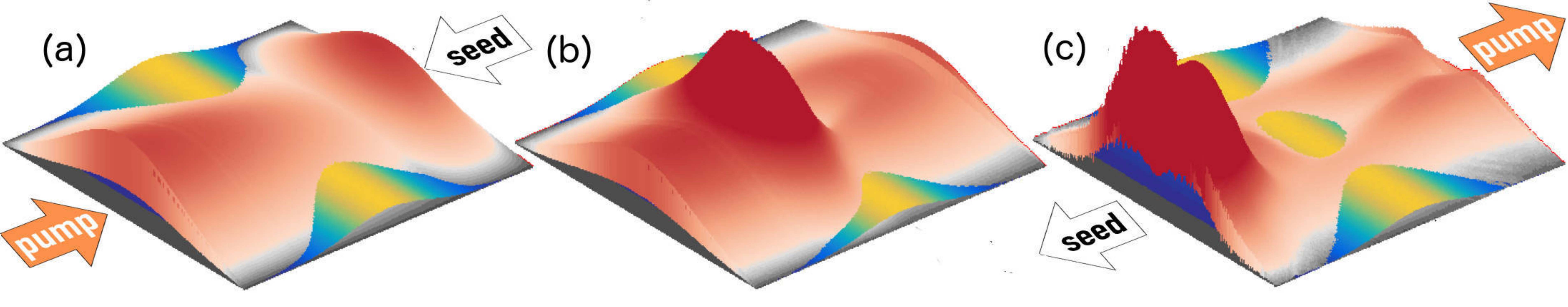}
    \caption{Evolution of the pump and seed intensities in the case of 2 pulse head-on collision at: (a) $t=5.8~{\rm ps}$, (b) $t=7.6~{\rm ps}$ and (c) $t=9.6~{\rm ps}$. The blue-yellow maps correspond to the plasma density while the white-red maps correspond to the lasers intensity.}\label{SIM_A_ampl}
\end{figure}

Typical simulation results are presented in Fig.~\ref{SIM_A_ampl} showing pump and seed intensities at three different amplification stages. 
At $t=5.8~{\rm ps}$ (panel~a), the seed starts interacting with the pump.
At $t=7.6~{\rm ps}$ (panel~b), the seed reaches the middle of the simulation box.
At that time, the seed is still in the linear amplification regime, and the pump is not depleted yet. 
At $t=9.7~{\rm ps}$ (panel~c), the seed has travelled through the entire simulation box and the pump is depleted.
The final intensity of the seed is $I_{s}^{\rm out} \simeq 4.6 \times 10^{15}~{\rm W/cm^2}$, i.e. nearly $5\times$ its initial intensity.
The spot size and phase front are also well conserved, suggesting that such a beam could be further focused using plasma mirrors to reach even larger intensities.

The second simulation deals with an innovative plasma-laser configuration to further optimize SBS amplification.
The seed pulse is now interacting with two pump lasers, both with a $\cos^2$-temporal shape with duration 4.2~ps FWHM, and top intensity $I_p=10^{15}~{\rm W/cm}^2$ that are propagating with an angle of $\pm 6^{\circ}$ degrees with respect to the $x$-axis. 
Taking two pump pulses is an experimentally convenient configuration that has the advantage to increase the pump intensity in the 3-pulse-interaction region while keeping a relatively low pump intensity during propagation in the non-overlapping region (thus reducing spurious Raman losses). 
Moreover, this laser-plasma configuration allows to separate the Raman backscattering of the pump from the amplified signal (as will be shown in what follows). 
The transverse size of the pump pulses is, for this simulation, reduced to $30~{\rm \mu m}$ FWHM and the plasma has a constant density profile with electron density $n=0.05~n_c$.
 
 \begin{figure}
\centering
 \includegraphics[width=0.8\textwidth]{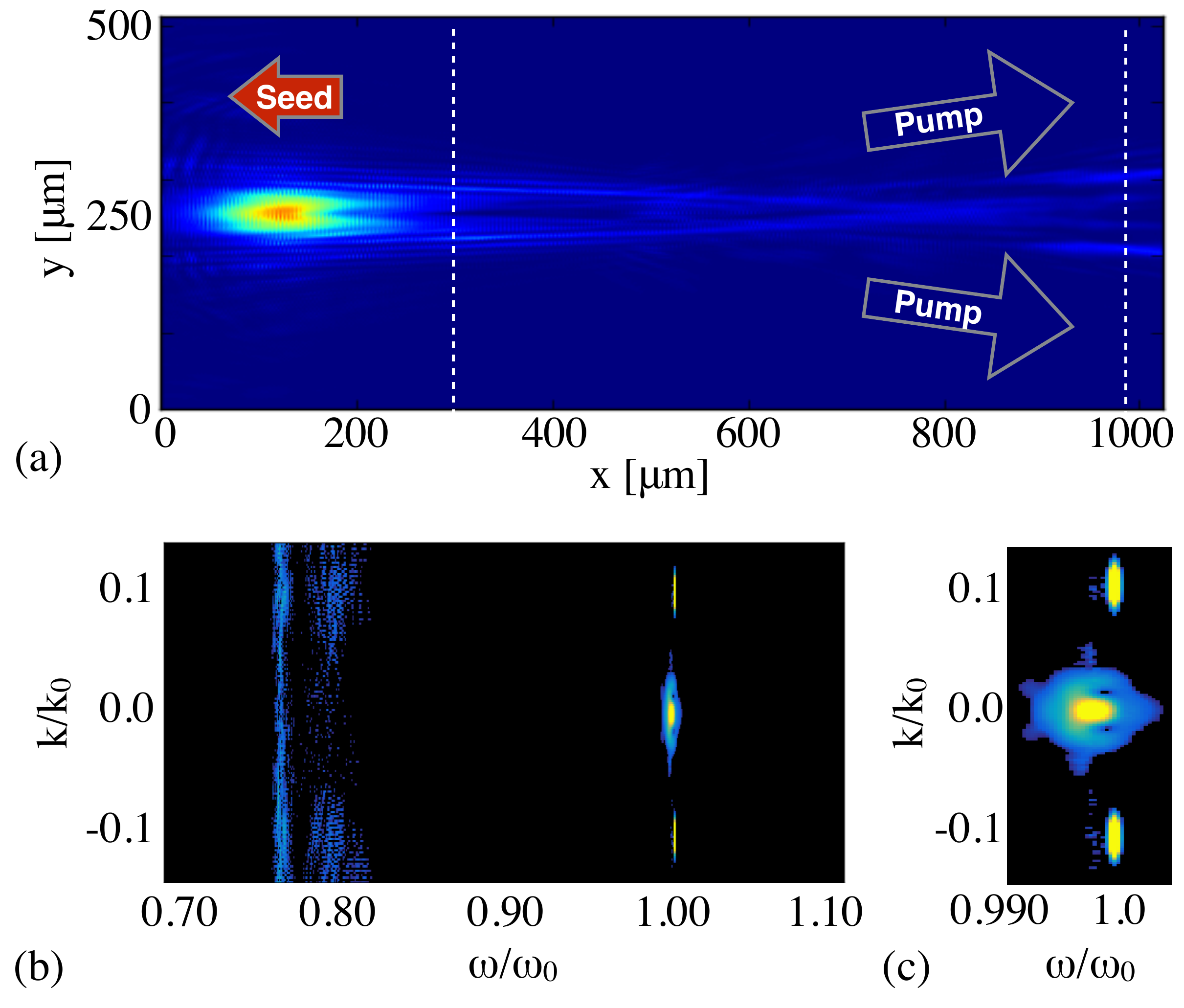}
 \caption{a) Pump and seed intensity at the end of amplification at $t=10~{\rm ps}$. 
 The final intensity of the seed is $I_{s}^{\rm out}\simeq 3 \times 10^{15}~{\rm W/cm^2}$ ($3\times$ its initial intensity). 
 b) Spectrum (in terms of wave number $k/k_0$ and frequency $\omega/\omega_0$, where
$k_0$ and $\omega_0$ are the nominal wavenumber and frequency of the pump lasers) of the electric
field recorded on the entire length of the left side of the simulation box. 
c) Zoom of the spectrum for $\omega/\omega_0= [0.98,1.02]$.}
 \label{simb_out}
\end{figure}
 
The typical interaction set-up and simulation results are shown in Fig.~\ref{simb_out}(a). 
The vertical white-dashed lines delimit the constant plasma, and the amplified seed exiting the simulation box at $t=10~{\rm ps}$ reaches a final intensity $I_{s}^{\rm out}\simeq 3 \times 10^{15}$ ($3\times$ the initial intensity). 
Of outmost interest is the spatio-temporal $(\omega,k)$ spectrum of the light recorded on the left-boundary of the simulation box presented in Fig.~\ref{simb_out}(b). 
As expected, this set-up allows the Raman signal (at $\omega \simeq 0.76\,\omega_0$ with $\omega=2\pi c/\lambda_0$ the laser angular frequency) originating from the backscattering of the pump to propagate mostly in the opposite pump directions (the signal is mainly at $k\simeq \pm 0.11\,k_0$, with $k_0=2\pi/\lambda_0$), thus angularly separating its contribution from the seed. 
Furthermore, this spectrum confirms the dominant role of SBS amplification in the seed amplification. 
Indeed, both broadening and red-shift (toward small temporal frequencies $\omega<\omega_0$) shown in Fig.~\ref{simb_out}(b) [and insert~(c)] are signatures of SBS amplification. 
The signal at $\omega \simeq \omega_0$ and $k \simeq \pm 0.11\,k_0$ correspond to the (forward-propagating) pump lasers. 
Notice that, at the end of the amplification, the transverse focal spot size of the seed at FWHM in intensity is of the order of $28~{\rm \mu m}$, i.e. of the same order than the initial one.\\

Both simulations have been performed on the IDRIS/Turing (BlueGene/Q) super-computer using 1.8~million CPU-hours on 32768 MPI processes, 
and 4 OpenMP threads per core to take best advantage of the architecture.
The average time to push a particle was $\sim 1.9 {\rm \mu s}$, 5\% of which were devoted to diagnostics.
The typical memory footprint for these simulations was of the order of 1~Tb, and each simulation generated over 1~Tb of output data.
On the CINES/Occigen (Bullx) machine, we obtained an average time of $0.43~{\rm \mu s}$ to push one particle (without diagnostics).

\subsection{Magnetic reconnection at the Earth magnetopause}

Magnetic reconnection at the Earth magnetopause regulates the transport of matter, momentum and energy from the solar wind to the internal magnetosphere. Because of their different origins, the properties of the plasma and magnetic field on both side of the magnetopause are quite different. The solar wind plasma temperature is typically one tenth that of the magnetospheric plasma, but its density is about ten times larger. The magnetic field is typically 2-3 times larger on the magnetospheric side than on the solar wind side. This asymmetry makes the reconnection dynamics vastly more complex than in symmetric environments, and has only been studied for a decade via numerical simulations and spacecraft observations~\cite{hesse2013,nasaMMS2016}. Among all possible asymmetries, those in the particle density and magnetic field amplitude have by far the most important impact on the reconnection rate.

Following times of strong magnetospheric activity, very dense and cold plasma from the plasmasphere can be transported all the way up to the Earth magnetopause, forming an elongated tongue of dense material. As it impacts the magnetopause, it drastically changes the asymmetry described above. If it reaches the magnetopause at a location where magnetic reconnection is already on-going with a typical asymmetry, the filling of the reconnection site with cold plasma, which density can even exceed the solar wind density, should affect importantly the reconnection dynamics, first by lowering the reconnection rate. 

Studying the impact of a plasmaspheric plume on magnetopause reconnection via kinetic numerical simulation is difficult. Indeed, the simulation first needs to reach a quasi-steady state reconnection with a typical magnetopause asymmetry, see the arrival of the plume and then last longer for a quasi-steady state plume reconnection regime to settle. Due to the large particle density of plumes, the transition and last phases have substantially longer time scales than the early phase, which makes the simulation heavy. The domain must be long enough in the downstream direction for the plasma, expelled during the early and transition phases, to be evacuated from the reconnection region. Otherwise, upstream plasma would not inflow, thereby stopping reconnection.

We designed a simulation so that typical magnetopause reconnection can proceed and form a reconnection exhaust of about $100\,c/\omega_{pi}$, where $\omega_{pi}$ is the ion plasma frequency corresponding to the reference (solar wind) density $n_0$, long before the plume reaches the reconnection site. Using the Cassak-Shay estimate of the inflow velocity~\cite{cassak2007}, we need to position the plume on the magnetospheric side at about $20\,c/\omega_{pi}$ from the initial magnetopause position. Three ion populations are present. The solar wind and magnetospheric populations have densities equal to $n_0$ and $n_0/10$, respectively, on their side of the current sheet, and fall to zero on the other side. The plume population increases from 0 to $2\,n_0$ at $20\,c/\omega_{pi}$ from the initial current sheet on the magnetospheric side. The magnetic field amplitude goes from $2\,B_0$ in the magnetosphere to $B_0=m_e\omega_{pe}/e$ in the solar wind and is totally in the simulation plane. The temperature is initially isotropic and its profile is calculated to balance the total pressure.

The domain size is $1280\,c/\omega_{pi} \times 256\,c/\omega_{pi}$  for $25600\times10240$ cells, in the $x$ (downstream) and $y$ (upstream) directions. 
The total simulation time is $800\,\Omega_{ci}^{-1}$ with a time step $0.00084\,\Omega_{ci}^{-1}$, where $\Omega_{ci}=e B_0/m_i$ is the ion gyrofrequency. 
We used a reduced ion to electron mass ratio $m_i/m_e = 25$, and a ratio $c/V_A = 50$ of the speed of light by the Alfv\'en velocity. 
There are initially $8.6$ billion quasi-protons for the three populations, and 13 billion electrons.

\begin{figure}
\centering
\includegraphics[width=.8\linewidth]{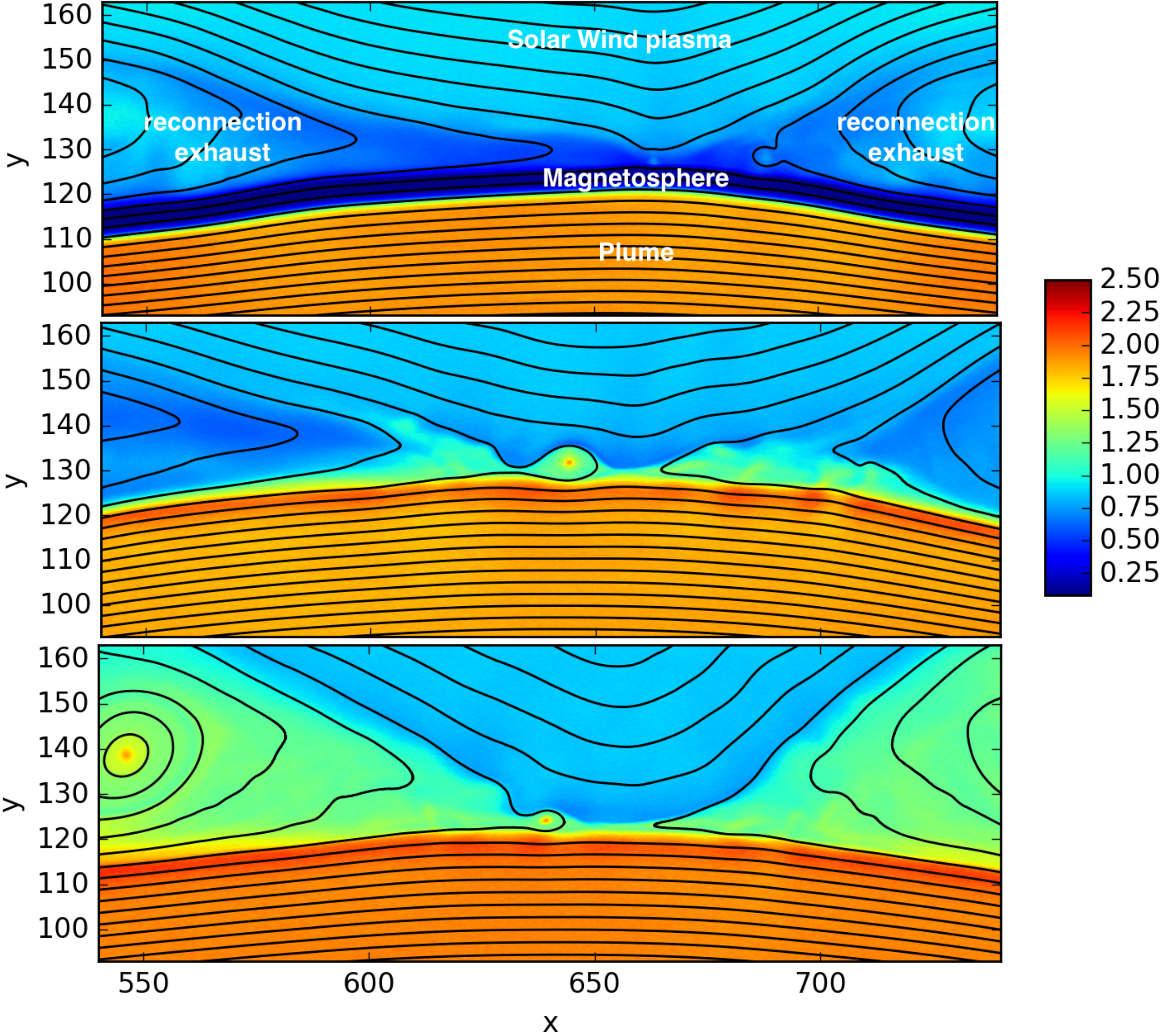}
\caption{Magnetopause reconnection simulation results: electron density color coded at different times ($t=220, 370$ and $800\Omega_{ci}^{-1}$ from top to bottom) in a region zoomed around the reconnection site. Solid black lines are in-plane magnetic field lines.} \label{fig:densities}
\end{figure}

Figure~\ref{fig:densities} presents some of the simulation results: the electron density at three different times. In the top panel, reconnection is in steady state between the solar wind plasma of density $\simeq n_0$ and the magnetosphere plasma of density $\simeq 0.1~n_0$. At this time, the exhaust is filled with mixed solar wind/hot magnetospheric plasma as the plume (of density $\simeq 2~n_0$) is still located at $\simeq 10~c/\omega_{pi}$ from the magnetospheric separatrix. The reconnection rate during this period has a typical value around $0.1~\Omega_{ci}^{-1}$, with important fluctuations caused by plasmoid formation. The plume, originally at $20~c/\omega_{pi}$ from the magnetopause, is slowly advected towards the magnetosphere separatrix and finally touches the reconnection site at about $t=300~\Omega_{ci}^{-1}$. The second panel at $t=370~\Omega_{ci}^{-1}$ shows the plume starting to fill the exhaust after reaching the reconnection site and mixing with solar wind plasma. At this time, the reconnection rate collapses to about half its previous value. The transition phase lasts for about $100~\Omega_{ci}^{-1}$ before a plume reconnection regime reaches a quasi-steady state. The third panel shows the electron density at the end of the simulation, where the exhaust is filled with plume and solar wind plasma.

This large-scale simulation has run for a total of 14 million CPU-hours on 16384 cores of the CINES/Occigen (Bullx) supercomputer within a GENCI-CINES special call.
Overall, the characteristic (full) push-time for a single particle was of the order of $1.6~{\rm \mu s}$, 31\% of which were devoted to diagnostics. 
Note that no dynamic load balancing was used for this simulation.

\subsection{Collisionless shock in pair plasmas}\label{sec:physShock}

Relativistic collisionless shocks play a fundamental role
in various astrophysical scenarios
(active galactic nuclei, micro-quasars, pulsar wind nebulae and gamma-ray bursts)
where they cause high-energy radiation and particle acceleration related to the cosmic-ray spectrum~\cite{kirk1999}.  
The long-standing problem of describing collisionless shock formation has gained renewed interest
as PIC simulations provide insight into the micro-physics of these non-linear structures~\cite{spitkovsky2008,haugbolle2011,sironi2013}. 

In the absence of particle collisions, the shock is mediated by collective plasma processes, 
produced by electromagnetic plasma instabilities, taking place at the shock front. 
In particular, we study the Weibel (or current filamentation) instability~\cite{weibel1959,bret2010,grassi2016} 
that is observed in most of the astrophysical relativistic outflows interacting with the interstellar medium. 
It can be excited by counter-streaming unmagnetized relativistic flows, and it has been shown to 
dominate the instability spectrum for a wide range of parameters~\cite{bret2010}.
It converts part of the kinetic energy of the counter-propagating flows into small-scale magnetic fields, 
which are then amplified up to sub-equipartition levels of the total energy. 
The resulting strong magnetic turbulence can isotropize the incoming flow (in the center-of-mass frame), 
hence stopping it and leading to compression of the downstream (shocked plasma) and shock formation.

The density compression ratio between the upstream relativistic flow (with density $n_0$) and the downstream (with density $n_d$) plasma 
can be derived from macroscopic conservation laws giving the Rankine-Hugoniot (RH) jump conditions~\cite{blandford1976}. 
The shock is considered formed when the density jump becomes $n_d/n_0 =1+(\gamma_0+1)/[\gamma_0(\Gamma_{ad}-1)]$. 
Considering an ultra-relativistic incoming flow $\gamma_{0}\gg 1$ and adiabatic index $\Gamma_{ad}=3/2$ for a 2-dimensional downstream plasma at ultra-relativistic temperature,  we expect a compression factor $n_d/n_0 =3$. 
Another clear signature of the shock formation is the isotropization of the downstream plasma.
This physical picture has been confirmed by various PIC simulations using counter-penetrating relativistic flows~\cite{spitkovsky2008,haugbolle2011,sironi2013}. 

In what follows, we present a 2-dimensional PIC simulation of a Weibel-mediated collisionless shock driven in an initially unmagnetized electron-positron plasma.
The simulation relies on the ``piston'' method that consists in initializing the simulation with a single cold electron-positron plasma drifting in the $+x$-direction at a relativistic velocity $v_0 \simeq 0.995\,c$ ($\gamma_0=10$). 
A reflecting (for both fields and particles) boundary condition is applied at the right border of the simulation box, hence creating a counter-penetrating (reflected) flow, the reflected beam mimicking a flow with velocity $-v_0$.

The simulation box size is $2048\,\delta_e \times 128\,\delta_e$, $\delta_e = c/\omega_p$ being the (non-relativistic) electron skin-depth of the initial flow. 
The spatial resolution is set to $\Delta x=\Delta y=\delta_e/16$, the timestep to $c\Delta t = \Delta x/2$ and 16 particles-per-cell were used for each species leading to a total of $\simeq 2.15\times 10^9$ quasi-particles. 
Temporal Friedman filtering (with $\theta=0.1$) and binomial current filtering (using 3 passes) have been applied in order to avoid spurious effects (e.g. upstream heating) due to the grid-Cerenkov numerical instability (see Sec.~\ref{sec:filters}). 

\begin{figure}\centering
\includegraphics[width=0.9\textwidth]{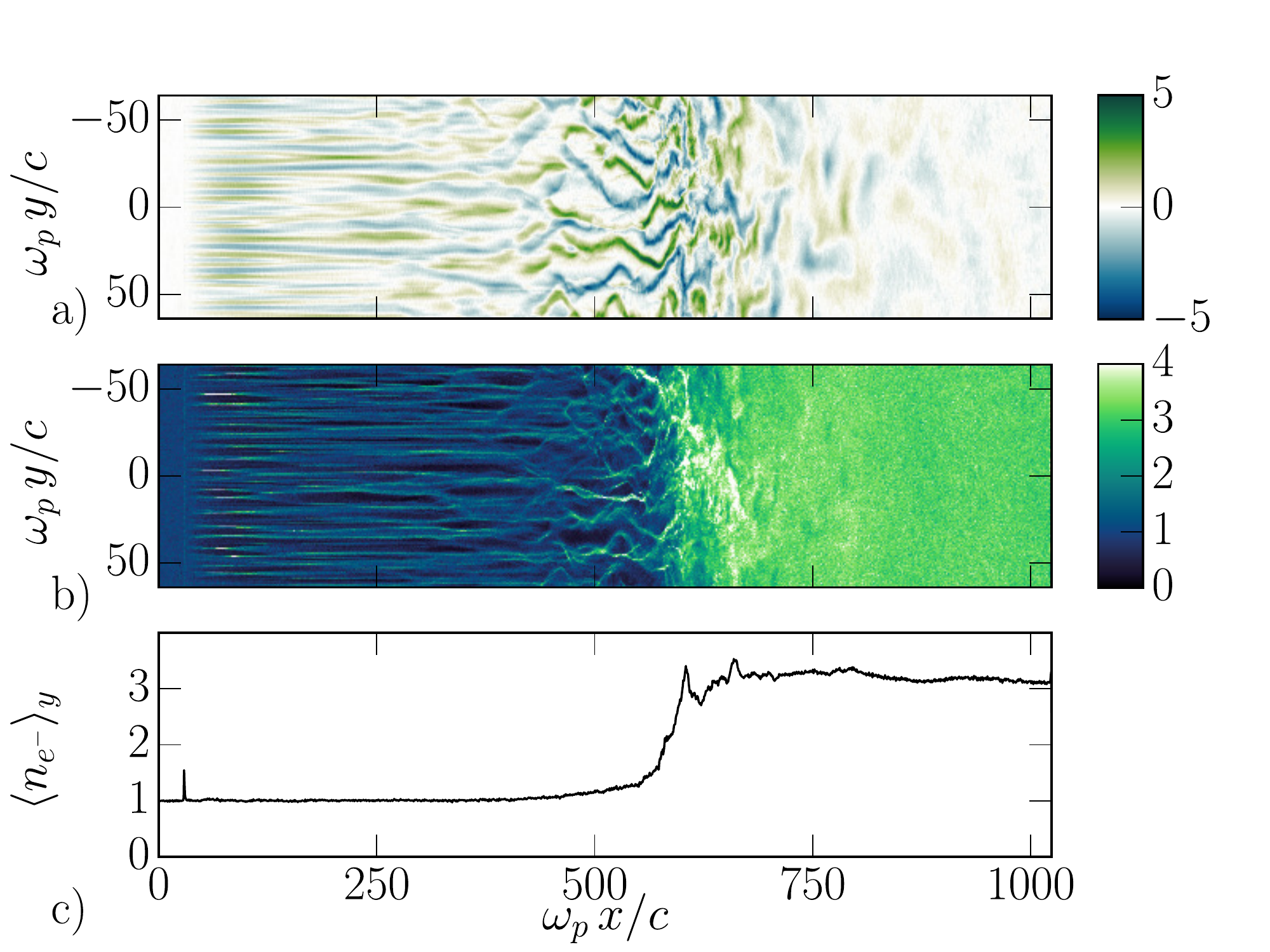}
\caption{\label{fig:shock1} Snapshot at $t=1000\,\omega_p^{-1}$. a) Weibel generated magnetic field $B_z$ in units of $B_0=m_e\,\omega_p/e$. b) Electron density in units of $n_0$. c) Electron density (in units of $n_0$) averaged along the $y$-direction.}
\end{figure}

Figure~\ref{fig:shock1} presents the characteristic simulation results at time $t=1000\,\omega_p^{-1}$.
The overlapping region of incoming and reflected flows is Weibel-unstable which results in the creation, before the shock ($ 50\,\delta_e<x< 400\,\delta_e$), of filamentary structures in both the magnetic field (panel a) and the total plasma density (panel b).
The magnetic field at the shock front ($400\,\delta_e<x<600\,\delta_e$) becomes turbulent and it is strong enough to stop the incoming particles leading to a pile-up of the plasma density up to $n_d\simeq 3.2~n_0$ (panel c), as predicted by the RH conditions.  
The simulation also indicates that the shock propagates toward the left with a velocity $v_{sh}\simeq (0.46\pm0.01)\,c$. 
The RH conditions predict a shock velocity $v_{sh}=c\,(\Gamma_{ad}-1)(\gamma_0-1)/(\gamma_0 v_0)\simeq 0.452\,c$, in excellent agreement with the value observed in the simulation. Isotropization and thermalization of the downstream distribution function was also observed (not shown), with a typical temperature close to that predicted from the RH conditions $T_d = \tfrac{1}{2}(\gamma_0-1)\,m_e c^2\simeq 4.5\,m_e c^2$.

\begin{figure}\centering
\includegraphics[width=0.5\textwidth]{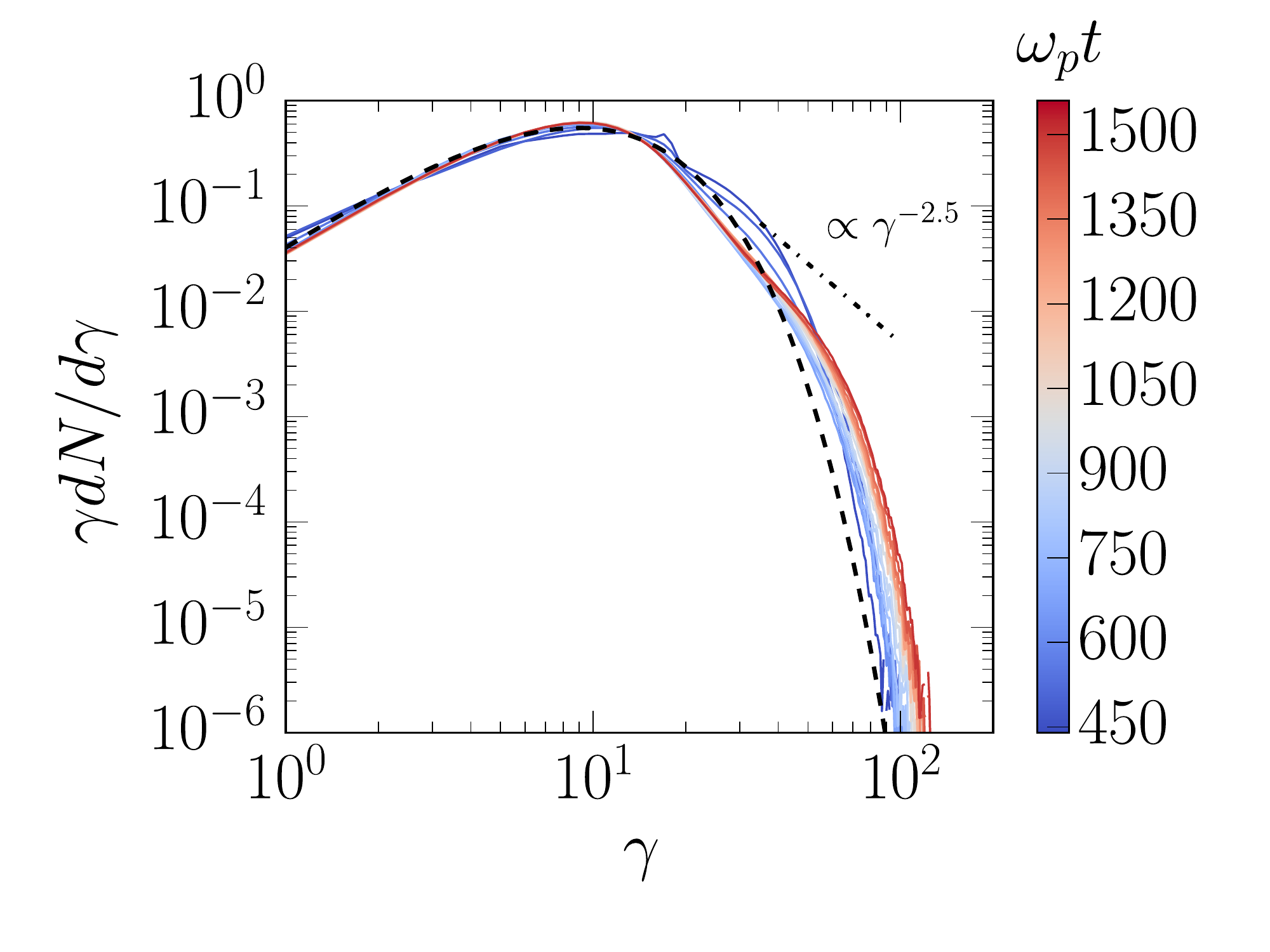}
\caption{\label{fig:shock2} Electron energy distribution in the downstream region ($800\,c/\omega_p <x< 900\,c/\omega_p$) for different times $t>450\,\omega_p^{-1}$. At time $t\simeq 500\,\omega_p^{-1}$, the shock is formed and the energy distribution closely follows the (2-dimensional) thermal Maxwell-Jüttner distribution with temperature $T_d=4.5\,m_e c^2$ expected from RH conditions (thick dashed line). At later times, a supra-thermal tail appears. The dot-dashed guide line goes as the power law~$\gamma^{-2.5}$.}
\end{figure}

Finally, the simulation also demonstrates the build-up of a supra-thermal tail in the downstream particle energy distribution, as shown in Fig.~\ref{fig:shock2}.
At the moment of shock formation ($t \simeq 500\,\omega_p^{-1}$), the particle distribution in the downstream (shocked plasma, here taken at $800\,c/\omega_p <x< 900\,c/\omega_p$) is found to have relaxed to an isotropic, quasi-thermal, distribution with a temperature initially slightly larger than that predicted from RH conditions $T_d\simeq 4.5\,m_e c^2$ (dashed line).  At later times, a supra-thermal tail, characteristic of first order Fermi acceleration at the shock front~\cite{spitkovsky2008}, appears following a $\gamma^{-2.5}$ power law.

This simulation run on the TGCC/Curie machine using 128 MPI $\times$ 8 OpenMP threads over a total of 18800~CPU-hours for 49780 timesteps. For this simulation, the characteristic (total) push time for a single quasi-particle was of $0.63~{\rm \mu s}$, 20\% of which were devoted to diagnostics.

\section{Conclusions}\label{secConclusions}

To summarize the capabilities of the open-source PIC code \Smilei,
we emphasize on its object-oriented (C++) structure
complemented by its user-friendly Python interface,
making it a versatile, multi-purpose tool for plasma simulation.

Co-developed by both physicists and HPC experts,
\Smilei benefits from a state-of-the-art parallelization technique
relying on a patch-based super-decomposition strategy.
This approach allows for an improved cache use, and provides a straightforward
implementation of dynamic load balancing.
This strategy is shown to manage efficiently any load imbalance and to scale well
over a very large number of computing elements (up to several hundred of thousands).
The code was tested on various super-computers, architectures
(Bullx and BlueGene/Q in particular) and processors (Intel Sandy Bridge, Broadwell and Haswell, and IBM Power A2).

Still a young project (development started in 2013), \Smilei benefits from
a wide range of additional modules, including a Monte-Carlo treatment of
binary collisions as well as collisional and field ionization.
\Smilei is currently used by a growing community, as illustrated by the presented
applications to both laser-plasma interaction and astrophysics.

An ongoing collaborative project, more enhancements are today being incorporated in \Smilei.
On the physics side, Monte-Carlo modules are being developed to account for various quantum-electrodynamics effects, from 
high-energy photon to electron-positron pair production.
Furthermore, ongoing work focuses on spectral maxwell solvers, various optimizations (including vectorization),  
providing the user with a complete documentation, etc.
Importantly, the maintainability of the code will greatly benefit from a full set of automated benchmarks (continuous integration).

\appendix
\section{Quasi-particles shape functions and interpolation/projection order}
 
The quasi-particles shape function $S(\vx)$ has the following properties: 
(i)~it is symmetric with respect to its argument $\vx$,
(ii)~it is non-zero in a region centered around $\vx=0$ that extends over a distance $n\,\Delta x^{\mu}$ in $x^{\mu}=(x,y,z)$-direction, with $\Delta x^{\mu}$ the size of a cell in this direction, hence defining a so-called quasi-particle volume $V_p = \Pi_{\mu} n\Delta\mu$, where the integer $n$, henceforth referred to as the interpolation/projection order is discussed in what follows, and
(iii)~it is normalized so that $\int d\vx S(\vx) = 1$. 

In what follows, we will consider different shape functions all of which can be written as a product over the $D$ spatial dimensions of the simulation: 
\begin{eqnarray}\label{eq_Sfct}
S(\vx) = \prod_{\mu=1}^{D}\,s^{(n)}(x^{\mu})\,,
\end{eqnarray}
where $n$ denotes the previously introduced interpolation/projection order. 
The one-dimensional shape functions $s^{(n)}(x)$ used in \Smilei can be written in a recursive way.
The interpolation/projection order $n=0$ corresponds to a point-like quasi-particle and $s^{(0)}(x)=\delta(x)$, with $\delta(x)$ the Dirac distribution.
The shape functions of higher order $n>0$ are then obtained recursively:
\begin{eqnarray}\label{eq_snfct}
s^{(n)}(x) = \Delta x^{-1}P(x) \otimes s^{(n-1)}(x) \equiv \Delta x^{-1}\int_{-\infty}^{\infty}\!\!dx'\,P(x'-x)\,s^{(n-1)}(x')\,,
\end{eqnarray}
with the crenel function $P(x)=1$ if $\vert x \vert \le \Delta x/2$ and $P(x)=0$ otherwise. In what follows, we write explicitly the shape-functions $\hat{s}^{(n)}=\Delta x\,s^{(n)}$ for $n$ up to $4$ (i.e. up to fourth order):
\begin{eqnarray}
\hat{s}^{(0)}(x) = \Delta x\,\delta(x)\,,\\
\hat{s}^{(1)}(x) = \left\{ 
\begin{array}{ll} 
1 & {\rm if}\,\, \vert x \vert \le \frac{1}{2}\,\Delta x\,,\\
0                                         & {\rm otherwise,}
\end{array}\right.\\
\hat{s}^{(2)}(x) = \left\{ 
\begin{array}{ll} 
\left(1-\left\vert\frac{x}{\Delta x}\right\vert\right) & {\rm if}\,\, \vert x \vert \le \Delta x\,,\\
0                                         & {\rm otherwise,}
\end{array}\right.\\
\hat{s}^{(3)}(x) = \left\{ 
\begin{array}{ll} 
\frac{3}{4}\,\left[1-\frac{4}{3}\,\left(\frac{x}{\Delta x}\right)^2\right] & {\rm if}\,\, \vert x \vert \le \frac{1}{2}\,\Delta x\,,\\
\frac{9}{8}\,\left(1-\frac{2}{3}\,\left\vert\frac{x}{\Delta x}\right\vert\right)^2 & {\rm if}\,\, \frac{1}{2}\,\Delta x < \vert x \vert \le \frac{3}{2}\,\Delta x\,,\\
0                                         & {\rm otherwise,}
\end{array}\right.\\
\hat{s}^{(4)}(x) = \left\{ 
\begin{array}{ll} 
\frac{2}{3}\,\left[1-\frac{3}{2}\,\left(\frac{x}{\Delta x}\right)^2+\frac{3}{4}\left\vert\frac{x}{\Delta x}\right\vert^3\right] & {\rm if}\,\, \vert x \vert \le \,\Delta x\,,\\
\frac{4}{3}\,\left(1-\frac{1}{2}\,\left\vert\frac{x}{\Delta x}\right\vert\right)^3 & {\rm if}\,\, \Delta x < \vert x \vert \le 2\,\Delta x\,,\\
0                                         & {\rm otherwise.}
\end{array}\right.
\end{eqnarray}

\subsection*{Field interpolation at the particle position}

In Sec.~\ref{sec:quasiPart}, it is shown that the quasi-particles are subject to the electric and magnetic fields interpolated at their positions, these interpolated fields being given by Eqs.~\eqref{eq:interpE} and~\eqref{eq:interpB}, respectively. For the simplest case of a one-dimensional grid, the field (either electric or magnetic) seen by the quasi-particle at position $\vx_p = x_p\,\hat{\bf x}$ can thus be written in the form:
\begin{eqnarray}\label{eq:fieldpart}
F_p = \int\!\!dx\,s^{(n)}(x-x_p)\,F(x)\,,
\end{eqnarray}
where field $F(x)$ can be reconstructed from the grid as:
\begin{eqnarray}\label{eq:fieldgrid}
F(x) = \sum_{i} F_{i}\,P(x-x_i)\,,
\end{eqnarray}
$i$ denoting the grid point index and $x_i$ the location of the $i^{th}$ grid point.
Injecting Eq.~\eqref{eq:fieldgrid} in Eq.~\eqref{eq:fieldpart}, and using the recursive definition of the shape-function Eq.~\eqref{eq_snfct}, one obtains a simple way to interpolate the field at the quasi-particle position:
\begin{eqnarray}\label{eq:fieldpartFinal}
F_p = \sum_i F_i\,\hat{s}^{(n+1)}(x_p-x_i)\,.
\end{eqnarray}
The generalisation to an arbitrary number of spatial dimension is straightforward.

\subsection*{Direct projection of the charge and current densities onto the grid}

Direct projection of the charge and/or current densities onto a grid point $x_i$ can be performed considering the projected quantity [$Q=(\rho,J)$] as the amount of charge and/or current contained in the cell located around this grid point:
\begin{eqnarray}
Q_i = \int\!\! dx\,Q_p\,s^{(n)}(x-x_p)\,P(x-x_i)\,.
\end{eqnarray}
Using the recursive definition of the shape-function Eq.~\eqref{eq_snfct}, one obtains:
\begin{eqnarray}
Q_i = Q_p\,\hat{s}^{(n+1)}(x_i-x_p)\,.
\end{eqnarray}

For the sake of completeness, it is worth noting that using the same shape-function for both interpolation and projection is mandatory to avoid unphysical self-force acting on the quasi-particles.

\label{appendixA}

\section*{Acknowledgements}

The authors are grateful to L. Gremillet, R. Nuter and A. Sgattoni for fruitful discussions, and Ph. Savoini for feedback on the code.
MG and GB thank F. Quéré and H. Vincenti for sharing physics insights.
Financial support from the {\it Investissements d'Avenir} of the PALM LabEx (ANR-10-LABX-0039-PALM, Junior Chair SimPLE) 
and from the Plas@Par LabEx (ANR-11-IDEX-0004-02) are acknowledged.
AG acknowledges financial support from the {\it Universit\'e Franco-Italienne} through the Vinci program.
NA and JDa thank the ANR (project ANR-13-PDOC-0027) for funding their research.
This work was performed using HPC resources from GENCI-IDRIS {\it Grands Challenges 2015}, 
GENCI-IDRIS/TGCC (Grant 2016-x2016057678), GENCI-CINES (Grant 2016-c2016067484) and GENCI-CINES (Special allocation n$^{\circ}$ t201604s020).

\bibliographystyle{unsrt}
\bibliography{SmileiBiblio}

\begin{thebibliography}{10}

\bibitem{harlow1955}
F.~H. Harlow.
\newblock A machine calculation for hydrodynamic problems.
\newblock Technical report, Los Alamos Scientific Laboratory report LAMS-1956,
  November 1956.

\bibitem{birdsall1985}
C.~K. Birsall and A.~B. Langdon.
\newblock {\em Plasma physics via computer simulation}.
\newblock McGraw-Hill, New York, 1985.

\bibitem{tajima1979}
T.~Tajima and J.~M. Dawson.
\newblock Laser electron accelerator.
\newblock {\em Phys. Rev. Lett.}, 43:267--270, Jul 1979.

\bibitem{pukhov2002}
A.~Pukhov and J.~Meyer-ter Vehn.
\newblock Laser wake field acceleration: the highly non-linear broken-wave
  regime.
\newblock {\em Applied Physics B}, 74(4):355--361, 2002.

\bibitem{mangles2004}
S.~P.~D. Mangles, C.~D. Murphy, Z.~Najmudin, A.~G.~R. Thomas, J.~L. Collier,
  A.~E. Dangor, E.~J. Divall, P.~S. Foster, J.~G. Gallacher, C.~J. Hooker,
  D.~A. Jaroszynski, A.~J. Langley, W.~B. Mori, P.~A. Norreys, F.~S. Tsung,
  R.~Viskup, B.~R. Walton, and K.~Krushelnick.
\newblock Monoenergetic beams of relativistic electrons from intense
  laser-plasma interactions.
\newblock {\em Nature}, 431(7008):535--538, 09 2004.

\bibitem{geddes2004}
C.~G.~R. Geddes, Cs. Toth, J.~van Tilborg, E.~Esarey, C.~B. Schroeder,
  D.~Bruhwiler, C.~Nieter, J.~Cary, and W.~P. Leemans.
\newblock High-quality electron beams from a laser wakefield accelerator using
  plasma-channel guiding.
\newblock {\em Nature}, 431(7008):538--541, 09 2004.

\bibitem{faure2004}
J.~Faure, Y.~Glinec, A.~Pukhov, S.~Kiselev, S.~Gordienko, E.~Lefebvre, J.~P.
  Rousseau, F.~Burgy, and V.~Malka.
\newblock A laser-plasma accelerator producing monoenergetic electron beams.
\newblock {\em Nature}, 431(7008):541--544, 09 2004.

\bibitem{macchi2013}
A~Macchi, M~Borghesi, and M~Passoni.
\newblock Ion acceleration by superintense laser-plasma interaction.
\newblock {\em Rev. Mod. Phys.}, 85:751--793, May 2013.

\bibitem{thaury2010}
C~Thaury and F~Qu{\'e}r{\'e}.
\newblock High-order harmonic and attosecond pulse generation on plasma
  mirrors: basic mechanisms.
\newblock {\em Journal of Physics B: Atomic, Molecular and Optical Physics},
  43(21):213001, 2010.

\bibitem{chen2009}
Hui Chen, Scott~C. Wilks, James~D. Bonlie, Edison~P. Liang, Jason Myatt,
  Dwight~F. Price, David~D. Meyerhofer, and Peter Beiersdorfer.
\newblock Relativistic positron creation using ultraintense short pulse lasers.
\newblock {\em Phys. Rev. Lett.}, 102:105001, Mar 2009.

\bibitem{sarri2015}
G.~Sarri, K.~Poder, J.~M. Cole, W.~Schumaker, A.~Di~Piazza, B.~Reville,
  T.~Dzelzainis, D.~Doria, L.~A. Gizzi, G.~Grittani, S.~Kar, C.~H. Keitel,
  K.~Krushelnick, S.~Kuschel, S.~P.~D. Mangles, Z.~Najmudin, N.~Shukla, L.~O.
  Silva, D.~Symes, A.~G.~R. Thomas, M.~Vargas, J.~Vieira, and M.~Zepf.
\newblock Generation of neutral and high-density electron--positron pair
  plasmas in the laboratory.
\newblock {\em Nature Communications}, 6:6747 EP --, 04 2015.

\bibitem{cros2014}
B.~Cros, B.S. Paradkar, X.~Davoine, A.~Chanc{\'e}, F.G. Desforges,
  S.~Dobosz-Dufr{\'e}noy, N.~Delerue, J.~Ju, T.L. Audet, G.~Maynard, M.~Lobet,
  L.~Gremillet, P.~Mora, J.~Schwindling, O.~Delferri{\`e}re, C.~Bruni,
  C.~Rimbault, T.~Vinatier, A.~Di Piazza, M.~Grech, C.~Riconda, J.R.
  Marqu{\`e}s, A.~Beck, A.~Specka, Ph. Martin, P.~Monot, D.~Normand,
  F.~Mathieu, P.~Audebert, and F.~Amiranoff.
\newblock Laser plasma acceleration of electrons with multi-pw laser beams in
  the frame of \{CILEX\}.
\newblock {\em Nuclear Instruments and Methods in Physics Research Section A:
  Accelerators, Spectrometers, Detectors and Associated Equipment}, 740:27 --
  33, 2014.
\newblock Proceedings of the first European Advanced Accelerator Concepts
  Workshop 2013.

\bibitem{dipiazza2012}
A.~Di~Piazza, C.~M\"uller, K.~Z. Hatsagortsyan, and C.~H. Keitel.
\newblock Extremely high-intensity laser interactions with fundamental quantum
  systems.
\newblock {\em Rev. Mod. Phys.}, 84:1177--1228, Aug 2012.

\bibitem{taflove2005}
Allen Taflove.
\newblock {\em Computation electrodynamics: The finite-difference time-domain
  method, 3rd Ed.}
\newblock Artech House, Norwood, 2005.

\bibitem{nuter2014}
Rachel Nuter, Mickael Grech, Pedro Gonzalez~de Alaiza~Martinez, Guy Bonnaud,
  and Emmanuel d'Humi{\`e}res.
\newblock Maxwell solvers for the simulations of the laser-matter interaction.
\newblock {\em The European Physical Journal D}, 68(6):177, 2014.

\bibitem{wright1975}
Thomas~P. Wright and G.~Ronald Hadley.
\newblock Relativistic distribution functions and applications to electron
  beams.
\newblock {\em Phys. Rev. A}, 12:686--697, Aug 1975.

\bibitem{zenitani2015}
Seiji Zenitani.
\newblock Loading relativistic maxwell distributions in particle simulations.
\newblock {\em Physics of Plasmas}, 22(4):042116, 2017/01/20 2015.

\bibitem{numericalRecipies}
Brian~P. Flannery, Saul Teukolsky, William~H. Press, and William~T. Vetterling.
\newblock {\em Numerical Recipies, 3rd Ed.}
\newblock Cambridge University Press, 2007.

\bibitem{boris1970}
J.~P. Boris.
\newblock Relativistic plasma simulation - optimization of a hybrid code.
\newblock {\em Proceeding of the 4th Conference on Numerical Simulation of
  Plasmas}, pages 3--67, 1970.

\bibitem{vay2008}
J.~L. Vay.
\newblock Simulation of beams or plasmas crossing at relativistic velocity.
\newblock {\em Physics of Plasmas}, 15(5):056701, 2017/01/20 2008.

\bibitem{esirkepov2001}
T.~Zh. Esirkepov.
\newblock Exact charge conservation scheme for particle-in-cell simulation with
  an arbitrary form-factor.
\newblock {\em Computer Physics Communications}, 135(2):144 -- 153, 2001.

\bibitem{spohn1991}
H.~Spohn.
\newblock {\em Large scale dynamics of interacting particles}.
\newblock Springer-Verlag, Berlin Heidelberg, 1991.

\bibitem{barucq1997}
H.~Barucq and B.~Hanouzet.
\newblock Asymptotic behavior of solutions to maxwell's system in bounded
  domains with absorbing silver--m{\"u}ller's condition on the exterior
  boundary.
\newblock {\em Asymptotic Analysis}, 15(1):25, 1997.

\bibitem{fonseca2002}
{\em Lecture Notes in Computer Science}, volume 2331. Springer, Heidelberg,
  2002.

\bibitem{lifschitz2009}
A.F. Lifschitz, X.~Davoine, E.~Lefebvre, J.~Faure, C.~Rechatin, and V.~Malka.
\newblock Particle-in-cell modelling of laser--plasma interaction using fourier
  decomposition.
\newblock {\em Journal of Computational Physics}, 228(5):1803 -- 1814, 2009.

\bibitem{haugbolle2013}
Troels Haugb{\o}lle, Jacob~Trier Frederiksen, and {\AA}ke Nordlund.
\newblock photon-plasma: A modern high-order particle-in-cell code.
\newblock {\em Physics of Plasmas}, 20(6):062904, 2013.

\bibitem{stantchev2008}
George Stantchev, William Dorland, and Nail Gumerov.
\newblock Fast parallel particle-to-grid interpolation for plasma \{PIC\}
  simulations on the \{GPU\}.
\newblock {\em Journal of Parallel and Distributed Computing}, 68(10):1339 --
  1349, 2008.
\newblock General-Purpose Processing using Graphics Processing Units.

\bibitem{decyk2011}
Viktor~K. Decyk and Tajendra~V. Singh.
\newblock Adaptable particle-in-cell algorithms for graphical processing units.
\newblock {\em Computer Physics Communications}, 182(3):641 -- 648, 2011.

\bibitem{germaschewski2016}
Kai Germaschewski, William Fox, Stephen Abbott, Narges Ahmadi, Kristofor
  Maynard, Liang Wang, Hartmut Ruhl, and Amitava Bhattacharjee.
\newblock The plasma simulation code: A modern particle-in-cell code with
  patch-based load-balancing.
\newblock {\em Journal of Computational Physics}, 318:305 -- 326, 2016.

\bibitem{hilbert1891}
D.~Hilbert.
\newblock {\"U}ber die stetige abbildung einer linie auf ein
  fl{\"a}chenst{\"u}ck.
\newblock {\em Math. Ann.}, 38:459, 1891.

\bibitem{frederiksen2015}
J.~Trier~Frederiksen, G.~Lapenta, and M.~E. Pessah.
\newblock Particle control in phase space by global k-means clustering.
\newblock {\em arXiv:1504.03849}, 2015.

\bibitem{beck2016}
A.~Beck, J.T. Frederiksen, and J.~D{\'e}rouillat.
\newblock Load management strategy for particle-in-cell simulations in high
  energy particle acceleration.
\newblock {\em Nuclear Instruments and Methods in Physics Research Section A:
  Accelerators, Spectrometers, Detectors and Associated Equipment}, 829:418 --
  421, 2016.
\newblock 2nd European Advanced Accelerator Concepts Workshop - \{EAAC\} 2015.

\bibitem{godfrey1974}
B.~B. Godfrey.
\newblock Numerical cherenkov instabilities in electromagnetic particle codes.
\newblock {\em Journal of Computational Physics}, 15(4):504 -- 521, 1974.

\bibitem{lehe2014}
R~Lehe.
\newblock {\em Improvement of the quality of laser-wakefield accelerators:
  towards a compact free-electron laser}.
\newblock PhD thesis, Ecole Polytechnique, 2014.

\bibitem{nuter2016}
Rachel Nuter and Vladimir Tikhonchuk.
\newblock Suppressing the numerical cherenkov radiation in the yee numerical
  scheme.
\newblock {\em Journal of Computational Physics}, 305:664 -- 676, 2016.

\bibitem{greenwood2004}
Andrew~D. Greenwood, Keith~L. Cartwright, John~W. Luginsland, and Ernest~A.
  Baca.
\newblock On the elimination of numerical cerenkov radiation in \{PIC\}
  simulations.
\newblock {\em Journal of Computational Physics}, 201(2):665 -- 684, 2004.

\bibitem{vay2011}
J.-L. Vay, C.G.R. Geddes, E.~Cormier-Michel, and D.P. Grote.
\newblock Numerical methods for instability mitigation in the modeling of laser
  wakefield accelerators in a lorentz-boosted frame.
\newblock {\em Journal of Computational Physics}, 230(15):5908 -- 5929, 2011.

\bibitem{nuter2011}
R.~Nuter, L.~Gremillet, E.~Lefebvre, A.~L{\'e}vy, T.~Ceccotti, and P.~Martin.
\newblock Field ionization model implemented in particle in cell code and
  applied to laser-accelerated carbon ions.
\newblock {\em Physics of Plasmas}, 18(3):033107, 2017/01/20 2011.

\bibitem{umstadter1996}
D.~Umstadter, J.~K. Kim, and E.~Dodd.
\newblock Laser injection of ultrashort electron pulses into wakefield plasma
  waves.
\newblock {\em Phys. Rev. Lett.}, 76:2073--2076, Mar 1996.

\bibitem{perelomov1966}
A.~M. Perelomov, V.~S. Popov, and M.~V. Terent'ev.
\newblock Ionization of atoms in an alternating electric field.
\newblock {\em Sov. Phys. JETP}, 23:924, 1966.

\bibitem{perelomov1967}
A.~M. Perelomov, V.~S. Popov, and M.~V. Terent'ev.
\newblock Ionization of atoms in an alternating electric field: {II}.
\newblock {\em Sov. Phys. JETP}, 24:207, 1967.

\bibitem{ammosov1986}
M.~V. Ammosov, N.~B. Delone, and V.~P. Krainov.
\newblock Tunnel ionization of complex atoms and of atomic ions in an
  alternating electromagnetic field.
\newblock {\em Sov. Phys. JETP}, 64:1191, 1986.

\bibitem{mulser1998}
P.~Mulser, F.~Cornolti, and D.~Bauer.
\newblock Modeling field ionization in an energy conserving form and resulting
  nonstandard fluid dynamics.
\newblock {\em Physics of Plasmas}, 5(12):4466--4475, 2017/01/20 1998.

\bibitem{perez2012}
F.~P{\'e}rez, L.~Gremillet, A.~Decoster, M.~Drouin, and E.~Lefebvre.
\newblock Improved modeling of relativistic collisions and collisional
  ionization in particle-in-cell codes.
\newblock {\em Physics of Plasmas}, 19(8):083104, 2017/01/20 2012.

\bibitem{nanbu1997}
K.~Nanbu.
\newblock Theory of cumulative small-angle collisions in plasmas.
\newblock {\em Phys. Rev. E}, 55:4642--4652, Apr 1997.

\bibitem{nanbu1998}
K.~Nanbu and S.~Yonemura.
\newblock Weighted particles in coulomb collision simulations based on the
  theory of a cumulative scattering angle.
\newblock {\em Journal of Computational Physics}, 145(2):639 -- 654, 1998.

\bibitem{NRL}
J.~D. Huba.
\newblock {\em NRL plasma formulary}.
\newblock Office of Naval Research, Naval Research Laboratory (U.S.), 2013.

\bibitem{frankel1979}
N.~E. Frankel, K.~C. Hines, and R.~L. Dewar.
\newblock Energy loss due to binary collisions in a relativistic plasma.
\newblock {\em Phys. Rev. A}, 20:2120--2129, Nov 1979.

\bibitem{kim2000}
Yong-Ki Kim, Jos\'e~Paulo Santos, and Fernando Parente.
\newblock Extension of the binary-encounter-dipole model to relativistic
  incident electrons.
\newblock {\em Phys. Rev. A}, 62:052710, Oct 2000.

\bibitem{rohrlich1954}
F.~Rohrlich and B.~C. Carlson.
\newblock Positron-electron differences in energy loss and multiple scattering.
\newblock {\em Phys. Rev.}, 93:38--44, Jan 1954.

\bibitem{thevenet2016}
M.~Thevenet, A.~Leblanc, S.~Kahaly, H.~Vincenti, A.~Vernier, F.~Quere, and
  J.~Faure.
\newblock Vacuum laser acceleration of relativistic electrons using plasma
  mirror injectors.
\newblock {\em Nat. Phys.}, 12(4):355--360, 04 2016.

\bibitem{stuart1995}
B.~C. Stuart, M.~D. Feit, A.~M. Rubenchik, B.~W. Shore, and M.~D. Perry.
\newblock Laser-induced damage in dielectrics with nanosecond to subpicosecond
  pulses.
\newblock {\em Phys. Rev. Lett.}, 74:2248--2251, Mar 1995.

\bibitem{ristau2014}
D.~Ristau.
\newblock {\em Laser-Induced Damage in Dielectrics with Nanosecond to
  Subpicosecond Pulses}.
\newblock Taylor \& Francis Inc., 2014.

\bibitem{forslund1975}
D.~W. Forslund, J.~M. Kindel, and E.~L. Lindman.
\newblock Theory of stimulated scattering processes in laser-irradiated
  plasmas.
\newblock {\em Physics of Fluids}, 18(8):1002--1016, 1975.

\bibitem{cohen1979}
Bruce~I. Cohen and Claire~Ellen Max.
\newblock Stimulated scattering of light by ion modes in a homogeneous plasma:
  Space-time evolution.
\newblock {\em Physics of Fluids}, 22(6):1115--1132, 1979.

\bibitem{andreev2006}
A.~A. Andreev, C.~Riconda, V.~T. Tikhonchuk, and S.~Weber.
\newblock Short light pulse amplification and compression by stimulated
  brillouin scattering in plasmas in the strong coupling regime.
\newblock {\em Physics of Plasmas}, 13(5):053110, 2006.

\bibitem{weber2013}
S.~Weber, C.~Riconda, L.~Lancia, J.-R. Marqu\`es, G.~A. Mourou, and J.~Fuchs.
\newblock Amplification of ultrashort laser pulses by brillouin backscattering
  in plasmas.
\newblock {\em Phys. Rev. Lett.}, 111:055004, 2013.

\bibitem{chiaramello2016}
M.~Chiaramello, C.~Riconda, F.~Amiranoff, J.~Fuchs, M.~Grech, L.~Lancia, J.~R.
  Marqu{\`e}s, T.~Vinci, and S.~Weber.
\newblock Optimization of interaction conditions for efficient short laser
  pulse amplification by stimulated brillouin scattering in the strongly
  coupled regime.
\newblock {\em Physics of Plasmas}, 23(7):072103, 2016.

\bibitem{chiaramello2016b}
M.~Chiaramello, F.~Amiranoff, C.~Riconda, and S.~Weber.
\newblock Role of frequency chirp and energy flow directionality in the strong
  coupling regime of brillouin-based plasma amplification.
\newblock {\em Phys. Rev. Lett.}, 117:235003, 2016.

\bibitem{fuchs2014}
J.~Fuchs, A.~A. Gonoskov, M.~Nakatsutsumi, W.~Nazarov, F.~Qu{\'e}r{\'e}, A.~M.
  Sergeev, and X.~Q. Yan.
\newblock Plasma devices for focusing extreme light pulses.
\newblock {\em The European Physical Journal Special Topics},
  223(6):1169--1173, 2014.

\bibitem{wilson2016}
R.~Wilson, M.~King, R.~J. Gray, D.~C. Carroll, R.~J. Dance, C.~Armstrong, S.~J.
  Hawkes, R.~J. Clarke, D.~J. Robertson, D.~Neely, and P.~McKenna.
\newblock Ellipsoidal plasma mirror focusing of high power laser pulses to
  ultra-high intensities.
\newblock {\em Physics of Plasmas}, 23(3):033106, 2016.

\bibitem{lancia2010}
L.~Lancia, J.-R. Marqu\`es, M.~Nakatsutsumi, C.~Riconda, S.~Weber, S.~H\"uller,
  A.~Man\ifmmode \check{c}\else \v{c}\fi{}i\ifmmode~\acute{c}\else \'{c}\fi{},
  P.~Antici, V.~T. Tikhonchuk, A.~H\'eron, P.~Audebert, and J.~Fuchs.
\newblock Experimental evidence of short light pulse amplification using
  strong-coupling stimulated brillouin scattering in the pump depletion regime.
\newblock {\em Phys. Rev. Lett.}, 104:025001, Jan 2010.

\bibitem{lancia2016}
L.~Lancia, A.~Giribono, L.~Vassura, M.~Chiaramello, C.~Riconda, S.~Weber,
  A.~Castan, A.~Chatelain, A.~Frank, T.~Gangolf, M.~N. Quinn, J.~Fuchs, and
  J.-R. Marqu\`es.
\newblock Signatures of the self-similar regime of strongly coupled stimulated
  brillouin scattering for efficient short laser pulse amplification.
\newblock {\em Phys. Rev. Lett.}, 116:075001, Feb 2016.

\bibitem{hesse2013}
Michael Hesse, Nicolas Aunai, Seiji Zenitani, Masha Kuznetsova, and Joachim
  Birn.
\newblock Aspects of collisionless magnetic reconnection in asymmetric systems.
\newblock {\em Physics of Plasmas}, 20(6):061210, 2013.

\bibitem{nasaMMS2016}
First results from nasa's magnetospheric multiscale (mms) mission.
\newblock
  \url{http://agupubs.onlinelibrary.wiley.com/hub/issue/10.1002/(ISSN)1944-8007.NASAMMS1/},
  2016.
\newblock [Online; accessed 14 February 2016].

\bibitem{cassak2007}
P.~A. Cassak and M.~A. Shay.
\newblock Scaling of asymmetric magnetic reconnection: General theory and
  collisional simulations.
\newblock {\em Physics of Plasmas}, 14(10):102114, 2007.

\bibitem{kirk1999}
J~G Kirk and P~Duffy.
\newblock Particle acceleration and relativistic shocks.
\newblock {\em Journal of Physics G: Nuclear and Particle Physics}, 25(8):R163,
  1999.

\bibitem{spitkovsky2008}
Anatoly Spitkovsky.
\newblock Particle acceleration in relativistic collisionless shocks: Fermi
  process at last?
\newblock {\em The Astrophysical Journal Letters}, 682(1):L5, 2008.

\bibitem{haugbolle2011}
Troels Haugb{\o}lle.
\newblock Three-dimensional modeling of relativistic collisionless ion-electron
  shocks.
\newblock {\em The Astrophysical Journal Letters}, 739(2):L42, 2011.

\bibitem{sironi2013}
Lorenzo Sironi, Anatoly Spitkovsky, and Jonathan Arons.
\newblock The maximum energy of accelerated particles in relativistic
  collisionless shocks.
\newblock {\em The Astrophysical Journal}, 771(1):54, 2013.

\bibitem{weibel1959}
Erich~S. Weibel.
\newblock Spontaneously growing transverse waves in a plasma due to an
  anisotropic velocity distribution.
\newblock {\em Phys. Rev. Lett.}, 2:83--84, 1959.

\bibitem{bret2010}
A.~Bret, L.~Gremillet, and M.~E. Dieckmann.
\newblock Multidimensional electron beam-plasma instabilities in the
  relativistic regime.
\newblock {\em Physics of Plasmas}, 17(12):120501, 2017/01/20 2010.

\bibitem{grassi2016}
A.~Grassi, M.~Grech, F.~Amiranoff, F.~Pegoraro, A.~Macchi, and C.~Riconda.
\newblock Electron weibel instability in relativistic counterstreaming plasmas
  with flow-aligned external magnetic fields.
\newblock {\em Phys. Rev. E}, 95:023203, Feb 2017.

\bibitem{blandford1976}
R.~D. Blandford and C.~F. McKee.
\newblock Fluid dynamics of relativistic blast waves.
\newblock {\em Physics of Fluids}, 19(8):1130--1138, 1976.

\end{thebibliography}

\end{document}